\documentclass[preprint,prd,aps,amsmath,amssymb,superscriptaddress,floatfix]{revtex4}
\usepackage{graphicx}
\usepackage{dcolumn}
\usepackage{bm}

\usepackage{subfigure}
\usepackage{epsfig}
\usepackage{float}
\usepackage{enumurate}
\usepackage{tabularx} 

\usepackage{color}

\usepackage{txfonts}
\usepackage{morefloats}
\newcommand{\no}{\nonumber}

\graphicspath{{//Users/mkh/Documents/PhD_Thesis/UoMThesis/figures/Sky/} {//Users/mkh/Documents/Sky_paper/Fit/}{//Users/mkh/Documents/Sky_prod/}{//Users/mkh/Documents/Sky_paper/Min/}{//Users/mkh/Documents/Sky_paper/Spin/}}

\begin{document}


\title{Classically isospinning Skyrmion solutions}

\author{Richard A.~Battye}
\email{richard.battye@manchester.ac.uk}
\affiliation{%
Jodrell Bank Centre for Astrophysics, University of Manchester, Manchester M13 9PL, U.K.\\
}%
\author{Mareike Haberichter}%
 \email{m.haberichter@kent.ac.uk}

\author{Steffen Krusch}%
                                                                                                                                                                                                       
 \email{s.krusch@kent.ac.uk}

\affiliation{%
School of Mathematics, Statistics and Actuarial Science,\\ University of Kent, Canterbury CT2 7NF, U.K.\\
}%

\date{\today}

\begin{abstract}
We investigate how isospin affects the geometrical shape and energy of classical soliton solutions of topological charges $B=1-4,8$ in the Skyrme model. The novel approach in our work is that we study classically isospinning Skyrmions beyond the rigid-body approximation; that is, we explicitly allow the soliton solutions to deform and to break the symmetries of the static configurations. Our fully three-dimensional relaxation calculations reveal that the symmetries of isospinning Skyrme solitons can differ significantly from the ones of the static configurations. In particular, isospinning Skyrmion solutions can break up into lower-charge Skyrmions, can deform into new solution types that do not exist at vanishing angular frequency $\omega$ or energy degeneracy can be removed. These types of deformations have been largely ignored in previous work on modeling nuclei by quantized Skyrmion solutions.



\end{abstract}

\pacs{Valid PACS appear here}
\maketitle

\section{Introduction}

In the $S\!U(2)$ Skyrme model \cite{Skyrme:1961vq,Skyrme:1962vh}, atomic nuclei of nucleon (or baryon) number $B$ can be identified with soliton solutions of conserved topological charge $B$ which are known as Skyrmions. By numerically relaxing initial Skyrme configurations created with the rational map ansatz \cite{Houghton:1997kg} and its multilayer versions \cite{Feist:2012ps}, minimal-energy Skyrmion solutions have been constructed with various baryon numbers up to $B=108$ \cite{Battye:2001qn,Feist:2012ps}. In order to make contact with nuclear physics experiments it is necessary to semiclassically quantize these Skyrmion solutions. Traditionally this is done by treating Skyrmions as rigid bodies that can rotate in both space and isospace. This means that one quantizes just the rotational and isorotational zero modes  of each Skyrmion solution of a given $B$ and then determines the spin and isospin quantum numbers \cite{Irwin:1998bs,Krusch:2002by,Krusch:2005iq,Lau:2014sva} which are compatible with the symmetries of the static, classical soliton. This approach neglects any deformations and symmetry changes due to centrifugal effects. This rigid-body-type approximation resulted in qualitative and encouraging quantitative agreement with experimental nuclear physics data: for even $B$ the allowed quantum states for each Skyrmion often match the experimentally observed states of nuclei, and the energy spectra of a number of light nuclei have been reproduced  to a quite good degree of accuracy \cite{Manko:2007pr,Battye:2009ad}. However, nuclei of odd mass number are not well described by this approach. Many spin and isospin states do not appear in the right order or are not even predicted by the rigid-body quantization of the Skyrmion. One promising way to improve the agreement with experimental data is to allow Skyrmion solutions to deform when they spin and isospin. This can change the symmetries of the solutions and might result in different allowed quantum states \cite{Manko:2006dr}.

Classically isospinning soliton solutions have been studied recently in the Faddeev-Skyrme \cite{Battye:2013xf,Harland:2013uk} and the baby Skyrme model \cite{Battye:2013tka,Halavanau:2013vsa} beyond the rigid-body approximation. In the case of the fully $(3+1)$-dimensional Skyrme model a systematic, full numerical investigation of isospinning soliton solutions beyond the rigid-body approximation has not been performed yet. To our knowledge, numerical calculations of spinning Skyrmion configurations that take into account deformations originating from the kinematical terms have been carried out exclusively for baryon numbers $B=1$ \cite{Rajaraman:1985ty,Battye:2005nx,Houghton:2005iu,Fortier:2008yj} and $B=2$ \cite{Fortier:2008yj}. It is worth mentioning that Refs.~\cite{Rajaraman:1985ty,Battye:2005nx} consider classically spinning Skyrmions, whereas Refs.~\cite{Houghton:2005iu,Fortier:2008yj} first calculate the quantum Hamiltonian as a functional of the Skyrme fields and then minimize the Hamiltonian with respect to Skyrme fields for a given quantum state, see also Ref.~\cite{Jurciukonis:2013cda} for a related approach. It has been found that allowing for axial deformations drastically reduces the rotational energies and that in order to fit the energies of spinning Skyrmions to the nucleon and delta masses, the pion mass parameter $m_\pi$ of the Skyrme model has to be chosen much larger than its experimental value \cite{Battye:2005nx,Houghton:2005iu,Fortier:2008yj}. However, all these studies impose spherical or axial symmetry on the spinning Skyrme solitons to simplify the numerical computations. 

In this article, we perform numerical full field simulations of isospinning Skyrmion solutions with baryon numbers $B=1-4$ and $B=8$, without imposing any spatial symmetries. The main result of the present paper is that the symmetries and energies of  Skyrmion solutions at a given angular frequency $\omega$ and for a given mass value $\mu$  can significantly differ from the ones of the static soliton solutions. Classically isospinning Skyrmion solutions can break the symmetries of the static solutions and even split apart into lower charge solutions as the angular frequency increases further. A detailed study of the extent to which the inclusion of these classical deformations can result in an improved solitonic description of nuclei is beyond the scope of the present article and will form part of a forthcoming publication. 

This paper is organized as follows. Section~\ref{Sec_Sky} recalls briefly the $S\!U(2)$ Skyrme model and describes how we can construct isospinning Skyrmion solutions by solving energy minimization problems numerically. In Sec.~\ref{Sec_Sky_initial}, we review how suitable initial conditions for Skyrme configurations of a given nontrivial topological charge $B$ and of a specific symmetry $G$ can be created for our numerical relaxation simulations. By relaxing the initial Skyrme fields generated with the methods described in Sec.~\ref{Sec_Sky_initial} we find  Skyrme soliton solutions with topological charges $B=1-4,8$ for different values of the pion mass in Sec.~\ref{Sec_Skyrme_Static}. Then, in Sec.~\ref{Sec_Skyrme_Isospin} we investigate how the solitons' geometric shapes, energies, mean charge radii and critical frequencies are affected by the addition of classical isospin.  Furthermore, we show in Sec.~\ref{Sec_spin_induced} how the addition of classical isospin induces classical spin, consistent with the Finkelstein-Rubinstein constraints. This gives a better understanding why some of the isospinnning Skyrmion solutions constructed here tend to stay together, while others prefer to break up into lower-charge solutions. A brief summary and conclusion of our results is given in Sec.~\ref{Sec_Skyrme_Con}. For completeness, we explicitly list in the Appendix all diagonal and off-diagonal elements of the inertia tensors for the static $B=1-4,8$ Skyrmion solutions investigated in this article.

\section{Spinning and Isospinning Skyrmions}\label{Sec_Sky}

The Lagrangian density of the $(3+1)$-dimensional, massive Skyrme model \cite{Skyrme:1961vq,Skyrme:1962vh} is defined in $S\! U(2)$ notation by
\begin{align}
  \mathcal{L}&=-\frac{1}{2}\text{Tr}\left(R_\alpha R^\alpha\right)+\frac{1}{16}\text{Tr}\left([R_\alpha,R_\beta][R^\alpha,R^\beta]\right)+\mu^2\text{Tr}\left(U-1_2\right)\,,
\label{Lag_SU2}
\end{align}
where the Skyrme field $U(t,\boldsymbol{x})$ is an $S\! U(2)$-valued scalar, $R_\alpha=\left(\partial_\alpha U\right)U^\dagger$ its associated right-handed chiral current and $\mu$ is a rescaled pion mass parameter. In Lagrangian (\ref{Lag_SU2}) the energy and length units have been scaled away and  are given by  $F_\pi/4e$ and $2/eF_\pi$, respectively. Here $e$ is a dimensionless parameter and $F_\pi$ is the pion decay constant. The dimensionless pion mass $\mu$ is proportional to the tree-level pion mass $m_\pi$, explicitly $\mu=2m_\pi/eF_\pi$. Traditionally,  the Skyrme parameters $e$ and $F_\pi$ are calibrated so that the physical masses of the nucleon and delta resonance are reproduced when modeling them with a rigidly quantized Skyrmion solution, assuming the experimental value $m_\pi=138\,\text{MeV}$ for the pion mass \cite{Adkins:1983ya,Adkins:1983hy}. This approach yields the standard values $F_\pi=108\,\text{MeV}$, $e=4.84$ and $\mu=0.526$. Expressed in terms of standard ``Skyrme units'' \cite{Adkins:1983ya,Adkins:1983hy}, the energy  and length  units in Lagrangian (\ref{Lag_SU2}) are given by 5.58 MeV and 0.755 fm, respectively. Throughout this article  we consider pion values $\mu$ between 0.5 and 2. These parameter choices are motivated by Refs.~\cite{Battye:2005nx,Battye:2006tb,Manko:2007pr,Battye:2009ad} where it has been argued that a larger rescaled pion mass parameter $\mu$  (in particular, $\mu>0.526$) yields improved results when applying the Skyrme model to nuclear physics, while there are also studies which lead to lower values of $\mu$ see e.g. Ref.~\cite{Kopeliovich:2004pd}. In fact, the same reference \cite{Kopeliovich:2004pd} also considers the effect of a sixth order Skyrme term, which now has become important in so-called Bogomolny-Prasad-Sommerfield Skyrme models \cite{Adam:2013wya,Adam:2013tda}. 

Skyrmions arise as static solutions of minimal potential energy in the Skyrme model (\ref{Lag_SU2}). They can be characterized by their conserved, integer-valued topological charge $B$ which is given by the degree of the mapping $U: \mathbb{R}^3\rightarrow S\!U(2)$. To ensure fields have finite potential energy and a well-defined integer degree $B$ the Skyrme field $U(t,\boldsymbol{x})$ has to approach the vacuum configuration  $U(\boldsymbol{x})=1_2$  at spatial infinity for all $t$. Therefore, the domain can be formally compactified to a 3-sphere $S^3_{\text{space}}$ and the Skyrme field $U$ is then given by a mapping $S^3_{\text{space}}\rightarrow S\!U(2)\sim S^3_{\text{iso}}$ labeled by the topological invariant $B=\pi_3(S^3)\in\mathbb{Z}$. The topological degree $B$ of a static Skyrme soliton solution is explicitly given by
\begin{align}
B=\int \mathcal{B}(\boldsymbol{x})\,\text{d}^3x\,,
\label{Sky_bary}
\end{align}
where the topological charge density is defined by 
\begin{align}
\mathcal{B}(\boldsymbol{x})=-\frac{1}{24\pi^2} \epsilon_{ijk}\text{Tr}\left(R_iR_jR_k\right)\,.
\label{Sky_bary_dens}
\end{align}

When modeling atomic nuclei by spinning and isospinning Skyrmion solutions, the topological charge (\ref{Sky_bary}) can be interpreted as the mass number or baryon number of the configuration. Throughout this article, the energies $M_B$ of minimal-energy solutions in the Skyrme model will be given in units of $12\pi^2$, so that the Skyrme-Faddeev-Bogomolny lower energy bound \cite{Skyrme:1962vh,Faddeev:1976pg,Bogomolny:1975de} for a charge $B$ Skyrmion takes the form $M_B\geq |B|$. Recently a stronger lower topological energy bound has been derived in Refs.~ \cite{Harland:2013rxa,Adam:2013tga}.

Note that the $S\!U(2)$ field $U$ can be associated to the scalar meson field $\sigma$ and the pion isotriplet $\boldsymbol{\pi}=(\pi_1,\pi_2,\pi_3)$ of the $O(4)$ $\sigma$-model representation $\boldsymbol{\phi}=(\sigma,\boldsymbol{\pi})$ via
\begin{align}
U(\boldsymbol{x})&=\sigma(\boldsymbol{x}) 1_2+ i \boldsymbol{\pi}(\boldsymbol{x})\cdot\boldsymbol{\tau}\,,
\end{align}
where $\boldsymbol{\tau}$ denotes the triplet of standard Pauli matrices, and the unit vector constraint $\boldsymbol{\phi}\cdot\boldsymbol{\phi}=1$ has to be satisfied.

The Skyrme Lagrangian (\ref{Lag_SU2}) is manifestly invariant under translations in $\mathbb{R}^3$ and rotations in space and isospace. Classically spinning and isospinning Skyrmion solutions are obtained within the collective coordinate approach \cite{Adkins:1983ya,Adkins:1983hy}: the 6-dimensional space of zero modes -- the space of energy-degenerate Skyrmion solutions which only differ in their orientations in space and isospace -- is parametrized by collective coordinates which are then taken to be time dependent. 
 Here, we are mostly interested in static Skyrmion properties so that we can ignore translational degrees of freedom. Hence, the dynamical ansatz is given by
\begin{align}
\widehat{U}\left(t,\boldsymbol{x}\right)&=A(t)U_0(D(A^\prime (t)) \boldsymbol{x})A^\dagger (t)\,,
\label{Ansatz_dyn}
\end{align}
where the matrices $A,A^\prime\in$ $S\!U(2)$  are the collective coordinates describing isorotations and rotations around a static minimal-energy solution $U_0(\boldsymbol{x})$. Substituting (\ref{Ansatz_dyn}) in (\ref{Lag_SU2}) yields the effective Lagrangian
\begin{align}
L&=\tfrac{1}{2}\omega_{i} U_{ij}\omega_j+\tfrac{1}{2}\Omega_i V_{ij}\Omega_j-\omega_i W_{ij} \Omega_j -M_B\,,
\label{Lag_eff}
\end{align}
where $M_B$ is the classical Skyrmion mass given by
\begin{align}
M_B&=\int\left\{\left(\partial_i\boldsymbol{\phi}\cdot\partial_i\boldsymbol{\phi}\right)+\tfrac{1}{2}\Big[\left(\partial_i\boldsymbol{\phi}\cdot\partial_i\boldsymbol{\phi}\right)^2-\left(\partial_i\boldsymbol{\
\phi}\cdot\partial_j\boldsymbol{\phi}\right)^2\Big]+2\mu^2\left(1-\sigma\right)\right\}\,\text{d}^3x\,,
\label{Sky_mass}
\end{align}
and $\Omega_k=-i \text{Tr}\left(\tau_k\dot{A}^\prime A^{\prime\dagger}\right)$ and $\omega_k=-i \text{Tr}\left(\tau_kA^\dagger\dot{A}\right)$ are the rotational and isorotational angular velocities, respectively. The inertia tensors $U_{ij},V_{ij},W_{ij}$ are given explicitly by the integrals \cite{Kopeliovich:2001yg,Lau:2014sva}
\begin{subequations}\label{Sky_Inertia}
\begin{align}
U_{ij}&=2\int\Bigg\{\left(\pi_d\pi^d\delta_{ij}-\pi_i\pi_j\right)\left(1+\partial_k\boldsymbol{\phi}\cdot\partial_k\boldsymbol{\phi}\right)-\epsilon_{ide}\epsilon_{jfg}\left(\pi^d \partial_k\pi^e\right)\left(\
\pi^f\partial_k\pi^g\right)\Bigg\}\,{\text{d}}^3x\,,\label{U_phi_not}\\
V_{ij}&=2\int\epsilon_{ilm}\epsilon_{jnp}x_lx_n\Bigg(\partial_m\boldsymbol{\phi}\cdot\partial_p\boldsymbol{\phi}-\left(\partial_k\boldsymbol{\phi}\cdot\partial_m\boldsymbol{\phi}\right)\left(\partial_k\
\boldsymbol{\phi}\cdot\partial_p\boldsymbol{\phi}\right)\no\\
&\quad\quad\quad\quad\quad\quad\quad\quad+\left(\partial_k\boldsymbol{\phi}\cdot\partial_k\boldsymbol{\phi}\right)\left(\partial_m\boldsymbol{\phi}\cdot\partial_p\boldsymbol{\phi}\right)\
\Bigg)\,{\text{d}}^3x\label{V_phi_not}\,,\\
W_{ij}&=2\int\epsilon_{jlm}x_l\Bigg(\epsilon_{ide}\pi^d\partial_m\pi^e\left(1+\partial_k\boldsymbol{\phi}\cdot\partial_k\boldsymbol{\phi}\right)-\left(\partial_k\boldsymbol{\phi}\cdot\partial_m\boldsymbol{\phi}\right)\left(\epsilon_{ifg}\pi^f\partial_k\pi^g\right)\Bigg)\,{\text{d}}^3x\,.
\end{align}
\end{subequations}
Recall that the moments of inertia (\ref{Sky_Inertia}) are given in units of $1/e^3 F_\pi$, that is the mass scale multiplied by the square of the length scale. The conjugate body-fixed spin and isospin angular momenta ${\bf L}$ and ${\bf K}$ are given by \cite{Braaten:1988cc,Manko:2007pr}
\begin{align}
\label{LK}
L_i &= - W_{ij}^{T} \omega_j + V_{ij} \Omega_j\,,\\
K_i &= U_{ij} \omega_j - W_{ij} \Omega_j\,.
\end{align}

In this article, we focus on the construction of isospinning Skyrmion solutions and consequently (\ref{Lag_eff}) simplifies to
\begin{align}
L&=\tfrac{1}{2}\omega_{i} U_{ij}\omega_j-M_B\,.
\label{Lag_eff_iso}
\end{align}
Uniformly isospinning soliton solutions in Skyrme models are obtained by solving one of the following equivalent variational problems \cite{Harland:2013uk} for $\boldsymbol{\phi}$:
\begin{enumerate}
\item[(1)] Extremize the pseudoenergy functional $F_\omega\left(\boldsymbol{\phi}\right)=-L$ for fixed $\boldsymbol{\omega}$\,,
\item[(2)] Extremize the Hamiltonian $H=M_B+\frac{1}{2}K_iU_{ij}^{-1}K_j$ for fixed isospin $K_i=U_{ij}\omega_j$.
\end{enumerate}

In this paper, we will use a hybrid of approach (1) and (2). We are considering isospinning Skyrmions, in the sense that we seek stationary Skyrme configurations of the form (\ref{Ansatz_dyn}) with $A^\prime(t)$ constant, i.e. $\Omega = 0$. We fix the isospin ${\bf K}$ to be constant. Then we consider the energy
\begin{equation}
\label{EOmega0}
E = M_B + \frac{1}{2} \omega_i U_{ij} \omega_j\,,
\end{equation}
which implies
\begin{equation}
\label{omegaK}
K_i = U_{ij} \omega_j\,.
\end{equation}
Setting $\Omega = 0$ in (\ref{LK}) imposes a constraint on ${\bf L}$, namely
\begin{equation}
\label{omegaL}
L_i = -W_{ij}^T \omega_j = -W_{ij} U_{jk}^{-1} K_k\,.
\end{equation}
Hence, in our approach, if $W_{ij}$ is non-zero, then the configuration will obtain \emph{classical} spin. We will discuss this further in Sec.~\ref{Sec_spin_induced}.

We could now express the energy (\ref{EOmega0}) as a function of ${\bf K}$ and then minimise the energy $E$. However, it is more convenient to calculate $\omega$ using (\ref{omegaK}) and then minimize the pseudoenergy
\begin{equation}
\label{F}
F_\omega = M_B - \frac{1}{2} \omega_i U_{ij} \omega_j\,.
\end{equation}
The minus sign in equation (\ref{F}) is a consequence of the identity $$
\delta A^{-1} = - A^{-1} \delta A A^{-1}\,,
$$
where $\delta$ is a derivative and $A$ an invertible matrix. Since we only fix ${\bf K}$ but not ${\bf L}$, the value of $\omega$ is not conserved during the minimization. Hence for each step, we recalculate $\omega$ using (\ref{omegaK}). 

As a numerical minimisation we use the approach described in Ref.~\cite{Battye:2001qn} namely second order dynamics with a friction term. We rewrite the variational equations derived from (\ref{F}) in terms of the following \emph{modified Newtonian flow equations}
\begin{align}
M\ddot{\boldsymbol{\phi}}-\boldsymbol{\alpha}\left(\dot{\boldsymbol{\phi}},\partial_i\boldsymbol{\phi},\partial_i\dot{\boldsymbol{\phi}},\partial_i\partial_j\boldsymbol{\phi}\right)-\lambda\boldsymbol{\phi}+\epsilon\dot{\boldsymbol{\phi}}=0\,,
\label{Num_Flow}
\end{align}
where $M$ is a symmetric matrix, and we included the time dependence of the Skyrme Lagrangian  (\ref{Lag_SU2}) to evolve the equations of motion. The dissipation $\epsilon$ in (\ref{Num_Flow}) is added to speed up the relaxation process and the Lagrange multiplier $\lambda$ imposes the unit vector constraint $\boldsymbol{\phi}\cdot\boldsymbol{\phi}=1$. We do not present the full field equations here since they are cumbersome and not particularly enlightening. As initial field configurations, we take the static solutions at $\omega=0$ (see next section)  and increase the angular momentum $\boldsymbol{|K|}$ stepwise. Relaxed solutions at lower $|\boldsymbol{K}|$ serve as initial conditions for higher $|\boldsymbol{K}|$. In order to avoid precession effects in our numerical simulations the Skyrmion solutions have to be oriented in isospace so that their principal axes are aligned with the chosen isorotation axes. The initial configuration is then evolved according to the flow equations (\ref{Num_Flow}). Kinetic energy is removed periodically by setting $\dot{\boldsymbol{\phi}}=0$ at all grid points every 50th timestep. 

Most of our simulations are performed on regular, cubic grids of $(200)^3$ grid points with a lattice spacing $\Delta x=0.1$ and time step size $\Delta t=0.01$. Only for our relaxation calculations of charge-8 Skyrmion solutions do we choose cubic grids of $(201)^3$ grid points with the same lattice spacing $\Delta x=0.1$. The finite difference scheme used is fourth order accurate in the spatial derivatives. The dissipation is set to $\epsilon=0.5$. Most of our numerical simulations are performed for a range of mass values $\mu$.

\section{Initial Conditions}\label{Sec_Sky_initial}
We use the rational map ansatz \cite{Houghton:1997kg} to create approximate Skyrme fields of nontrivial topological charge $B$ and of given symmetry type $G$. Relaxing these initial field configurations with a fully three-dimensional numerical relaxation algorithm \cite{Battye:2001qn} we obtain static, minimum energy solutions of the Skyrme model with pion mass $\mu$ and baryon number $B$.

The main idea of the rational map ansatz is to approximate charge $B$ Skyrme configurations $\boldsymbol{\phi}$ which can be seen as maps from a 3-sphere in space to a 3-sphere in the target $S\!U(2)$ by rational maps $R:\,S^2\mapsto S^2$ of degree $B$. Within the rational map ansatz the angular dependence of the Skyrme field $\boldsymbol{\phi}$ is described by a rational function
\begin{align}
R(z)=\frac{p(z)}{q(z)}\,,
\end{align}
where $p$ and $q$ are polynomials in the complex Riemann sphere coordinate $z$. The $z$-coordinate can be expressed via standard stereographic projection, $z=\tan\left(\theta/2\right)e^{i\phi}$, in terms of the conventional spherical polar coordinates $\theta$ and $\phi$. The radial dependence is encoded in the radial profile function $f(r)$ which has to satisfy $f(0)=\pi$ and $f(\infty)=0$ to ensure a well-defined behavior at the origin and finite energy.

The rational function $R(z)$ takes values on the target $S^2$, and its value is associated via stereographic projection  with the Cartesian unit vector
\begin{align}
\widehat{\boldsymbol{n}}_R&=\frac{1}{1+|R|^2}\left(R+\overline{R},i\left(\overline{R}-R\right),1-|R|^2\right)\,.
\end{align}
The rational map approximation for the Skyrme field is given in terms of the isoscalar $\sigma$ and the pion isotriplet $\boldsymbol{\pi}$ of the nonlinear sigma model notation by
\begin{align}
\sigma&=\cos f(r),\quad\quad\boldsymbol{\pi}=\sin f(r)\,\widehat{\boldsymbol{n}}_{R(z)}\,.                      
\label{Sky_rat}
\end{align}

Substituting (\ref{Sky_rat}) in the Skyrme energy functional (\ref{Sky_mass}) results in an angular integral $\mathcal{I}$ which depends on the rational map $R(z)$  and a radial part only dependent on the monotonic function $f(r)$. To find low energy Skyrmion solutions of a given topological charge $B$ and pion mass $\mu$ one minimizes $\mathcal{I}$ over all maps of algebraic degree $B$ and then solves the Euler-Lagrange equation for $f(r)$ with $\mu$, $B$ and the minimized $\mathcal{I}$ occuring as parameters. As starting point for our numerical relaxations we choose initial Skyrme fields generated with the rational maps $R(z)$ given in Refs.~\cite{Houghton:1997kg,Battye:2001qn,Battye:2006na,Manko:2007pr}. Note that the rational maps given in these references are the optimal maps for Skyrmions with massless pions. However, since the angular integral $\mathcal{I}$ is independent of the mass parameter $\mu$, the same rational maps are also the minimizing maps for nonzero $\mu$ and the main effect of the pion mass is to change the shape function $f(r)$.

\section{Static Skyrmion Solutions with Baryon numbers $B=1-4, 8$ }\label{Sec_Skyrme_Static}

\begin{figure}[!htb]
\begin{center}
\includegraphics[totalheight=3.0cm]{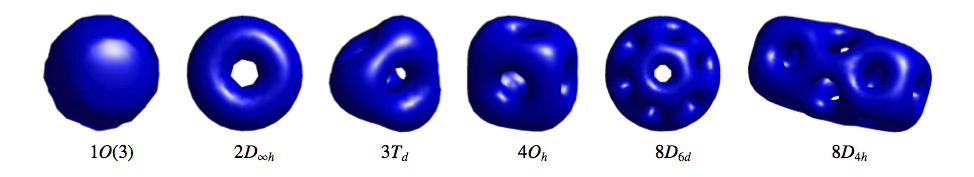}
\end{center}
\caption{Surfaces of constant baryon density (not to scale) of Skyrmion solutions with baryon number $B=1-4,\,8$ and with pion mass parameter $\mu$ set to 1. Each configuration is labeled by its baryon number and symmetry group.}
\label{Sky_bary_B1B8}
\end{figure}

In this section, we compute low energy static Skyrmion solutions with baryon numbers $B=1-4,8$ and with the rescaled pion mass set to $\mu=1$ by solving the full Skyrme field equations with a numerical three-dimensional relaxation algorithm \cite{Battye:2001qn}. Suitable initial Skyrme field configurations of given topological charge $B$ were created using the methods described in the previous section. For more detailed information on our relaxation procedure we refer the interested reader to the literature \cite{Battye:2001qn,Battye:2013xf} and to Sec.~\ref{Sec_Sky}. We list in Table~\ref{Tab_Stat_sky_low} the energies and the diagonal elements of the inertia tensors $U_{ij},V_{ij},W_{ij}$ for Skyrmions with baryon numbers $B=1-4, 8 $ of symmetry group $G$. All the corresponding off-diagonal elements and inertia tensor elements for different mass values are given in tabular form in the Appendix. The baryon density isosurfaces we obtained can be found in Fig.~\ref{Sky_bary_B1B8}. We can make a rough estimate of the numerical errors by computing the off-diagonal elements of the moment of inertia tensors \cite{Battye:2009ad}, which should be exactly zero for all the Skyrmion solutions investigated here. We find that in each case the off-diagonal entries are small and are of the order of $10^{-2}$ times the diagonal entries or less. In the following, all inertia tensor elements will be rounded to one decimal place.

\begin{table}[htb]
\caption{Skyrmions of baryon number $B=1-4,\,8$. We list the energies $M_B$, the energy per baryon $M_B/B$, the diagonal elements of the inertia tensors $U_{ij},V_{ij},W_{ij}$ and the symmetry group $G$ of the Skyrme solitons. Note that energies $M_B$ are given in units of $12\pi^2$ and that the mass parameter is chosen to be $\mu=1$. The calculated  configurations correspond to global energy minima for given baryon number $B$. For $B=8$ we are unable to decide within the limits of our numerical accuracy which configuration is of lower energy.}
\begin{tabularx}{\textwidth}{cXXXXXXXXXXXc}
\hline\hline
 $B$& $G$  & $M_B$   &$M_B/B$&$U_{11}$ &$U_{22}$ &$U_{33}$&$V_{11}$ &$V_{22}$ &$V_{33}$ &$W_{11}$ &$W_{22}$ &$W_{33}$\\
\hline
1& $O(3)$ &$1.415$  & 1.415& 47.5 &      47.5  &    47.5  &47.5 &47.5 &47.5&47.5 &47.5&47.5\\
2& $D_{\infty h}$  &$2.720$ &1.360  & 97.0 & 97.0&68.9 &153.8 &153.8&275.4 & 0.0 &0.0 & 137.7 \\
3& $T_d$           &$3.969$& 1.323 &124.1 &124.1 &124.1 &402.8 & 402.8 &402.8 &85.2 & 85.2 &85.2\\
4& $O_h$  &$5.177$ &1.294 & 148.2 & 148.2 & 177.4 & 667.6 &667.6 &667.6 & 0.0 & 0.0 & 0.3 \\
8 &$D_{6d}$&$10.235$&1.279& 296.3& 296.3 & 285.2 & 2261.4 &2261.4 &3036.3 &0.1 & 0.1 &137.9\\
&$D_{4h}$& 10.235&  1.279 & 298.4 & 292.1  &  326.9 &4093.9 & 4094.8 & 1381.3 & 0.1 &0.0 &0.1\\\hline\hline
\end{tabularx}
\label{Tab_Stat_sky_low}
\end{table}

\subsection{B=1}

The minimal-energy $B=1$ Skyrmion solution is spherically symmetric and substituting the rational map $R(z)=z$ in (\ref{Sky_rat}) reproduces the standard hedgehog form
\begin{align}
\sigma&=\cos f(r),\quad\quad\boldsymbol{\pi}=\sin f(r)\,\widehat{\boldsymbol{r}}\,,
\label{Sky_hedge}
\end{align}
where $r=|\boldsymbol{x}|$, $\widehat{\boldsymbol{r}} = \boldsymbol{x}/r$ and the radial profile function $f(r)$ satisfies the boundary conditions $f(0)=\pi$ and $f(\infty)=0$.

Solving the Skyrme field equation for a $B=1$ hedgehog Skyrmion (\ref{Sky_hedge}) gives an energy 
\begin{align}
M_1&=\frac{1}{3\pi}\int_0^\infty \left\{r^2f^{\prime 2}+2\sin^2f\left(1+f^{\prime 2 }\right)+\frac{\sin^4 f}{r^2}+2\mu^2\left(1-\cos f\right)r^2 \right\}\,\text{d}r=1.416\,.
\label{Sky_hedge_mass}
\end{align}
Here, we used the collocation method \cite{Ascher:1979,Ascher:1981} to determine the profile function $f(r)$ which minimizes $M_1$ (\ref{Sky_hedge_mass}). The rational map approximations for the higher charge Skyrmion solutions will be generated with the profile function $f(r)$ calculated in the charge-1 sector. Note that for an $O(3)$-symmetric Skyrme configuration (\ref{Sky_hedge}) the inertia tensors (\ref{Sky_Inertia}) take the simple form  $U_{ij}=V_{ij}=W_{ij}=\Lambda\delta_{ij}$ with 
\begin{align}
\Lambda=\frac{16\pi}{3}\int r^2\sin^2f\left(1+f^{\prime2}+\frac{\sin^2f}{r^2}\right)\,\text{d}r=47.625\,,
\label{Sky_hedge_lambda}
\end{align}
in agreement with the moment of inertia $\Lambda=47.623$ calculated within the hedgehog approximation in Ref.~\cite{Lau:2014sva}. 

For comparison, the energy value $M_1=1.415$ calculated in our full 3D simulation differs by $0.07\%$ compared to the greater accuracy hedgehog ansatz (\ref{Sky_hedge}).  Within our numerical accuracy the inertia tensors (\ref{Sky_Inertia}) computed with our 3D relaxation code are all found to be proportional to the unit matrix with $\Lambda = 47.5$. This is in reasonable agreement with the value calculated within the hedgehog ansatz (\ref{Sky_hedge_lambda}) and with the moments of inertia $U_{ij}=V_{ij}=W_{ij}=\Lambda\delta_{ij}$ with  $\Lambda= 47.5$ computed in Ref.~\cite{Lau:2014sva} using a three-dimensional nonlinear conjugate gradient method. 

Note that we cannot confirm the energy value $M_1=1.465$ calculated in the recent article \cite{Feist:2012ps} for a $B=1$ Skyrmion of mass $\mu=1$. We double checked our results using two very diffent numerical approaches (a collocation method \cite{Ascher:1979,Ascher:1981} and a one-dimensional gradient flow method) to solve numerically the Skyrme field equation for the hedgehog ansatz (\ref{Sky_hedge}). In both cases we obtain an energy value $M_1=1.416$ \footnote{The authors in Ref.~\cite{Feist:2012ps} adjusted their numerical energy values to compensate for lattice spacing and box sizes. This resulted in an energy shift by $\approx 0.05$ for all their energy values. After redoing their numerics they confirmed our energy value $1.415$ for the $B=1$ Skyrmion with mass parameter $\mu$ set to 1 (private communication).}. 

\subsection{B=2}

The $B=2$ Skyrmion has toroidal symmetry $D_{\infty h}$, and it can be approximated by choosing the rational map $R(z)=z^2$ in (\ref{Sky_rat}). Relaxing this rational map generates an initial Skyrme field configuration. We verify that all inertia tensors are diagonal, with $U_{11}=U_{22}= 97.0$, $V_{11}=V_{22}=153.8$ and $W_{11}=W_{22}=0$. Our numerically calculated charge-2 configuration satisfies the relation $U_{33}=\frac{1}{2}W_{33}=\frac{1}{4}V_{33}$ \cite{Braaten:1988cc,Houghton:2005iu} -- a manifestation of the axial symmetry. The soliton's energy is $M_2=2.720$, which is reasonably close to the energy value $M_2=2.77$ given in Ref.~\cite{Feist:2012ps}. Note  that the moments of inertia $U_{33}=68.9, V_{33}=275.4$ and $W_{33}=137.7$ are found to be in close agreement with the corresponding values $U_{33}=68.67, V_{33}=274.59$ and $W_{33}=137.31$ stated in Ref.~\cite{Lau:2014sva}.

 Finally, we can check the results of our fully three-dimensional numerical relaxation with those obtained by minimizing the two-dimensional, total energy functional of an axially symmetric Skyrme configuration. An axially symmetric ansatz \cite{Krusch:2004uf} is given by
\begin{align}
  \sigma&=\psi_3,\quad\quad\pi_1=\psi_1 \cos n\theta,\quad\quad\pi_2=\psi_1 \sin n\theta,\quad\quad\pi_3=\psi_2\,,
\label{Sky_axial}
\end{align}
where $\boldsymbol{\psi}(\rho,z)=(\psi_1,\psi_2,\psi_3)$ is a unit vector that is dependent on the cylindrical coordinates $\rho$ and $z$. Here, the non-zero winding number $n\in \mathbb{Z}$ counts the windings of the Skyrme fields in the $(x_1,x_2)$-plane. Substituting (\ref{Sky_axial}) in (\ref{Sky_mass}) results in the classical soliton mass \cite{Kopeliovich:1987bt,Braaten:1988cc,Battye:2005nx,Fortier:2008yj}
\begin{align}\label{Axial_mass}
M_B&=2\pi\int_0^\infty{\rm{d}}\rho\int_{-\infty}^{+\infty}{\rm{d}}z\, \rho\left\{{\left(\partial_\rho{\boldsymbol\psi}\cdot\partial_\rho\boldsymbol\psi+\partial_z\boldsymbol\psi\cdot\partial_z\boldsymbol\psi\right)}\left(1+{\frac{n^2}{\rho^2}\psi_1^2}\right)\right.\\
&\left.\quad+{\left|\partial_\rho\boldsymbol\psi\times\partial_z\boldsymbol\psi\right|^2}+{\frac{n^2}{\rho^2}\psi_1^2}+2\mu^2\left(1-\psi_3\right)\right\}\,,\no
\end{align}
and the baryon number $B$  of an axially symmetric configuration $\boldsymbol{\psi}$ is given by  substituting (\ref{Sky_axial}) in (\ref{Sky_bary}) to obtain 
\begin{align}
B=\frac{n}{\pi}\int_0^\infty{\rm{d}}\rho\int_{-\infty}^{+\infty}{\rm{d}}z\Big\{\psi_1\boldsymbol\psi\left|\partial_\rho\boldsymbol\psi\times\partial_z\boldsymbol\psi\right|\Big\}\,.
\end{align}
To ensure a configuration of finite energy $M_B$ the unit vector $\boldsymbol{\psi}$ has to satisfy the boundary condition $\boldsymbol{\psi}\rightarrow (0,0,1)$ as $\rho^2+z^2\rightarrow\infty$ together with $\psi_1=0$ and $\partial_\rho\psi_2=\partial_\rho\psi_3=0$ at $\rho=0$. A suitable start configuration with baryon number $B=n$ is given in Ref.~\cite{Krusch:2004uf} by 
\begin{align}
\psi_1&=\frac{\rho}{r}\sin f(r)\,,\quad\quad\psi_2=\frac{z}{r}\sin f(r)\,,\quad\quad\psi_3=\cos f(r)\,,
\end{align}
where $r=\sqrt{\rho^2+z^2}$ and $f(r)$ denotes, as usual, a monotonically decreasing profile function satisfying the boundary conditions $f(0)=\pi$ and $f(\infty)=0$. For $B=2$ we minimize the static energy functional (\ref{Axial_mass}) with a simple gradient flow algorithm on a rectangular grid in the $(\rho,z)$- plane. The grid contains $(401)^2$ points with a lattice spacing $\Delta x=0.05$, so that the range covered is $(\rho,z)\in[0,20]\times[-10,10]$. Our two-dimensional relaxation calculation results in $M_2=2.720$ which agrees with our 3D results. The nonvanishing components of the isospin inertia tensor $U_{ij}$ (\ref{U_phi_not}) are given for the axially symmetric ansatz (\ref{Sky_axial}) by Refs. \cite{Kopeliovich:1988np,Braaten:1988cc,Fortier:2008yj}
\begin{subequations}
\label{inertia_axial}
\begin{align}
U_{33}&=4\pi \int_0^\infty{\rm{d}}\rho\int_{-\infty}^{+\infty}{\rm{d}}z\, \rho\left\{\psi_1^2\left(\partial_\rho{\boldsymbol\psi}\cdot\partial_\rho\boldsymbol\psi+\partial_z\boldsymbol\psi\cdot\partial_z\boldsymbol\psi+1\right)\right\}\,,\label{inertia_axial_33}\\
U_{11}&=U_{22}=2\pi  \int_0^\infty{\rm{d}}\rho\int_{-\infty}^{+\infty}{\rm{d}}z\, \rho\left\{\psi_1^2+2\psi_2^2+\left(\partial_\rho\psi_3\right)^2+\left(\partial_z\psi_3\right)^2\right.\\
&\quad\left.+\left(\partial_\rho\boldsymbol\psi\cdot\partial_\rho\boldsymbol\psi+\partial_z\boldsymbol\psi\cdot\partial_z\boldsymbol\psi+n^2\frac{\psi_1^2}{\rho^2}\right)\psi_2^2+n^2\frac{\psi_1^4}{\rho^2}\right\}\,.\no
\end{align}
\end{subequations}
Our 2D gradient flow simulation gives $U_{11}=U_{22}=103.1$ and $U_{33}=71.5$ for an axially symmetric charge-2 Skyrmion solution of mass $\mu=1$. Note that classically bound toroidal states have been found in Ref.~\cite{Kopeliovich:1987bt} also for the case of Skyrme solitons stabilized by a term proportional to the baryon density squared, i.e. by the sixth order term. 

\subsection{B=3}

The minimal-energy $B=3$ Skyrmion has $T_d$ symmetry and can be created with the rational map \cite{Manko:2007pr}
\begin{align}
R(z)&=\frac{\sqrt{3}iz^2-1}{z^3-\sqrt{3}iz}\,.
\label{Rat_B3}
\end{align}
Relaxing the tetrahedrally symmetric $B=3$ Skyrme configuration (\ref{Rat_B3}), we verify that the inertia tensors (\ref{Sky_Inertia}) are all diagonal: $U_{ij}=u\delta_{ij}$, $V_{ij}=v\delta_{ij}$ and $W_{ij}=w\delta_{ij}$ with $u=124.1$, $v=402.8$ and $w=85.2$. We find for the total energy $M_3=3.969$ which is slightly lower than the numerical value $M_3=4.02$ given in Ref.~\cite{Feist:2012ps}.

\subsection{B=4}

The minimal-energy Skyrmion solution with $B=4$ has octahedral symmetry $O_h$ and can be approximated by the rational map \cite{Manko:2007pr}
\begin{align}
R(z)&=\frac{z^4+2\sqrt{3}iz^2+1}{z^4-2\sqrt{3}iz^2+1}\,.
\label{Rat_B4}
\end{align}
The inertia tensors $U_{ij}$, $V_{ij}$ and $W_{ij}$ for the cubically symmetric, numerically relaxed charge-4  configuration (\ref{Rat_B4}) are determined to be diagonal, satisfying $U_{11}=U_{22}=148.2$ and $V_{ij}=v\delta_{ij}$ with $v=667.6$ and with the cross term $W_{ij}$ vanishing within the limits of our numerical accuracy. We obtain for the total energy $M_4=5.177$ which agrees with the value $M_4=5.18$ stated in Ref.~\cite{Feist:2012ps}.

\subsection{B=8}
For baryon number $B=8$, we calculate two very different Skyrmion solutions  -- one with $D_{4h}$ and the other with $D_{6d}$ symmetry. We confirm the results in Ref.~\cite{Battye:2006na}. For $\mu=1$ the energies of both solutions are indeed identical up to four significant digits with $M_8=10.2356$. Within the limits of our numerical accuracy we are unable to conclude which of the two soliton solutions has the lower energy value. 

To generate suitable initial conditions for our numerical relaxation simulations  we approximate an initial field configuration with $D_{6d}$ symmetry by the rational map \cite{Manko:2007pr}
\begin{align}
R(z)&=\frac{z^6-ia}{z^2\left(iaz^6-1\right)}\,,
\label{Rat_B8_D6d}
\end{align}
where the free parameter is set to $a=0.14$. The relaxed Skyrme field resembles a hollow polyhedron, namely a ring of twelve pentagons with a hexagon at the top and at the bottom (see baryon density isosurface plot in Fig.~\ref{Sky_bary_B1B8}). The inertia tensors of the relaxed $D_{6d}$-symmetric Skyrme configuration satisfy $U_{11}= U_{22}= 296.3$, $U_{33}= 285.2$ and  $V_{11}= V_{22}= 2261.4$, $V_{33}= 3036.3$.

 Recall that in the massive pion model there exists not only a $D_{6d}$ symmetric, polyhedral Skyrmion solution but also a bound configuration of two $B=4$ cubes \cite{Battye:2006na}. This double cube Skyrmion solution is obtained by relaxing a perturbed, $D_{4h}$-symmetric starting configuration approximated with the rational map \cite{Manko:2007pr}
\begin{align}
R(z)&=\frac{z^8+bz^6-az^4+bz^2+1}{z^8-bz^6-az^4-bz^2+1}\,,
\label{Rat_B8_D4h}
\end{align}
where we choose $a=5$ and $b=2$. The numerically relaxed Skyrme configuration can be thought of being constructed from two touching cubic $B=4$ Skyrmions with one of the cubes twisted by $90^\circ$ relative to the other  around the axis joining them. We confirm that our computed isospin inertia tensors with $U_{11}=298.4$, $U_{22}=292.1$ and $U_{33}=326.9$ approximately satisfy the double cube approach ($U^{(B=8)}_{11}= U^{(B=8)}_{22}= 2U^{(B=4)}_{11}$, $U^{(B=8)}_{33}= 2U^{(B=4)}_{33}$) which was used in Ref.~\cite{Manko:2007pr} to determine the allowed spin, isospin and parity states for the $B=8$ Skyrmion within the rigid-body approach. Furthermore, our computed inertia tensors ($U_{11}=298.4$, $U_{22}=292.1$, $U_{33}=326.9$,$V_{11}= V_{22}=  4093.9$, $V_{33}=1381.3$, $W_{ij}=0$)  are in reasonable agreement with the values $U_{11}=299$, $U_{22}=291, U_{33}=326$, $V_{11}=V_{22}=4052, V_{33}=1390$ given in Ref.~\cite{Battye:2009ad}. Note that our energy value for the dihedral $B=8$ solution (\ref{Rat_B8_D4h}) is 0.9 \% lower than the one stated in Ref.~\cite{Feist:2012ps}.

\section{Numerical Results on Isospinning Skyrme Solitons}\label{Sec_Skyrme_Isospin} 

In this section, we investigate how the inclusion of isospin affects the geometrical shape and the total energy  of the classical Skyrmion solutions with baryon numbers $B=1-4,\,8$  computed in the previous section. Recall that in our numerical simulations we \emph{do not} impose any spatial symmetries on the isospinning Skyrme soliton solutions and we \emph{do not} assume that the solitons' shape is independent of the angular frequency $\omega$. Analogous calculations of isospinning solitons in the Skyrme-Faddeev model \cite{Faddeev:1975,Faddeev:1976pg} that go beyond the rigid-body-type approximation have been performed in Refs.~\cite{Battye:2013xf,Harland:2013uk}. 

We construct stationary isospinning soliton solutions by numerically solving the energy minimization problem formulated in Sec.~\ref{Sec_Sky}. Note that spinning Skyrmions with zero pion mass radiate away their energy, see Ref.~\cite{Schroers:1993yk} for a detailed discussion. For pion mass $\mu>0$ stationary solutions exist up to an angular frequency $\omega_{{\rm crit}} = \mu$. At $\omega_{\text{crit}}$ the values of the energy and angular momentum are finite, and therefore, the corresponding angular momenta $K_{\text{crit}}$ (and $L_{\text{crit}}$) is also finite, see Ref.~\cite{Battye:2005nx}. The situation is different for baby Skyrmions where energy and moment of inertia diverge at $\omega_{{\rm crit}}$ \cite{Piette:1994mh,Battye:2013tka,Halavanau:2013vsa} for $\mu<1$. From the point of view of numerics, this behavior is challenging. For $\omega < \omega_{\text{crit}}$ the problem is well posed, whereas for $\omega>\omega_{\text{crit}}$ the solutions become oscillatory which is difficult to detect in a finite box. Physically, this corresponds to pion radiation, and the fact that stationary solutions do not exist. Numerically, we can find energy minimizers for $\omega > \omega_{\text{crit}}$ but this is an artefact of the finite box approximation. By convention, throughout this paper, when displaying inertia tensors and energies as a function of isospin, we will cut our graphs at the critical isospin value $K_{\text{crit}}$.

\begin{figure}[!htb]
\centering
\includegraphics[totalheight=4.0cm]{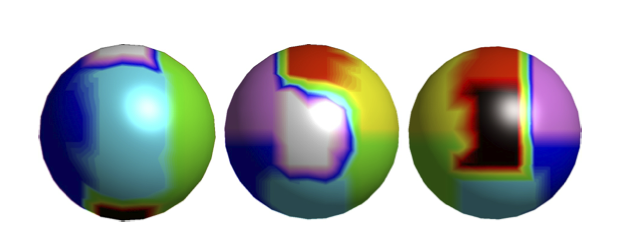}
\caption{Three different views of the baryon isodensity of a $B=1$ Skyrmion. The orientation in isospace is visualized using the field colouring scheme given in Refs.~\cite{Manton:2011mi,Feist:2012ps,Battye:2006na}.}
\label{Colourscheme_Manton}
\end{figure}

Recall that the different orientations of Skyrme solitons in isospace can be visualized using Manton and Sutcliffe's field colouring scheme described in detail in Ref.~\cite{Manton:2011mi}. We illustrate the colouring for a $B=1$ Skyrmion solution in Fig.~\ref{Colourscheme_Manton}: The points where the normalised pion isotriplet $\widehat{\boldsymbol{\pi}}$ takes the values $\widehat{\pi}_1=\widehat{\pi}_2=0$ and $\widehat{\pi}_3=+1$ are shown in white and those where $\widehat{\pi}_1=\widehat{\pi}_2=0$ and $\widehat{\pi}_3=-1$ are colored black.
The red, blue and green regions indicate where $\widehat{\pi}_1+i\widehat{\pi}_2$ takes the values $1$, $e^{2\pi i/3}$, $e^{4\pi i/3}$, respectively and the associated complementary colors in the RGB color scheme (cyan, yellow and magenta) show the segments where $\widehat{\pi}_1+i\widehat{\pi}_2=-1, e^{5\pi i/3}, e^{\pi i/3}$. 

Note that all our calculations are carried out in dimensionless Skyrme units which are related to natural units ($\hbar=c=1$) by rescaling the units of energy by $F_\pi/4e$ and those of length by $2/eF_\pi$, where the parameters $F_\pi$ and $e$ are the pion decay constant and the dimensionless Skyrme constant. The moment of inertia is given in units of $1/e^3F_\pi$ and consequently the isorotational (quantum) energy contributions are expressed in units of $e^3F_\pi$. Here, we are purely interested in classically isospinning Skyrmion solutions, hence we will use the classical energy scale  $F_\pi/4e$ to calculate the isosrotational energy contributions in Sec.~\ref{Sec_Skyrme_Con}. 
To estimate the value of $\hbar$ \cite{Leese:1994hb} for different Skyrme parameter sets we used  $\hbar=197.3\,\mbox{MeV fm}$. Thus, for the standard choice of Skyrme parameters we can relate Skyrme units to conventional units via $F_\pi/4e=5.58\,\mbox{MeV}$ and $2/eF_\pi=0.755\,\mbox{fm}$. It follows that $$\hbar=46.8\left(\frac{F_\pi}{4e}\right)\left(\frac{2}{eF_\pi}\right)\,.$$ Hence in standard Skyrme units we have $F_\pi/4e=2/eF_\pi=1$ and $\hbar=46.8$. 
Different parameter sets like the ones suggested by Ref.~\cite{Battye:2005nx} ($e=4.90$, $F_\pi=90.5\,\text{MeV}$) and by \cite{Manko:2007pr} ($e=3.26$, $F_\pi=75.2\,\text{MeV}$) yield $\hbar=48.1$ and $\hbar=21.3$, respectively.

\subsection{Lower Charge Skyrmions: $1\leq B\leq4$}\label{Sec_low}
In this section, we present our numerical results on isospinning soliton solutions of topological charges $B=1-4$. Our numerical simulations are performed for a range of mass values $\mu$.

\subsubsection{$B=1$}

\begin{figure}[!htb]
\centering
\subfigure[\,Total energy vs angular frequency ]{\includegraphics[totalheight=6.cm]{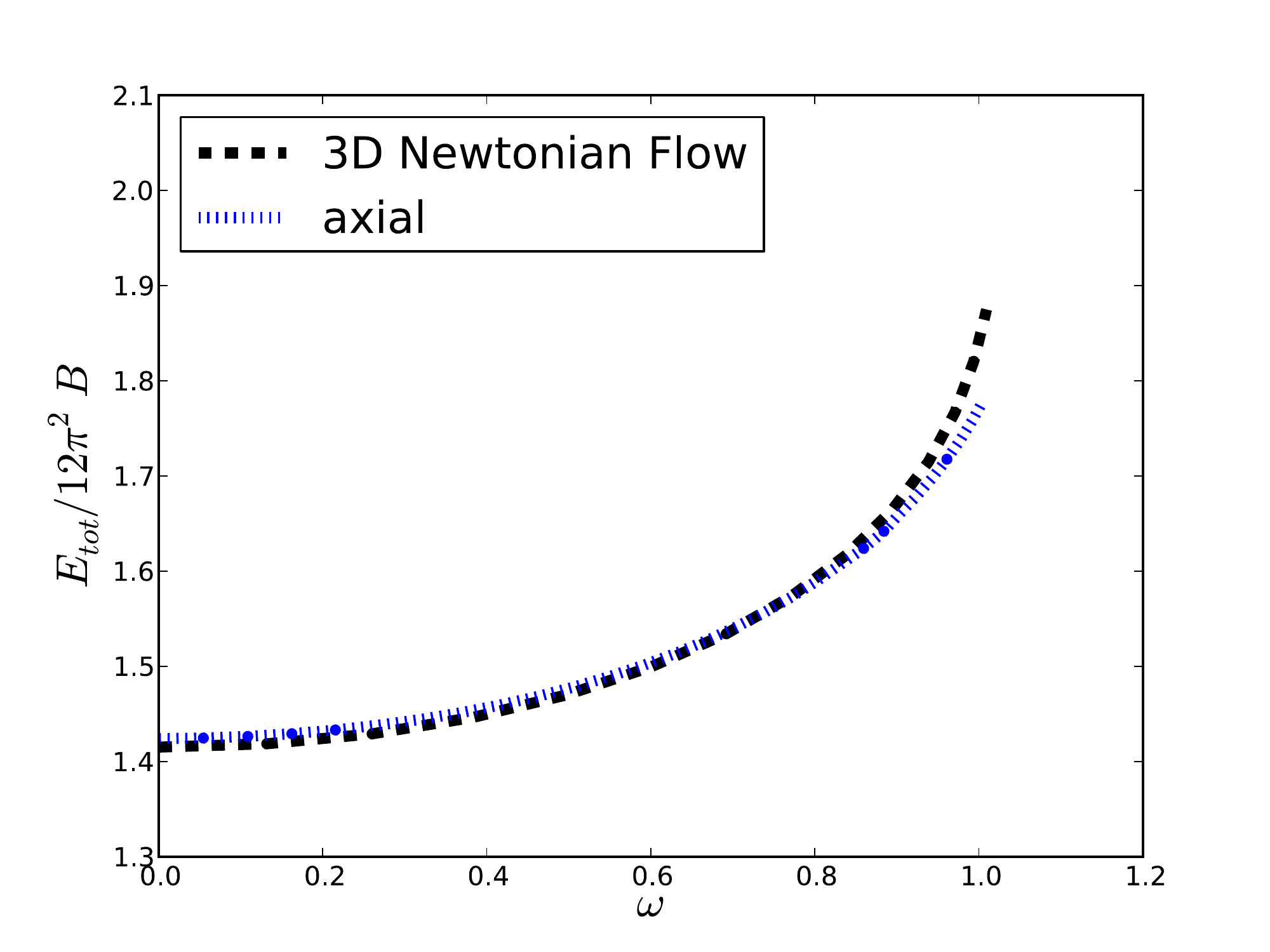}}
\subfigure[\,Mass-Spin relationship ]{\includegraphics[totalheight=6.cm]{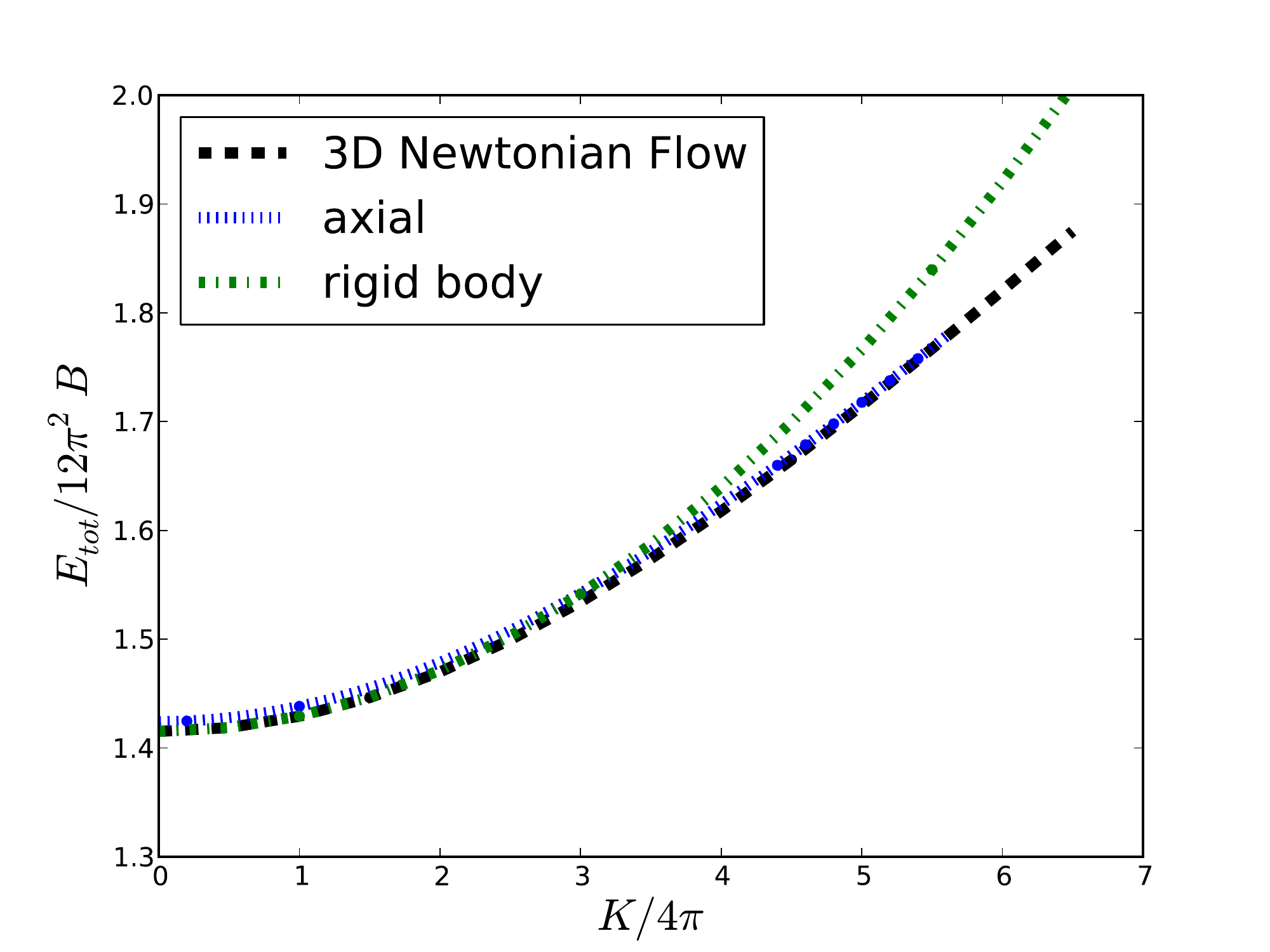}}\\
\subfigure[\,Inertia vs angular frequency]{\includegraphics[totalheight=6.cm]{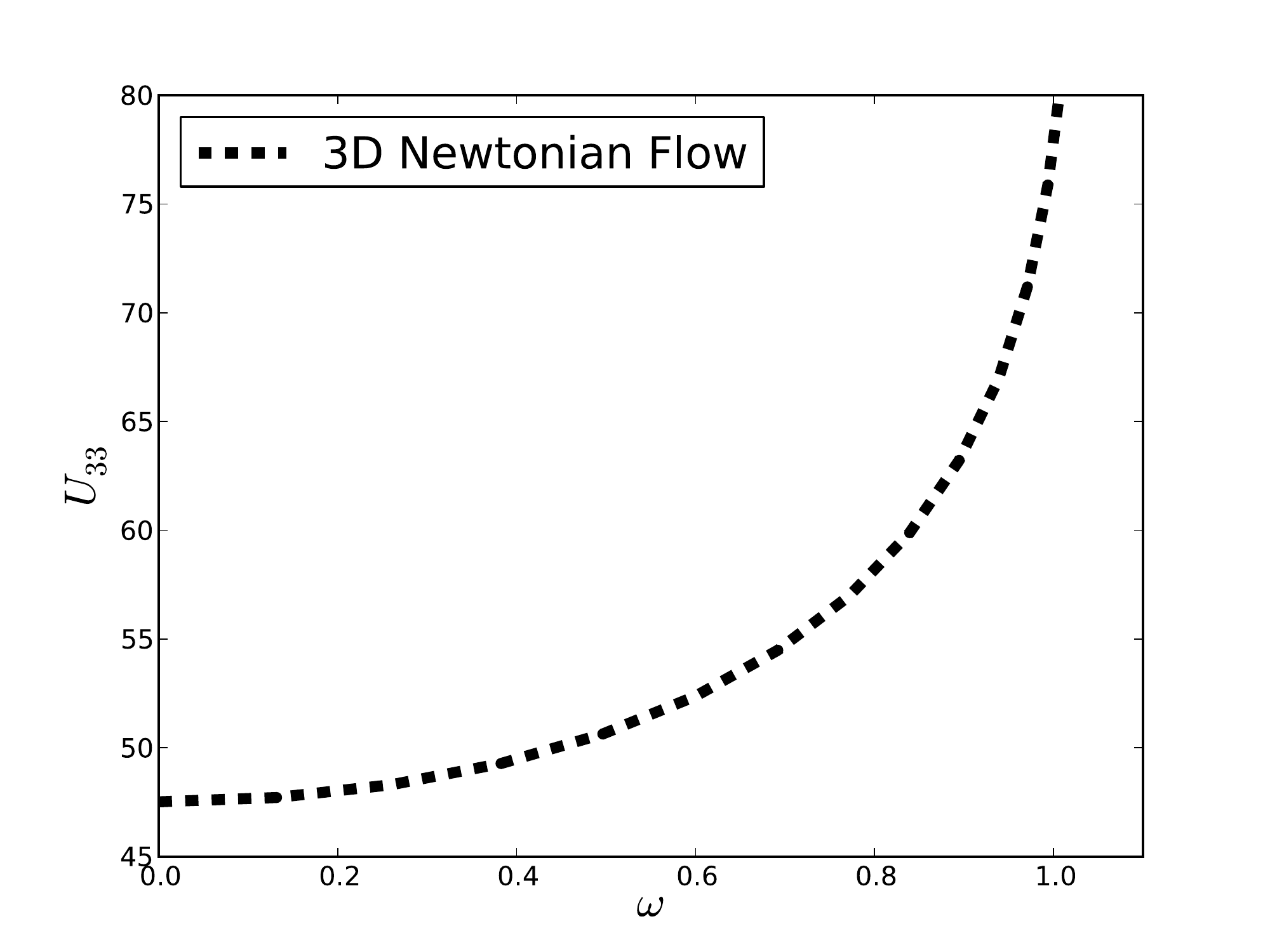}}
\subfigure[\,Inertia-Spin relationship]{\includegraphics[totalheight=6.cm]{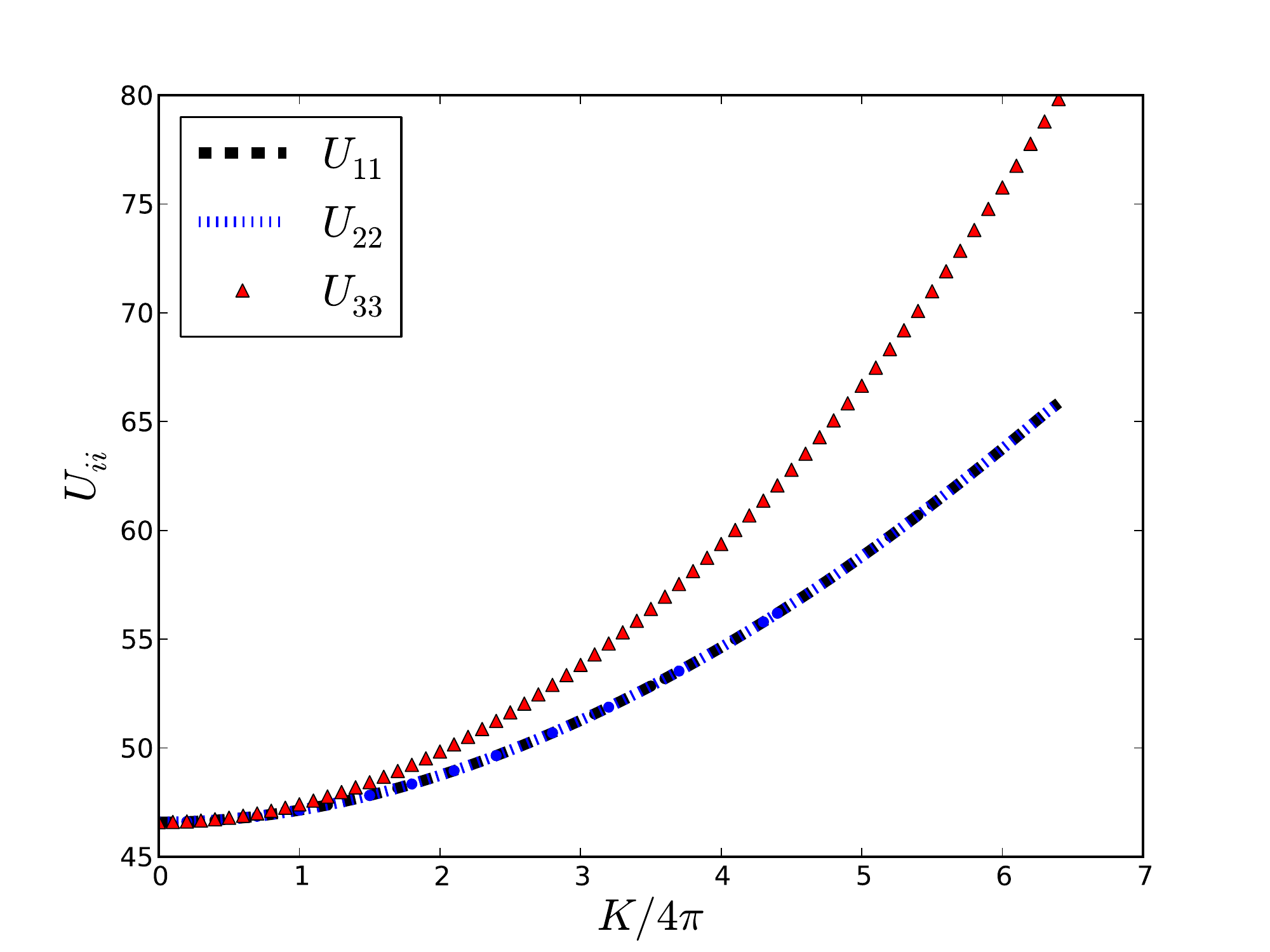}}
\caption[Total energy and inertia as function of $\omega$ and $K$ for isospinning $B=1$ Skyrmions.]{Isospinning $B=1$ Skyrmion $(\mu=1)$. A suitable start configuration of topological charge $1$ is numerically minimized using 3D modified Newtonian flow on a $(200)^3$ grid with a lattice spacing of $\Delta x=0.1$ and a time step size $\Delta t=0.01$. We choose the $z$ axis as our isorotation axis. Our results are compared with those obtained assuming an axially symmetric deforming $B=1$ Skyrme configuration (\ref{Sky_axial}). Furthermore, we include the energy curve for a rigidly isorotating Skyrme configuration.}
\label{Fig_B1_O3}
\end{figure}

For the $O(3)$ symmetric charge-1 Skyrmion solution we choose the $z$ axis as the axis of isorotation. This particular choice is motivated by Ref.~\cite{Manton:2011mi}, where it was argued that spin-polarised protons and neutrons are best modeled by hedgehog $B=1$ Skyrmions classically spinning relative to the $z$ axis. Note that in our numerical simulations  the $B=1$ Skyrmion is chosen to be in its standard position and orientation, that is the white-black axis (see Fig.~\ref{Colourscheme_Manton}) coincides with the isorotation axis, with white up and black down. The results of our fully three-dimensional numerical relaxation calculations of isospinning $B=1$ Skyrmion solutions with mass parameter $\mu=1$ are shown in Fig.~\ref{Fig_B1_O3}. The relaxation calculations have been performed on a $(200)^3$ grid with lattice spacing $\Delta x=0.1$ and a time step size $\Delta t=0.01$.

We verify that for $\mu=1$ the $E_{\text{tot}}(\omega)$ graph (calculated with our modified 3D Newtonian flow) shows the same behavior as predicted by an axially symmetric spinning, charge-1 Skyrme configuration (\ref{Sky_axial}).  As discussed in Sec.~\ref{Sec_Sky}, in our 3D simulations the soliton's energy is given by $E_\text{tot}(\omega)=M_1+\omega^2U_{33}/2$ and the energy curve terminates at $\omega_{\text{crit}} = \mu = 1$. Stable, internally spinnning solutions cease to exist beyond this critical value, but energy and moments of inertia remain finite at $\omega_{\text{crit}} = 1$. 

 For comparison, axially symmetric deforming, isospinning configurations are constructed by minimizing the total energy $E_{\text{tot}}=M_1+K^2/\left(2U_{33}\right)$ for fixed isospin $K$ with a 2D gradient flow algorithm, where the classical soliton mass $M_1$ is given by (\ref{Axial_mass}) and the relevant moment of inertia $U_{33}$ can be found in (\ref{inertia_axial_33}). Both energy curves agree within the limits of our numerical accuracy [see Fig.~\ref{Fig_B1_O3}(a)]. In Fig.~\ref{Fig_B1_O3}(b), we display the mass-isospin relationship $E_{\text{tot}}(K)$ for an isospinning $B=1$ Skyrmion solution calculated without imposing any symmetry constraints on the configuration. Imposing axial symmetry, we reproduce the same mass-isospin relation. As shown in Fig.~\ref{Fig_B1_O3}(b), the rigid-body formula proves to be a good approximation for small isospins ($K\le 3\times 4\pi$), whereas for higher isospin values $E_{\text{tot}}(K)$ deviates from the quadratic behavior. At the critical angular frequency $\omega_{\text{crit}}=1$ ($K_{\text{crit}}= 6.5\times 4\pi$) the rigid-body approximation gives an approximate  $7\%$ larger energy value for the isospinning soliton solution. The associated isospin inertia tensor $U_{ij}$ (\ref{U_phi_not}) is diagonal and its diagonal elements as a function of isospin $K$ are shown in Fig.~\ref{Fig_B1_O3}(d). For small isospin values the Skyrme configuration possesses, within our numerical accuracy, $O(3)$ symmetry ($U_{ij}=V_{ij}=W_{ij}=\Lambda\delta_{ij}$ where the moment of inertia $\Lambda$ is calculated to be $47.5$) and as $K$ increases the soliton solution deforms by breaking the spherical symmetry to an axial symmetry [the tensors of inertia (\ref{Sky_Inertia}) are all diagonal and satisfy $U_{11}=U_{22}=u$, $V_{11}=V_{22}=v$, $W_{11}=W_{22}=w$ and $U_{33}=V_{33}=W_{33}$]. At the critical angular frequency $\omega_{\text{crit}}=1$ ($K_{\text{crit}}= 6.5\times 4\pi$) we find numerically $u= 66.7$, $v=67.9$,  $w=58.5$  and $U_{33}= 80.9$. 

\begin{figure}[!htb]
\centering
\includegraphics[totalheight=7.cm]{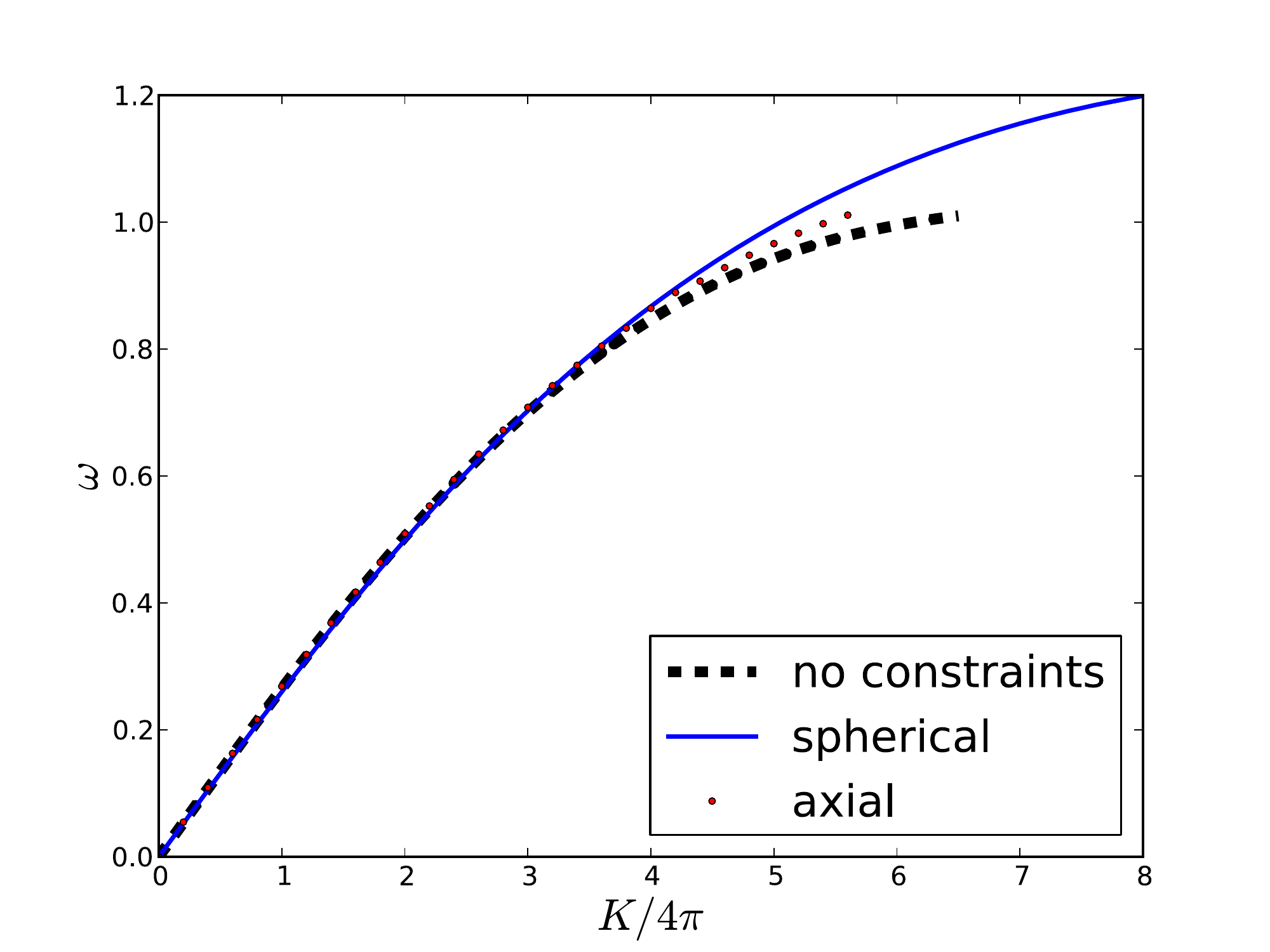}
\caption[Angular frequencies $\omega$ as function of isospin $K$ for $B=1$ Skyrmions ($\mu=1$).]{Angular frequencies $\omega$ as function of isospin $K$ for $B=1$ Skyrmions ($\mu=1$). We compare our results on arbitrarily deforming charge-1 Skyrme configurations with those obtained when allowing for  spherically symmetric and axially symmetric deformations, respectively.}
\label{Fig_B1_wK}
\end{figure}

In Fig.~\ref{Fig_B1_wK} we compare the frequency-isospin relation $\omega(K)$ for isospinning $B=1$ Skyrmions (with $\mu=1$) obtained when we \emph{do not} impose any constraints on the spatial symmetries of the isospinning soliton solution with those calculated when \emph{only} considering deformations within a spherically symmetric (\ref{Sky_hedge}) and an axially symmetric ansatz (\ref{Sky_axial}), respectively. For a spherically symmetric charge-1 Skyrme configuration ($\mu=1$) we solve the variational equation for the radial profile function $f(r)$ \cite{Rajaraman:1985ty} derived from the minimization of the pseudoenergy $F_\omega(f)=M_1(f)-\frac{1}{2}\Lambda(f)\,\omega^2$, where the classical soliton mass $M_1$ is given by (\ref{Sky_hedge_mass}) and the associated moment of inertia $\Lambda$ can be found in (\ref{Sky_hedge_lambda}). The underlying two point boundary value problem -- $f(0)=\pi$ and $f(\infty)=0$ -- is solved with the collocation method \cite{Ascher:1979} for fixed angular frequency $\omega$. Taking the asymptotic limit ($r\rightarrow\infty$, $f\rightarrow 0$) of the nonlinear equation for $f(r)$ reveals that a stable, isospinning soliton solution can only exist if the isorotation frequency satisfies $\omega\le \sqrt{3/2}\,\mu$. The classically isospinning Skyrmion will lower its isorotational energy by a spherical symmetric pion emission \cite{Battye:2005nx,Rajaraman:1985ty} when spinning faster than the critical angular frequency $\omega_{\text{crit}}= \sqrt{3/2}\,\mu$. Stationary solutions cease to exist at $\omega_{\text{crit}}$, but energy and moments of inertia remain finite at the critical frequency. Similarly, the linearization of the partial differential equations for an axially symmetrical deforming Skyrme soliton  (\ref{Sky_axial}) yields a critical frequency $\omega_{\text{crit}}=\mu$ \cite{Battye:2005nx}, beyond which the isospinning soliton solutions becomes unstable against pion emission. Clearly, the frequency-isospin relation $\omega(K)$ of the isospinning $B=1$ solution calculated without imposing any symmetry constraints agrees with the one of the axially symmetric deforming solution.

\begin{figure}[!htb]
\centering
\includegraphics[totalheight=9.5cm]{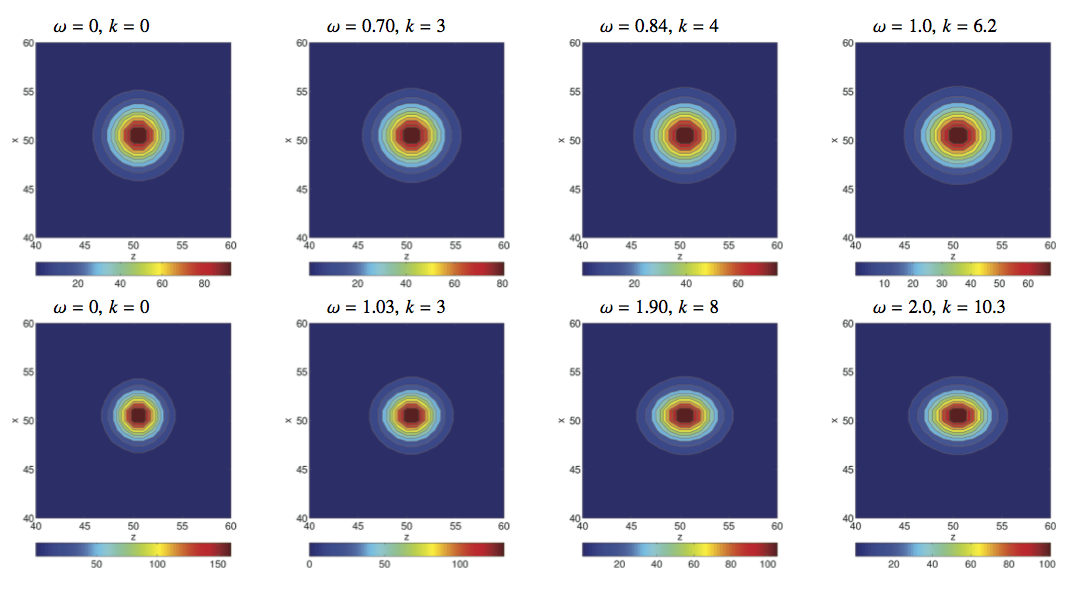}
\caption{Energy density contour plots of isospinning $B=1$ Skyrmion solutions as function of isospin $K$ with mass value chosen to be 1 (top row) and 2 (lower row). To visualize the axially symmetric deformation, we show a slice through the centre of the energy density distribution in the $xz$ plane. The isorotation axis is chosen to be the $z$ axis. Note that the angular momentum $K$ is given in units of $4\pi$, i.e. we define $k=K/4\pi$. The numerical calculations were performed with a full 3D relaxation algorithm. }
\label{Fig_B1_xz_contour}
\end{figure}

\begin{figure}[!htb]
\subfigure[\,$\mu=1.5$]{\includegraphics[totalheight=6.cm]{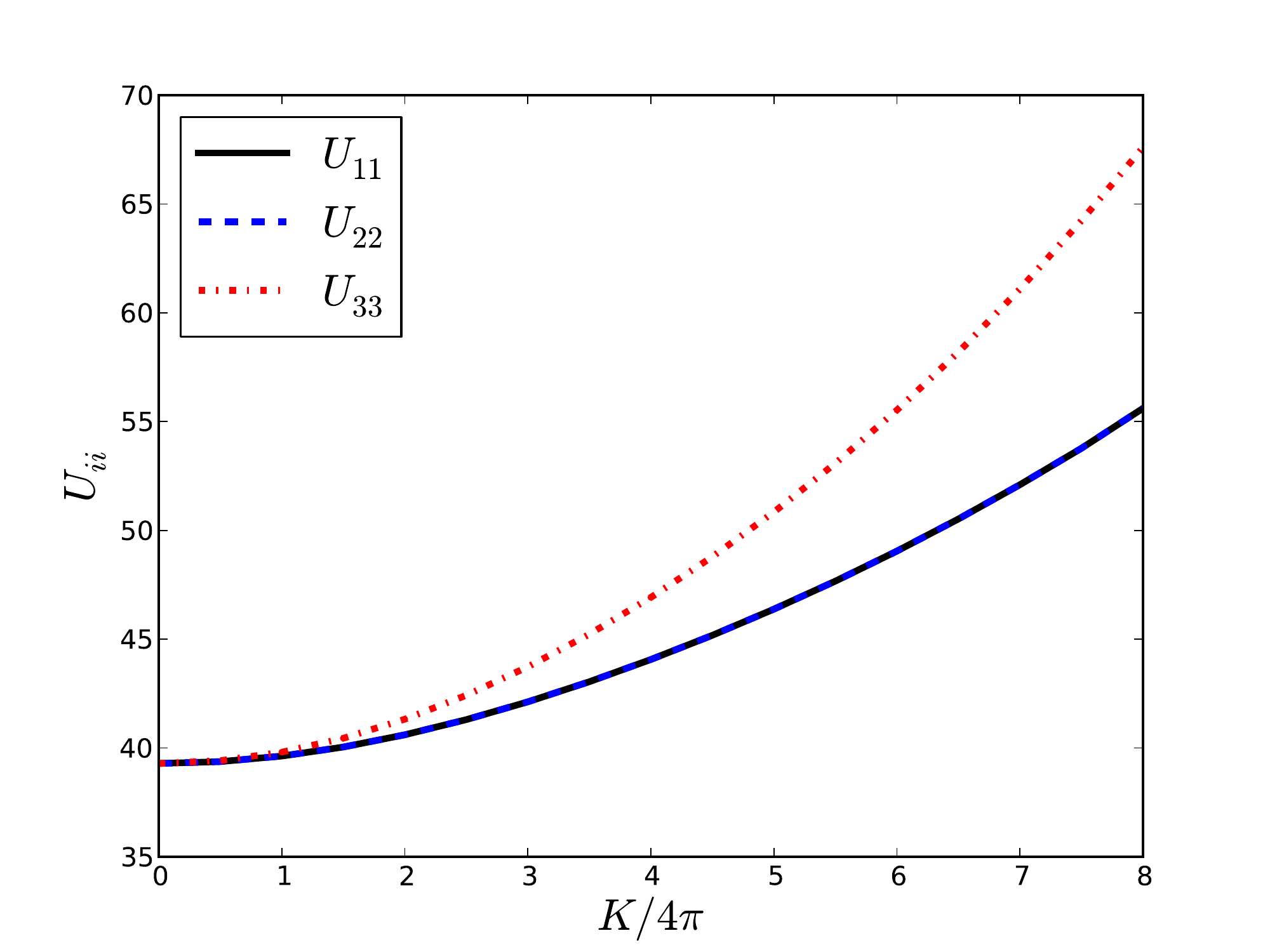}}
\subfigure[\,$\mu=2$]{\includegraphics[totalheight=6.cm]{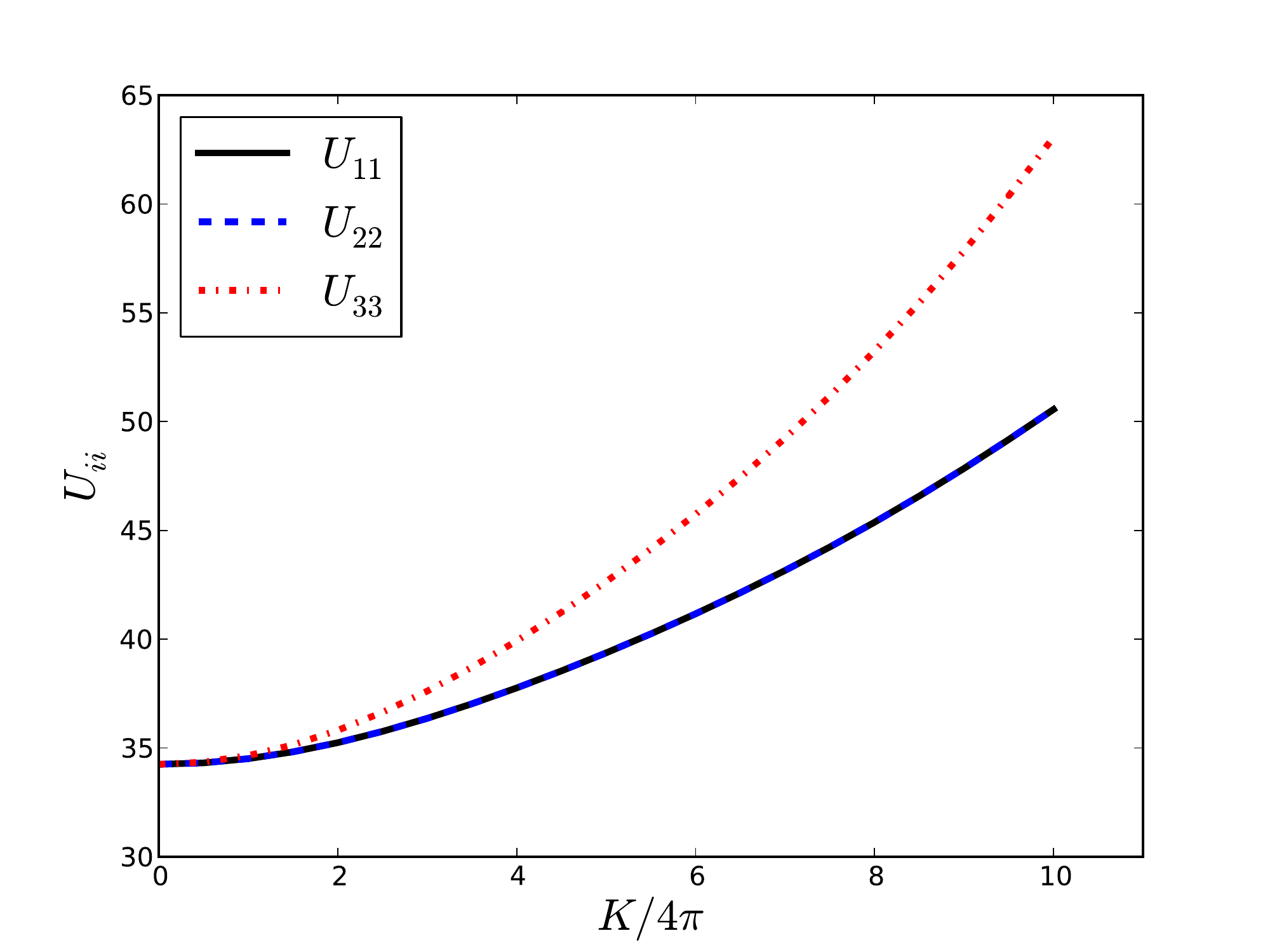}}
\caption{Diagonal elements of the isospin inertia tensor $U_{ij}$ as a function of isospin $K$ for $B=1$ Skyrme configurations with mass value $\mu=1.5,2$ and isospinning about the $z$ axis.}
\label{B1_Inertia_K_mass}
\end{figure}

\begin{figure}[!htb]
\centering
\subfigure[\,$B=1$]{\includegraphics[totalheight=6.cm]{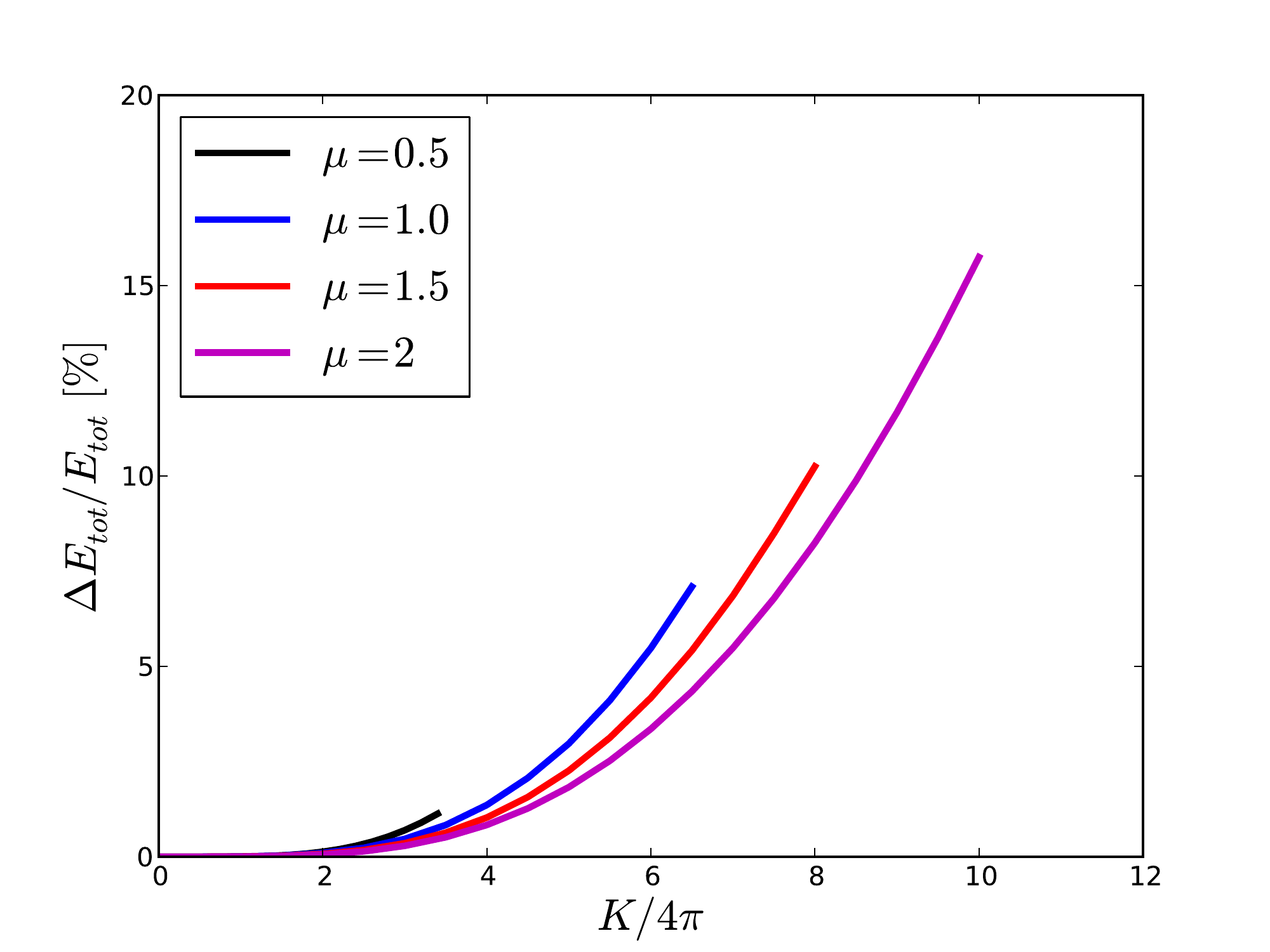}}
\subfigure[\,$B=2$]{\includegraphics[totalheight=6.cm]{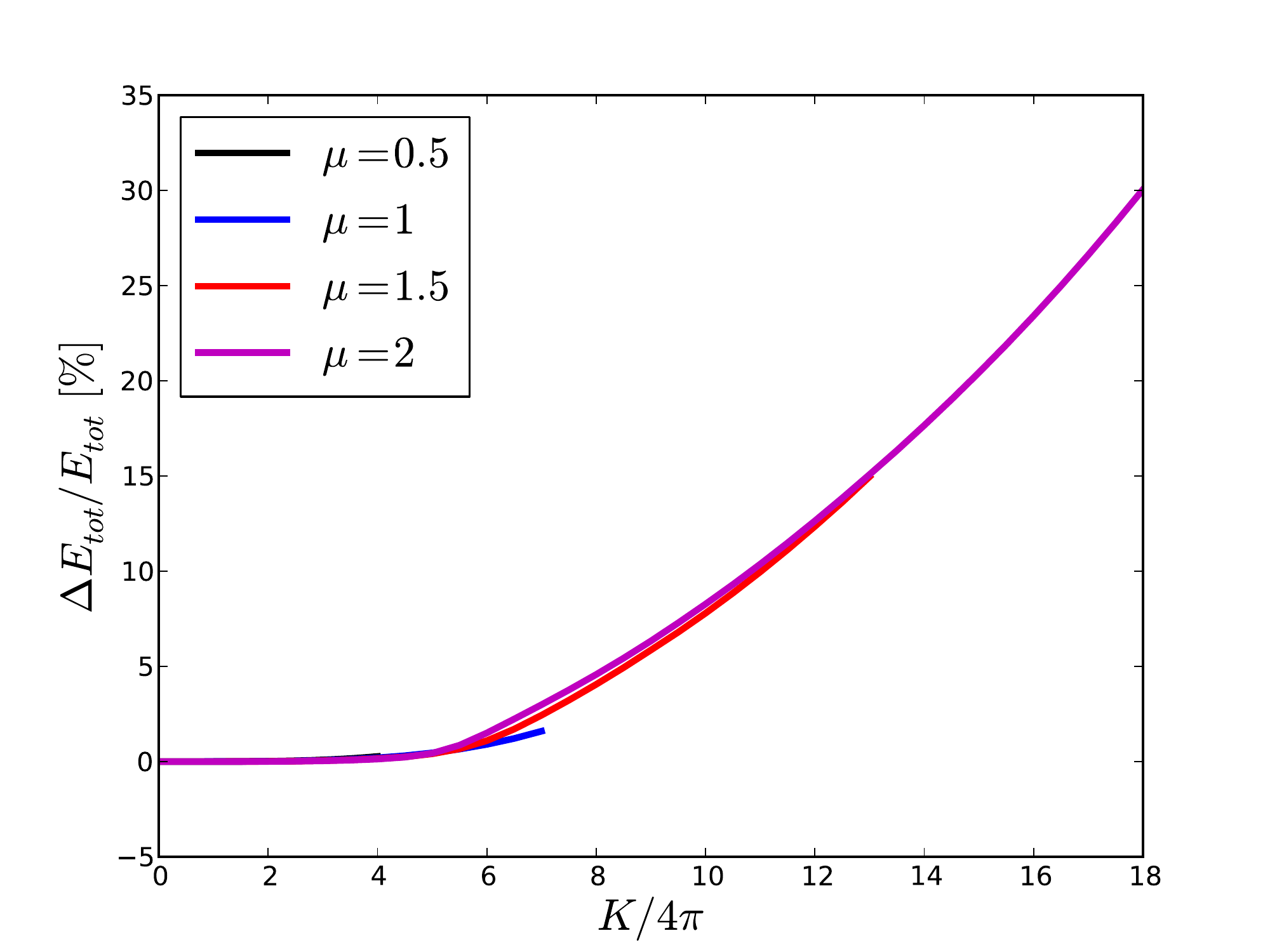}}
\caption{The deviation $\Delta E_{\text{tot}}/E_{\text{tot}}=\left(E_\text{Rigid}-E_{\text{tot}}\right)$ from the rigid body approximation for charge-1 (left) and charge-2 (right) Skyrmions as a function of isospin $K$ for various rescaled mass values $\mu$. The isorotation axis is chosen to be the $z$ axis.}
\label{B1B2_Rigid_compare}
\end{figure}

The energy density contour plots of $B=1$ Skyrmion solutions as a function of isospin $K$ and for two different choices of the mass parameter ($\mu=1,2$) are presented in Fig.~\ref{Fig_B1_xz_contour}. The  diagonal isospin inertia tensor elements for pion masses $\mu=1.5$ and $\mu=2$ are displayed in Fig.~\ref{B1_Inertia_K_mass} as functions of isospin $K$. As a measure of how much the Skyrme configuration deforms we calculate the deviation of the energy from the rigid-body approximation, namely 
$\Delta E_{\text{tot}}/E_{\text{tot}}=\left(E_\text{Rigid}-E_{\text{tot}}\right)$. This is displayed in Fig.~\ref{B1B2_Rigid_compare}(a) for various pion masses. Note that for a spherically symmetric, hedgehog  Skyrme configuration a rotation in physical space is equivalent to a rotation in isospace. Thus, we expect the same energy curves for a $B=1$ Skyrmion classically rotating about the $z$ axis.

\subsubsection{$B=2$}

For the toroidal $B=2$ Skyrmion solution we choose two different isospin axes \cite{Manton:2011mi}. One is the axis of symmetry -- the $z$ axis -- with the torus spinning in the $xy$ plane and the other is an axis orthogonal \cite{Braaten:1988cc,Leese:1994hb,Forest:1996kp} to it -- the $y$ axis --  so that the symmetry axis rotates in the $xz$ plane. It was argued in Ref.~\cite{Manton:2011mi} that these two spatial orientations are the relevant ones for describing the rotational states of the deuteron.  

\begin{figure}[!htb]
\centering
\subfigure[\,Total energy vs angular frequency ]{\includegraphics[totalheight=6.cm]{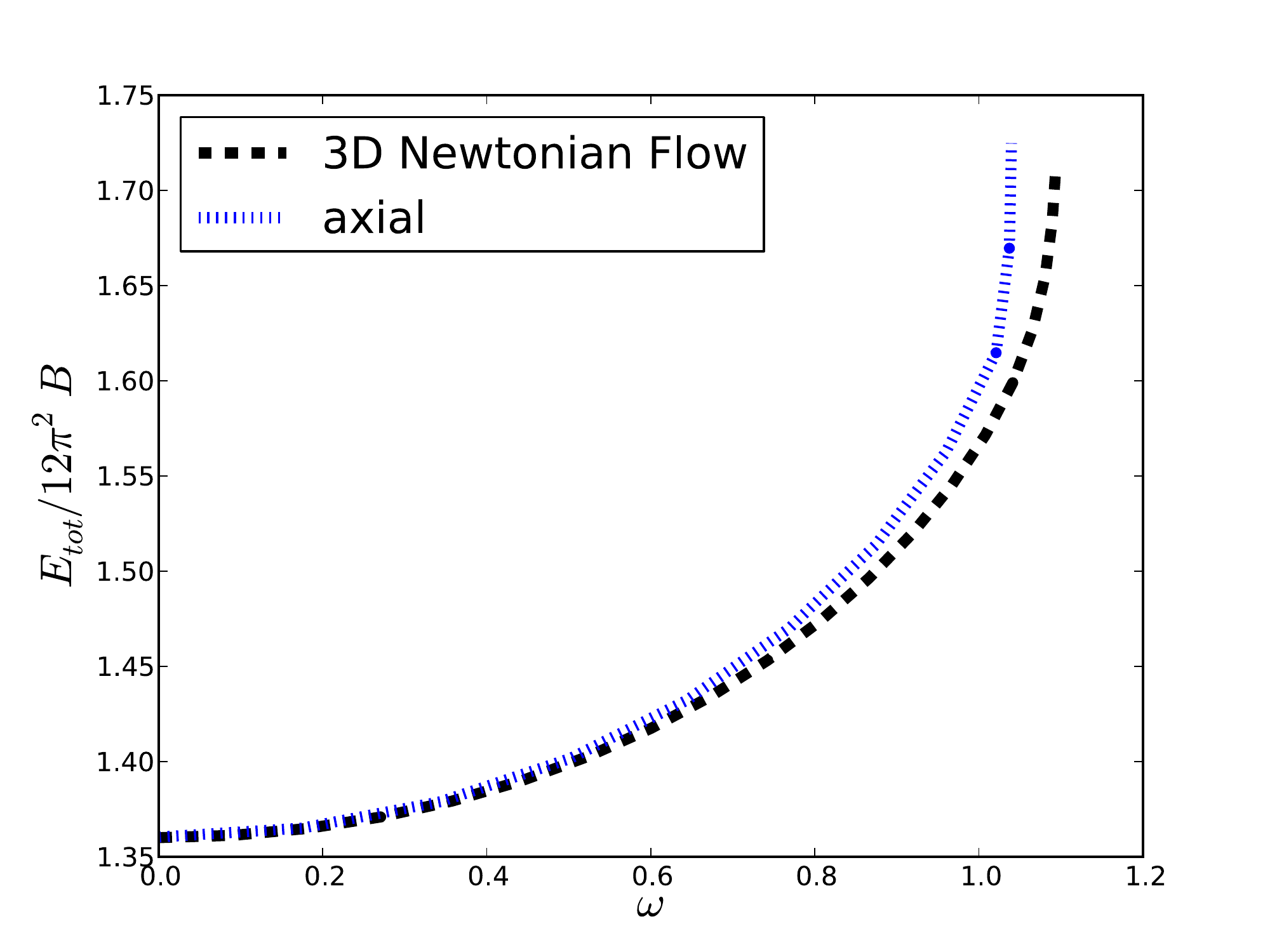}}
\subfigure[\, Mass-Spin relationship ]{\includegraphics[totalheight=6.cm]{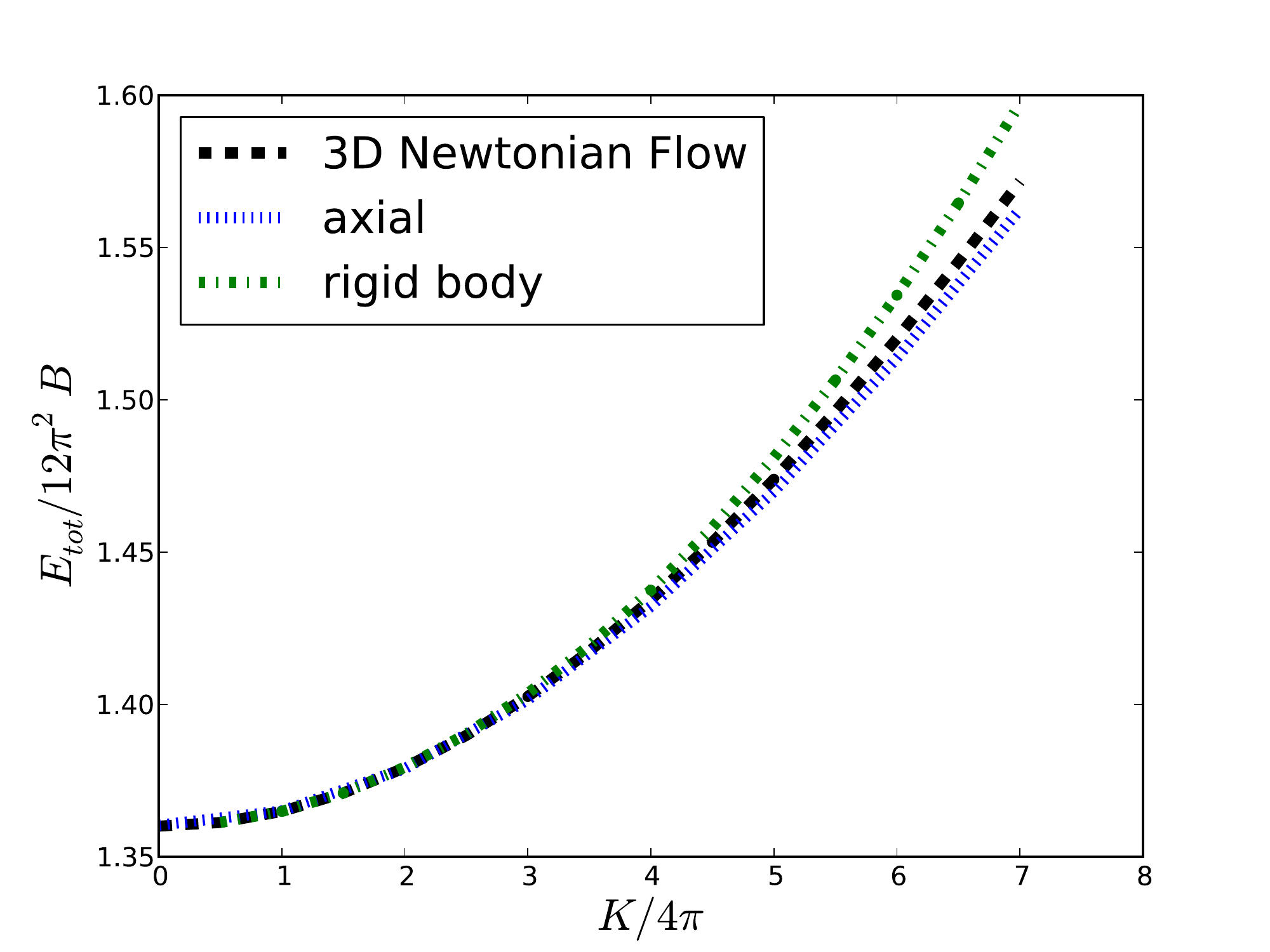}}\\
\subfigure[\, Inertia vs angular frequency]{\includegraphics[totalheight=6.cm]{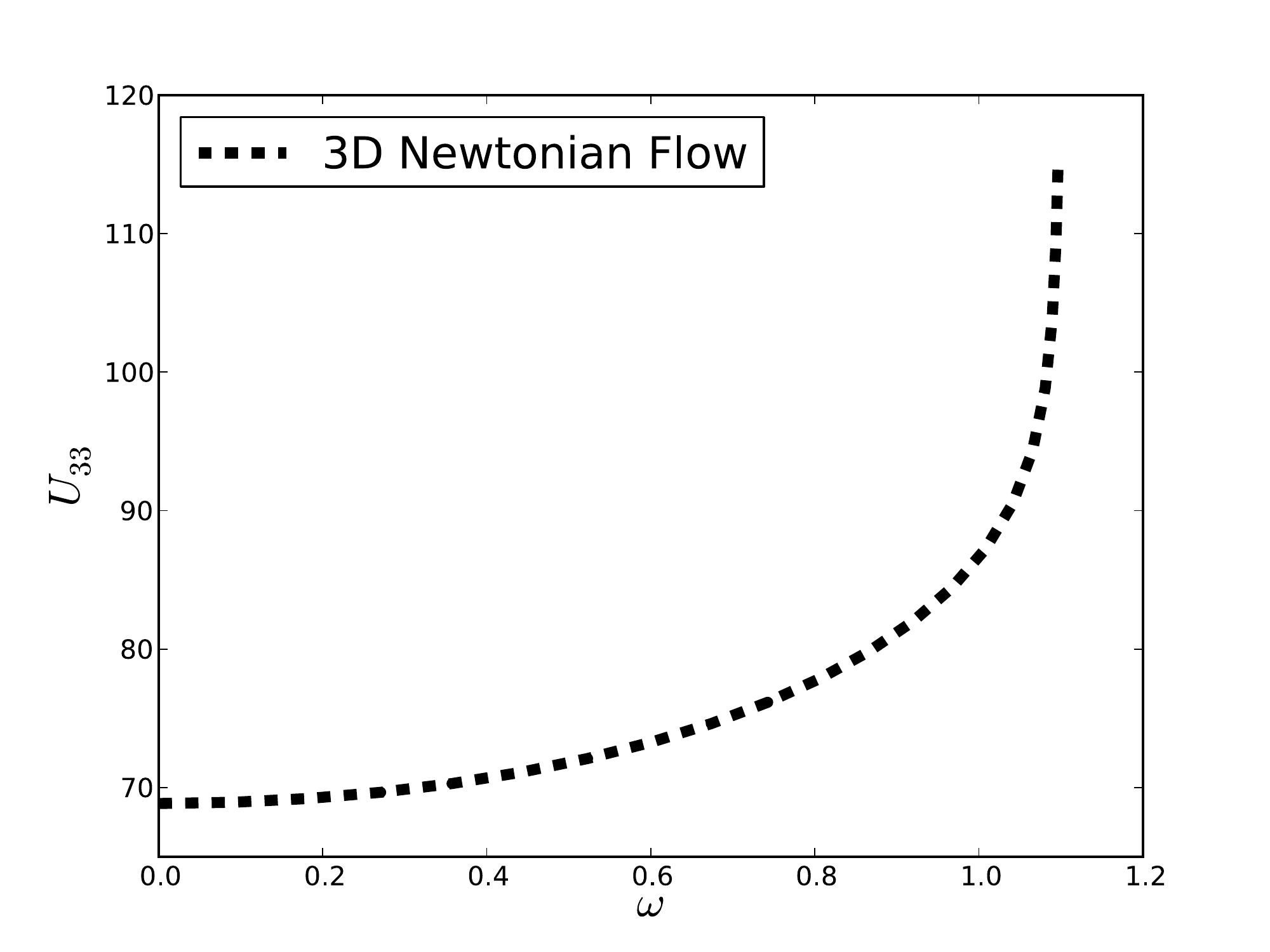}}
\subfigure[\, Inertia-Spin relationship]{\includegraphics[totalheight=6cm]{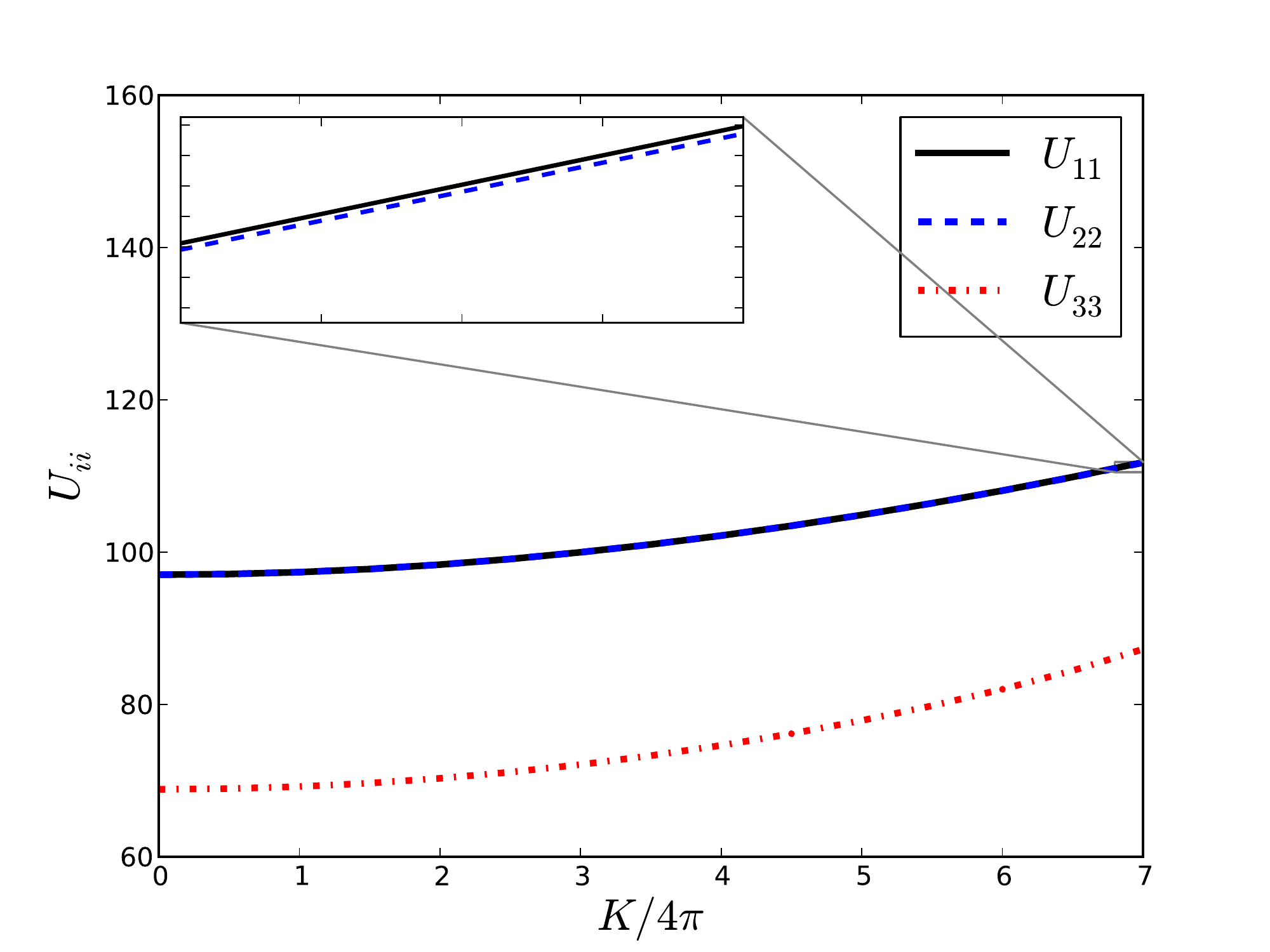}}
\caption{Isospinning $B=2$ Skyrmion $(\mu=1)$. A start configuration of topological charge 2 is numerically minimized using 3D modified Newtonian flow on a $(200)^3$ grid with a lattice spacing of $\Delta x=0.1$ and a time step size $\Delta t=0.01$. We choose the $\boldsymbol{\widehat{K}}=(0,0,1)$ as our isorotation axis. Our full 3-dimensional relaxation calculations are compared with the energy values for an axially symmetric deforming charge-2 Skyrme configuration (\ref{Sky_axial}). Additionally, we include the energy values obtained when assuming a rigid rotor.}
\label{Fig_B2_Kz}
\end{figure}

\begin{figure}[!htb]
\includegraphics[totalheight=20.0cm]{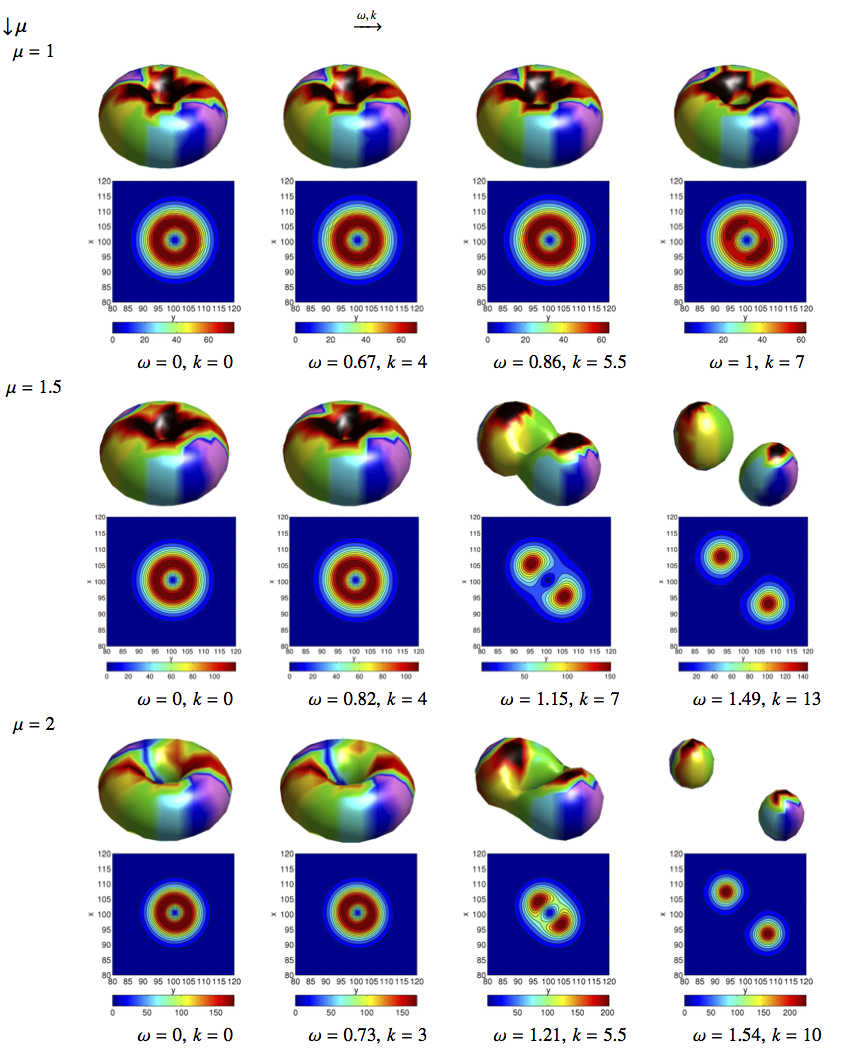}
\caption{Baryon density isosurfaces and energy density contour plots of isospinning $B=2$ Skyrmion solutions for a range of mass parameters $\mu$. The axis of isorotation is chosen to be the $\boldsymbol{\widehat{K}}=(0,0,1)$ axis. The energy minimization calculations have been performed on grids with $(200)^3$ grid points and a spacing $\Delta x=0.1$.}
\label{Fig_iso_B2_Skyrmion}
\end{figure}

With the $z$ axis chosen as our isorotation axis and the mass parameter $\mu$ set to 1, we obtain the energy and moment of inertia curves presented in Fig.~\ref{Fig_B2_Kz}. The corresponding energy density isosurface plots are displayed in Fig.~\ref{Fig_iso_B2_Skyrmion}. We verify in Figs.~\ref{Fig_B2_Kz}(a) and \ref{Fig_B2_Kz}(b) that the total energy $E_{\text{tot}}$ as a function of $\omega$ and $K$ follows in good approximation the behavior expected from an axially symmetric charge-2 Skyrme soliton for angular frequencies $\omega\lesssim \omega_{\text{crit}}$ ($K\lesssim K_{\text{crit}}$) with a small deviation near $\omega=\omega_{\text{crit}}$. 
An isospinning, axially symmetric Skyrme configuration can be computed by minimizing the two-dimensional energy functional $E_{\text{tot}}=M_2+K^2/\left(2U_{33}\right)$ for fixed isospin $K$,  where the classical soliton mass $M_2$ is given by (\ref{Axial_mass}) with winding number $n=2$ and the relevant moment of inertia $U_{33}$ can be found using (\ref{inertia_axial_33}). 
Again, we observe that stable isospinning soliton solutions cease to exist at $\omega_{\text{crit}}=\mu$, but energy and moments of inertia remain finite at $\omega_{\text{crit}}$. The rigid-body approximation is shown in Fig.~\ref{Fig_B2_Kz}(b) to be a valid simplification for small isospin values $K\le 4\times 4\pi$ ($\omega \le  0.6$). For higher angular frequencies the energy values for spinning soliton solutions predicted by the rigid-body formula are larger for a given angular momentum $K$. For example, at  the critical frequency  $\omega_{\text{crit}}=1.0$ ($K_{\text{crit}}= 7\times 4\pi$) we observe that the energy of the isospinning charge-2 solution is about $1.6\%$ smaller than the one calculated with the rigid-body formula. Finally we show in Fig.~\ref{Fig_B2_Kz}(d) the diagonal elements $U_{ii}$ of the isospin inertia tensor as function of isospin $K$. We verify numerically that the inertia tensors are all diagonal and satisfy for $K\lesssim K_{\text{crit}}$ the relations $U_{11}=U_{22}\neq U_{33}$, $V_{11}=V_{22}\neq V_{33}$, $W_{11}= W_{22}\neq W_{33}$ and $U_{33}=\frac{1}{2}W_{33}=\frac{1}{4}V_{33}$ which are consequences of the axial symmetry \cite{Houghton:2005iu,Braaten:1988cc}. As seen in the inset in Fig.~\ref{Fig_B2_Kz}(d) $U_{11}\neq U_{22}$ close to $K_{\text{crit}}$ which is consistent with the energy density contour plots in Fig.~\ref{Fig_iso_B2_Skyrmion} where a slight symmetry breaking occurs for $K= 7\times 4\pi$.

\begin{figure}[!htb]
\subfigure[\,$\mu=1.5$]{\includegraphics[totalheight=6.0cm]{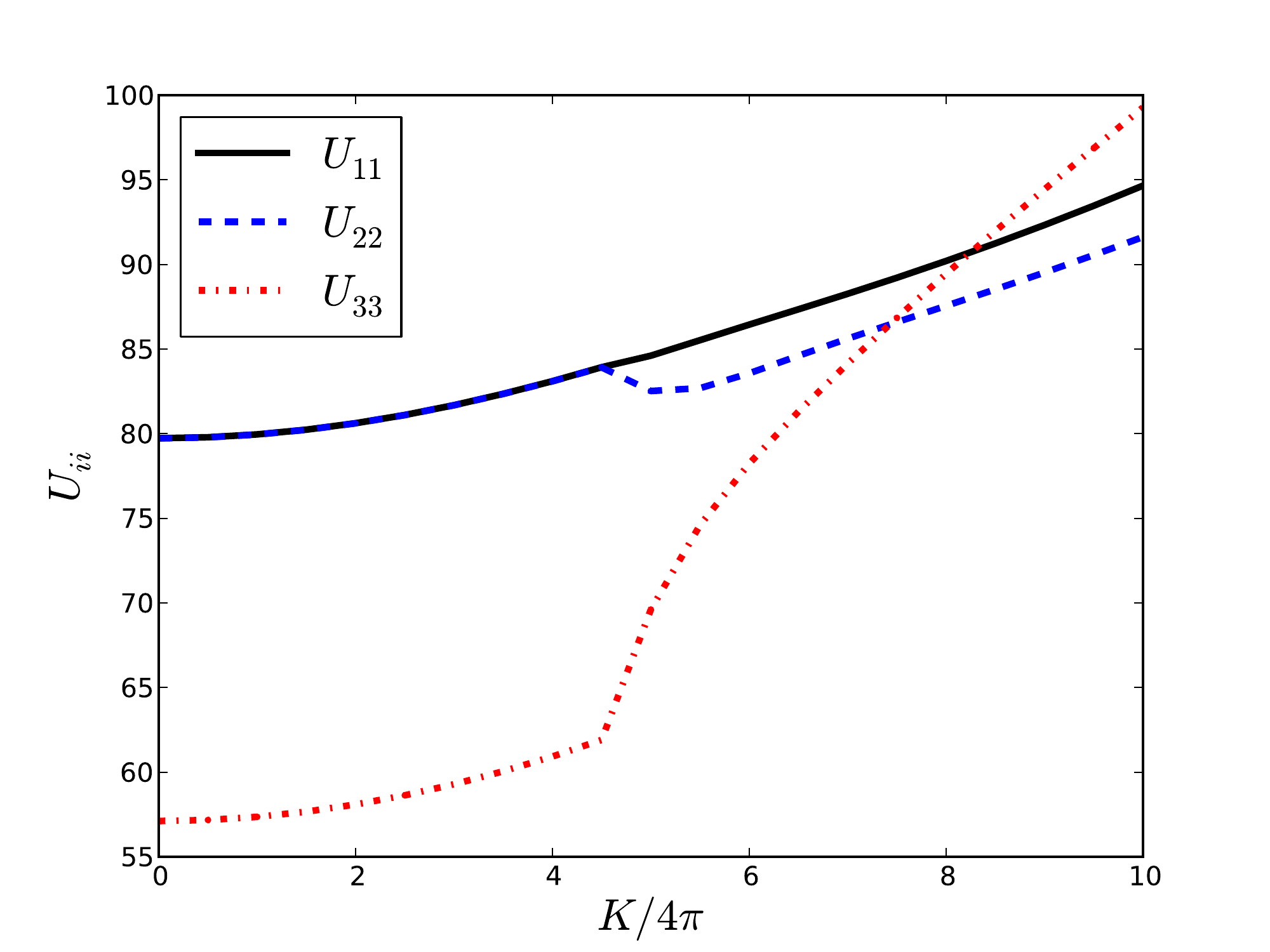}}
\subfigure[\,$\mu=2$]{\includegraphics[totalheight=6.0cm]{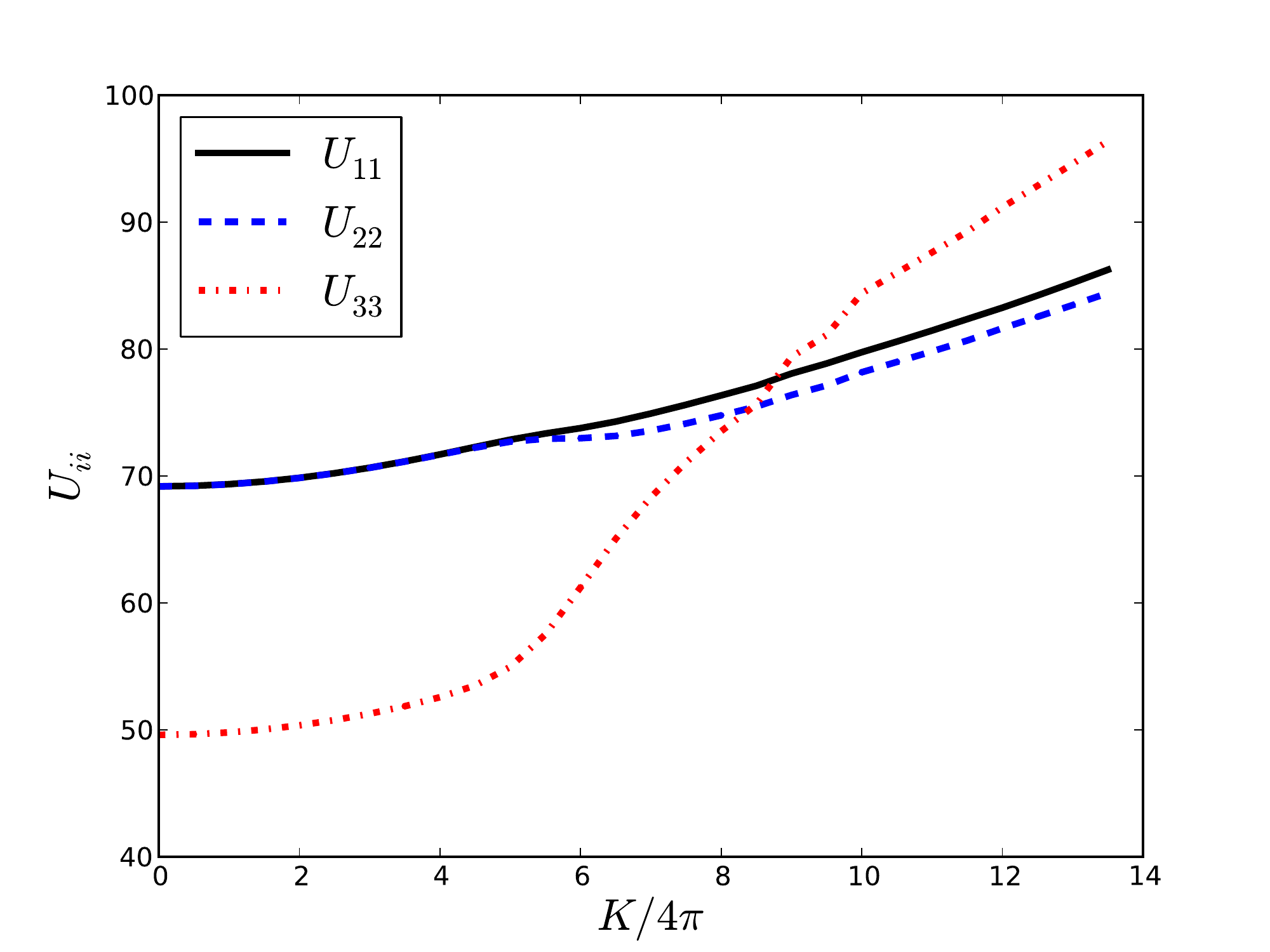}}
\caption{Diagonal elements of the isospin inertia tensor $U_{ij}$ as a function of isospin $K$ for $B=2$ Skyrme configurations with pion mass value $\mu=1.5,2$ and isospinning about $\boldsymbol{\widehat{K}}=(0,0,1)$.}
\label{B2Kz_Udiag_mass}
\end{figure}

When repeating our relaxation calculations for higher masses $\mu$, we find (analogous to observations in the conventional baby Skyrme model \cite{Battye:2013tka,Halavanau:2013vsa}) that the isospinning charge-2 Skyrmion solution breaks axial symmetry  at some critical value $\omega_{\text{SB}}$ and starts to split into its charge-1 constituents that move apart from each other and are orientated in the attractive channel. In Fig.~\ref{Fig_iso_B2_Skyrmion} we display baryon density isosurface and contour plots for a range of masses $\mu$. Note that the breakup of isospinning $B=2$ Skyrmion solutions into their charge-1 constituents is not as pronounced as the one observed for isospinning solutions in the conventional baby Skyrme model. For pion mass values $\mu=1.5$ and $\mu=2$ we find that the breaking of axial symmetry occurs at  $K_{\text{SB}}= 5.7\times 4\pi$ ($\omega_{\text{SB}}=1.1$) and $K_{\text{SB}}= 5.5\times 4\pi$  ($\omega_{\text{SB}}=1.2$), respectively (see the isospin diagonal elements shown as a function of $K$ in Fig.~\ref{B2Kz_Udiag_mass}). 

In Fig.~\ref{B1B2_Rigid_compare} the deviations from the rigid-body approximation are plotted as a function of the angular momentum $K$ for various mass values $\mu$  for charge-1 and charge-2 solutions isospinning around their $z$ axis. As the mass $\mu$ (or the topological charge $B$) increases, the rigid-body approximation provides more accurate results for the isospinning solutions of the model.

\begin{figure}[!htb]
\subfigure[\,$\mu=1.5$]{\includegraphics[totalheight=6.0cm]{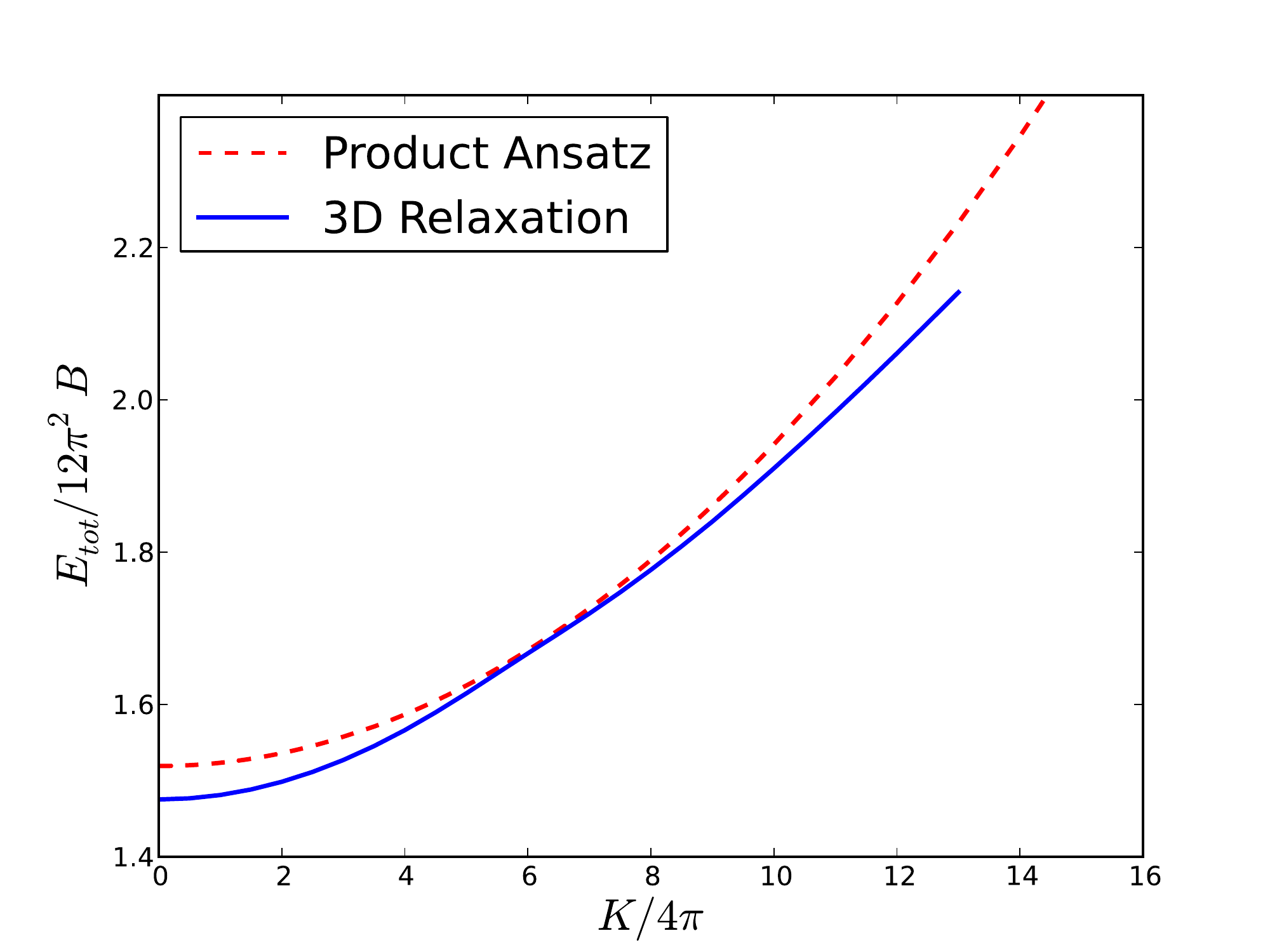}}
\subfigure[\,$\mu=2$]{\includegraphics[totalheight=6.0cm]{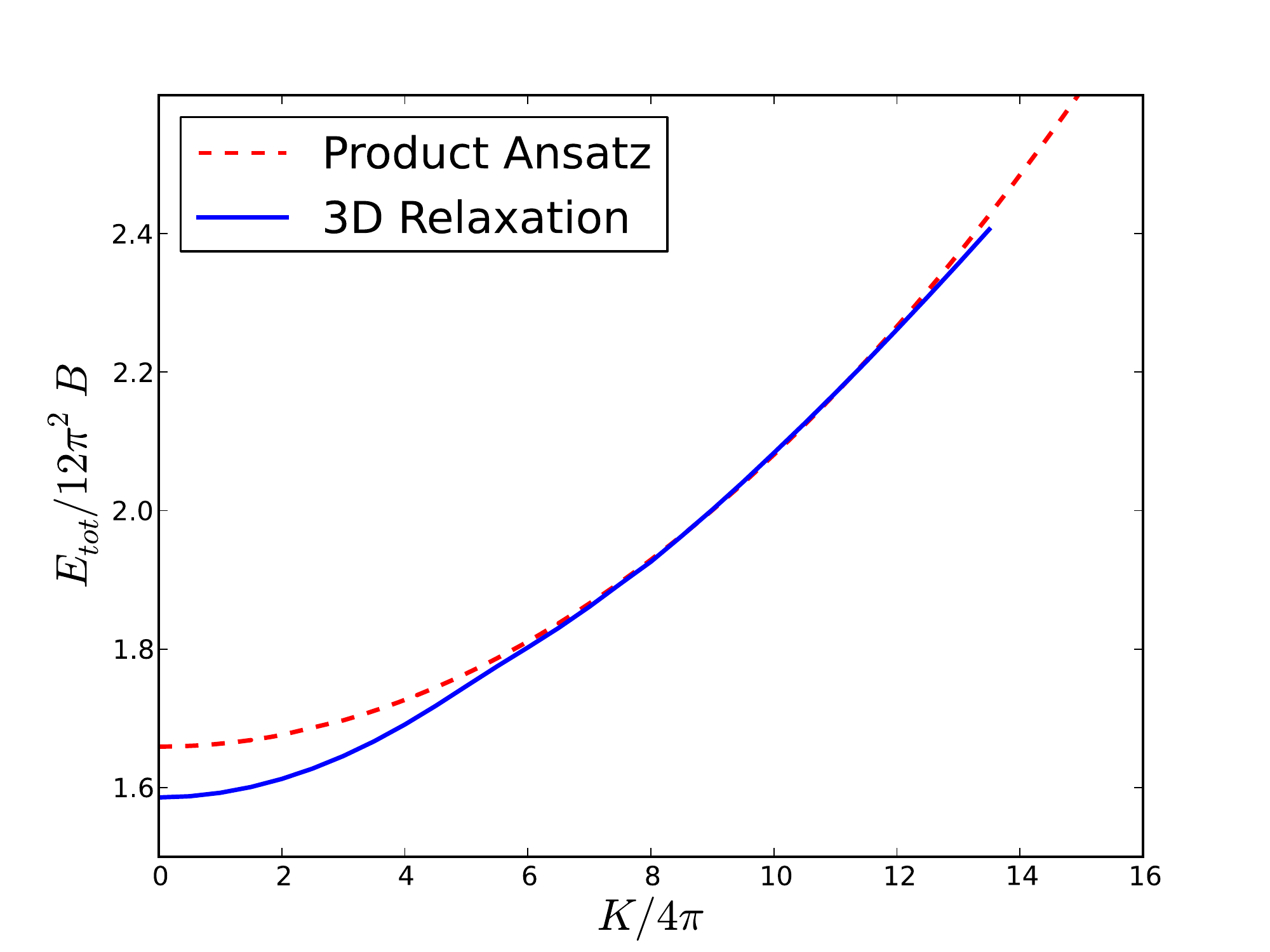}}
\caption{We compare the energy values calculated within the product ansatz approximation with those obtained when performing full three-dimensional numerical relaxations. Here we choose the isorotation axis   $\boldsymbol{\widehat{K}}=(0,0,1)$ and the pion mass takes the values $\mu=1.5,2$.}
\label{B2Kz_product}
\end{figure}

As a further check of our numerics, we calculate numerically the total energy $E_{\text{tot}}=M_2+K^2/(2U_{33})$ within the product ansatz approximation \cite{Skyrme:1961vq}. In analogy to Refs.~\cite{Jackson:1985bn,Stern:1989bc}, we generate a $B=2$ Skyrme configuration by superposing two $B=1$ solitons centered around $\boldsymbol{x_1}$ and $\boldsymbol{x_2}$:
\begin{align}
U_{B=2}(\boldsymbol{x},\boldsymbol{x_1},\boldsymbol{x_2})=U_H(\boldsymbol{x}-\boldsymbol{x_1})A(\boldsymbol{\alpha})U_H(\boldsymbol{x}-\boldsymbol{x_2})A^\dagger(\boldsymbol{\alpha}) \,,
\label{Prod_B2}
\end{align}
where  $U_H(\boldsymbol{x})$ is the hedgehog solution (\ref{Sky_hedge}) and $A(\boldsymbol{\alpha})=\exp{\left(i \boldsymbol{\tau}\cdot\boldsymbol{\alpha}/2\right)}$  with $\boldsymbol{\alpha}$ parametrizing the relative isospin orientation. The two individual Skyrmions are initially arranged so that the attraction is maximal. The Skyrmion-Skyrmion interaction \cite{Jackson:1985bn} is maximal when one Skyrmion is rotated in isospace relative to the other through angle $\pi$ about an axis perpendicular to the line of separation of the two Skyrmions. For a given isospin value $K$ we compute numerically  the total energy $E_{\text{tot}}=M_2+K^2/(2U_{33})$ for two maximally attractive $B=1$ Skyrmions as a function of their separation $r=|\boldsymbol{x_1}-\boldsymbol{x_2}|$ and determine for which separation $r$ the total energy takes its minimal value. The resulting energy values (dashed line) plotted as a function of $K$ are compared in  Fig.~\ref{B2Kz_product}  with the ones (solid line) obtained in our full 3D numerical minimizations. We observe that as $K$ increases the energy values of the non-rigidly isospinning $B=2$ Skyrmion tend to the values calculated within the product ansatz. That is, for these isospin values the separations $r$ are sufficiently large so that deformations of the Skyrmion fields can be neglected. Furthermore, the product ansatz approximation is found to provide more accurate results as the pion mass parameter $\mu$ increases.

\begin{figure}[!htb]
\centering
\subfigure[\,Total energy vs angular frequency ]{\includegraphics[totalheight=6.cm]{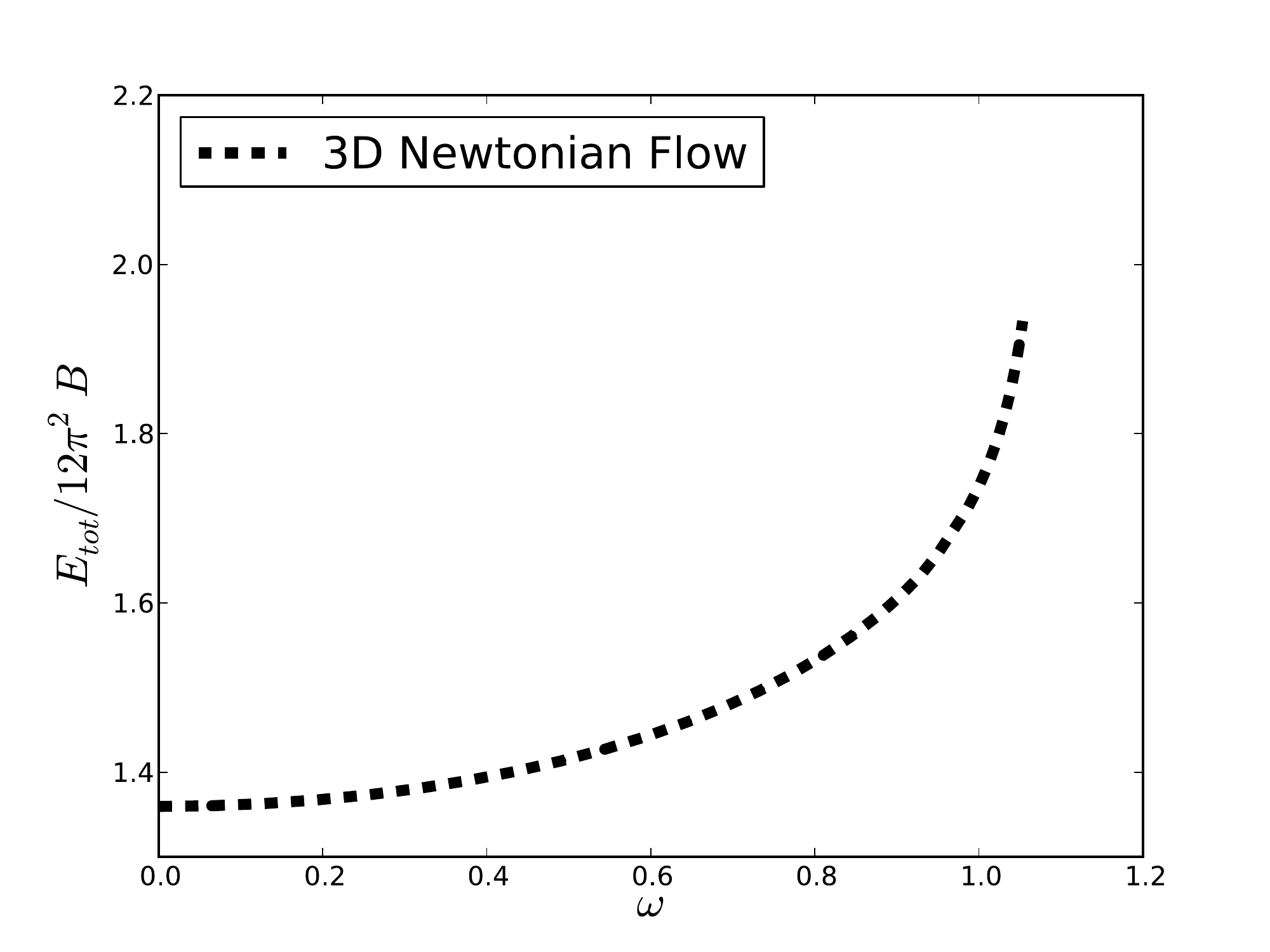}}
\subfigure[\,Mass-Spin relationship ]{\includegraphics[totalheight=6.cm]{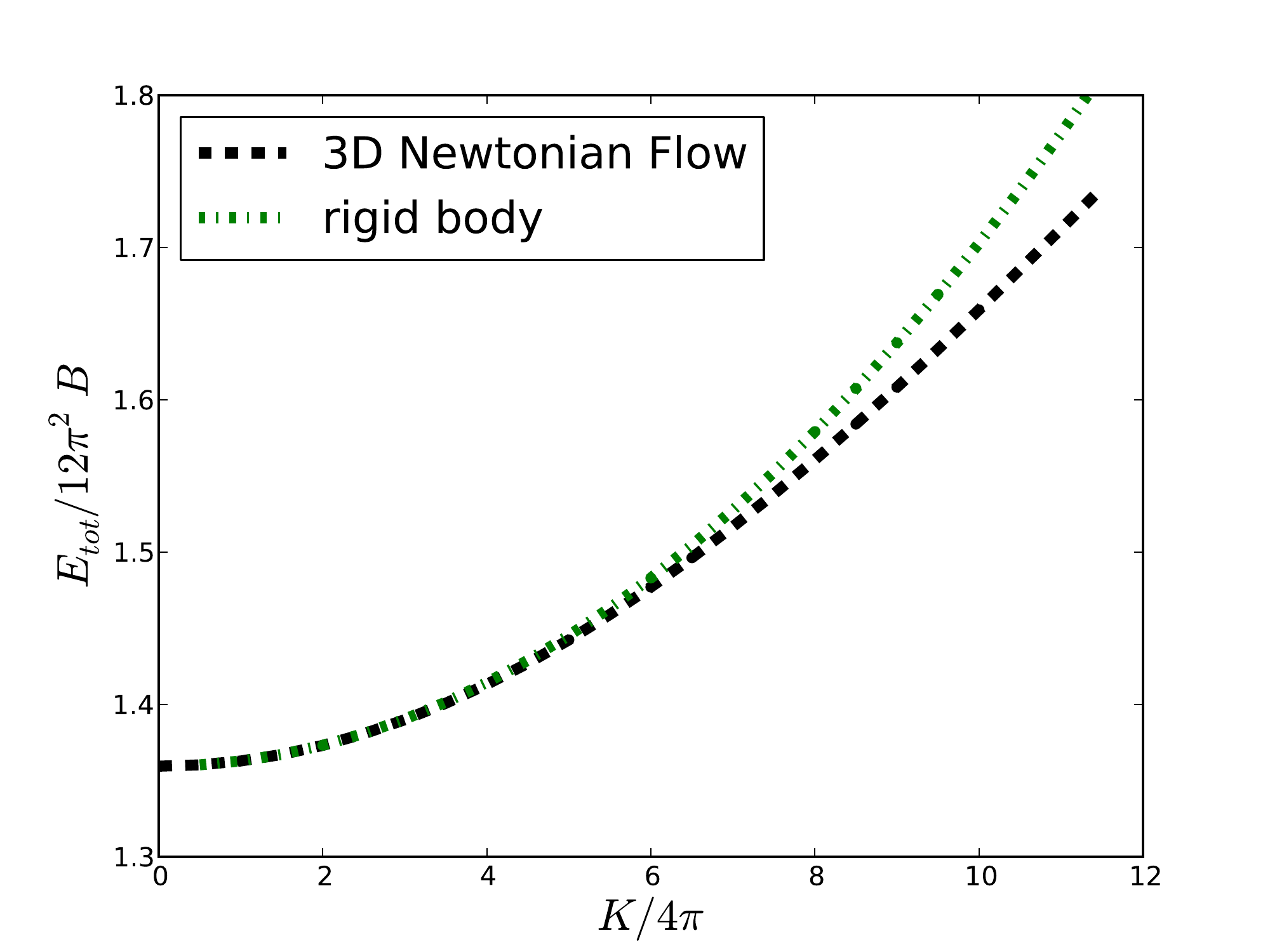}}\\
\subfigure[\,Inertia vs angular frequency]{\includegraphics[totalheight=6.cm]{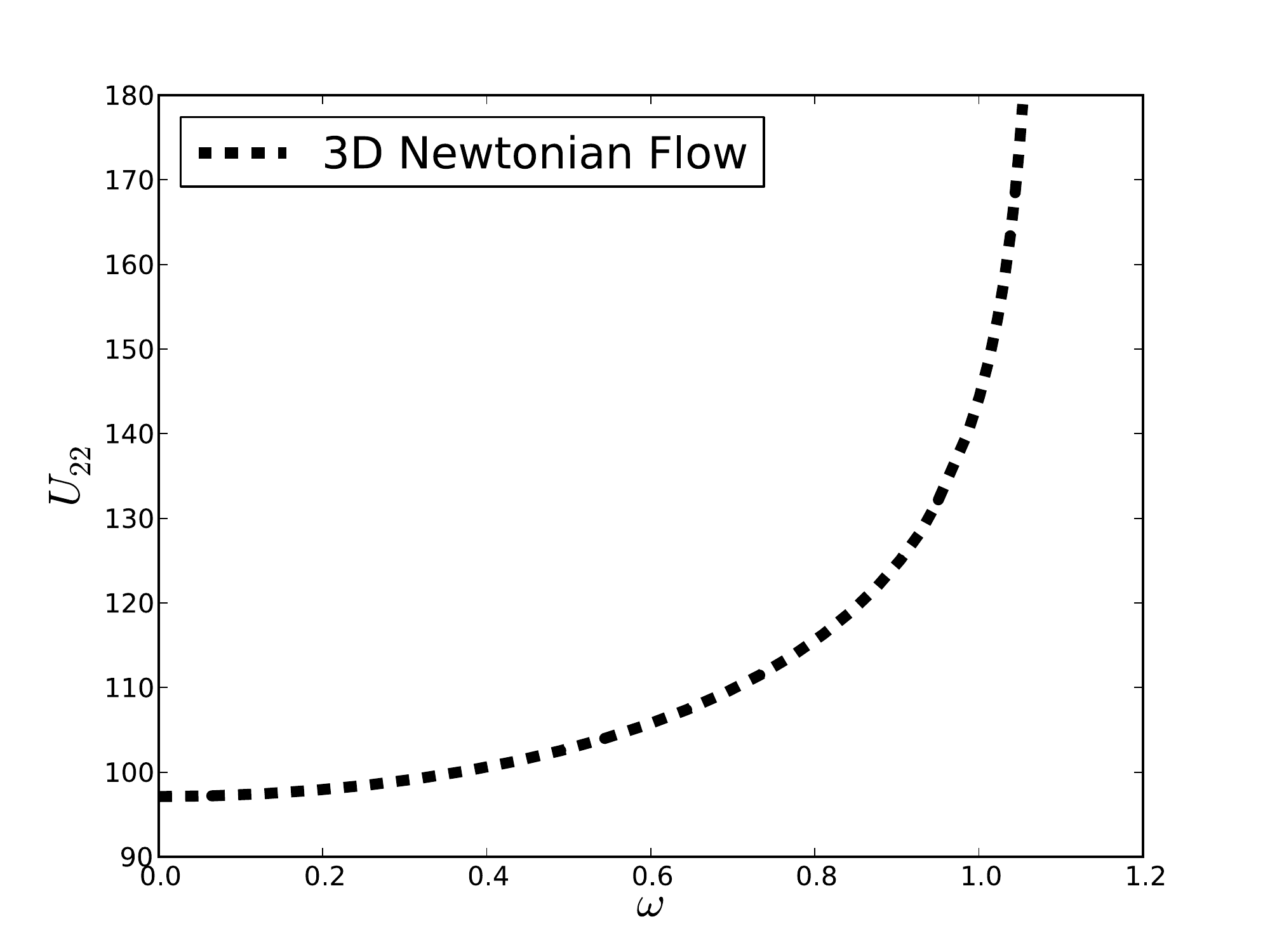}}
\subfigure[\,Inertia-Spin relationship]{\includegraphics[totalheight=6.cm]{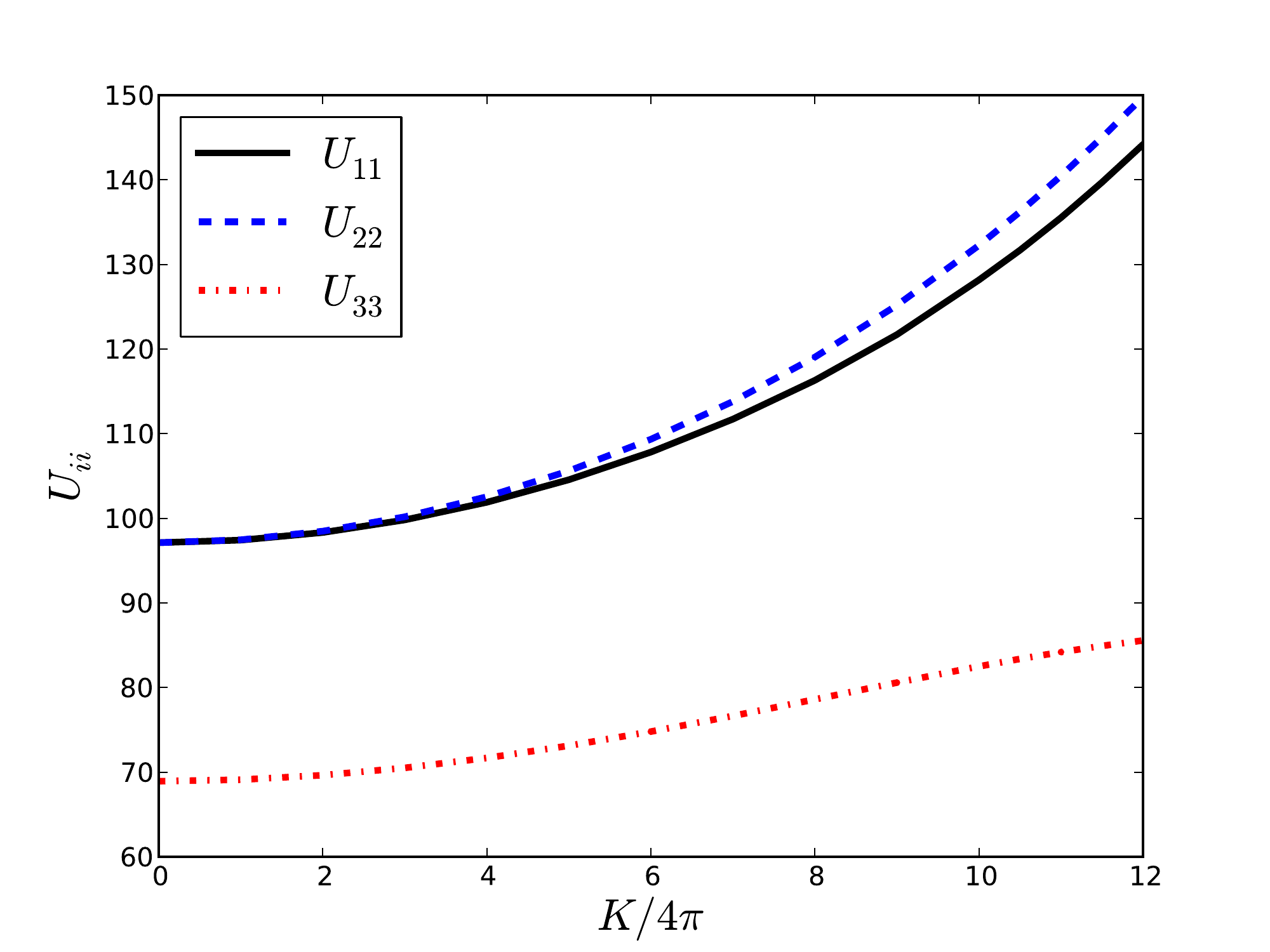}}
\caption{$B=2$ Skyrmion solution with pion mass $\mu=1$ and isospinning about  $\boldsymbol{\widehat{K}}=(0,1,0)$. We display the total energy $E_{\text{tot}}$ and moments of inertia $U_{ii}$ as function of isospin $K$ and angular frequency $\omega$. We compare in b) our full three-dimensional relaxation calculations with the rigid body approximation.}
\label{Fig_B2_Ky}
\end{figure}

\begin{figure}[!htb]
\subfigure[\,$\mu=1.5$]{\includegraphics[totalheight=6.0cm]{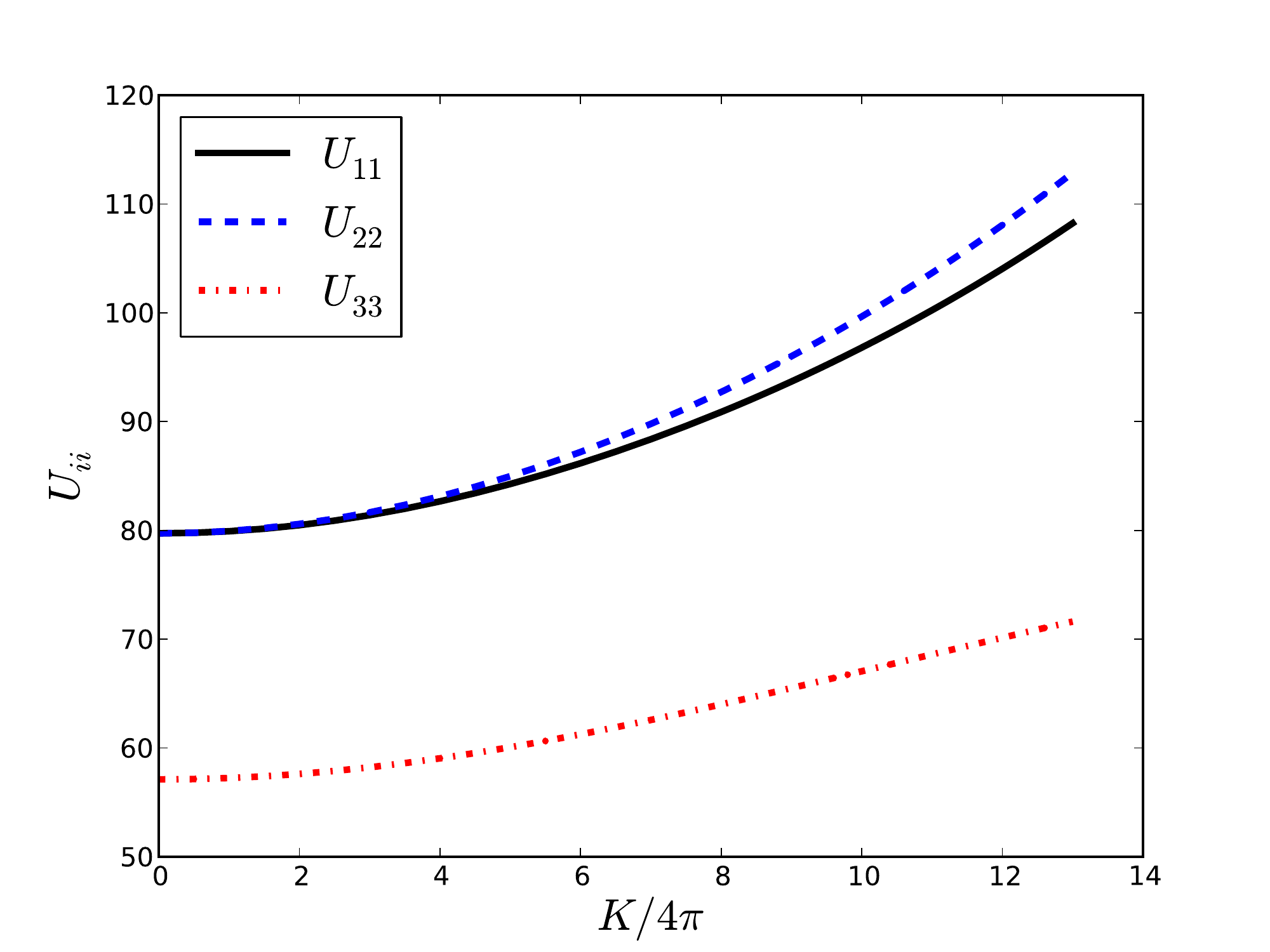}}
\subfigure[\,$\mu=2$]{\includegraphics[totalheight=6.0cm]{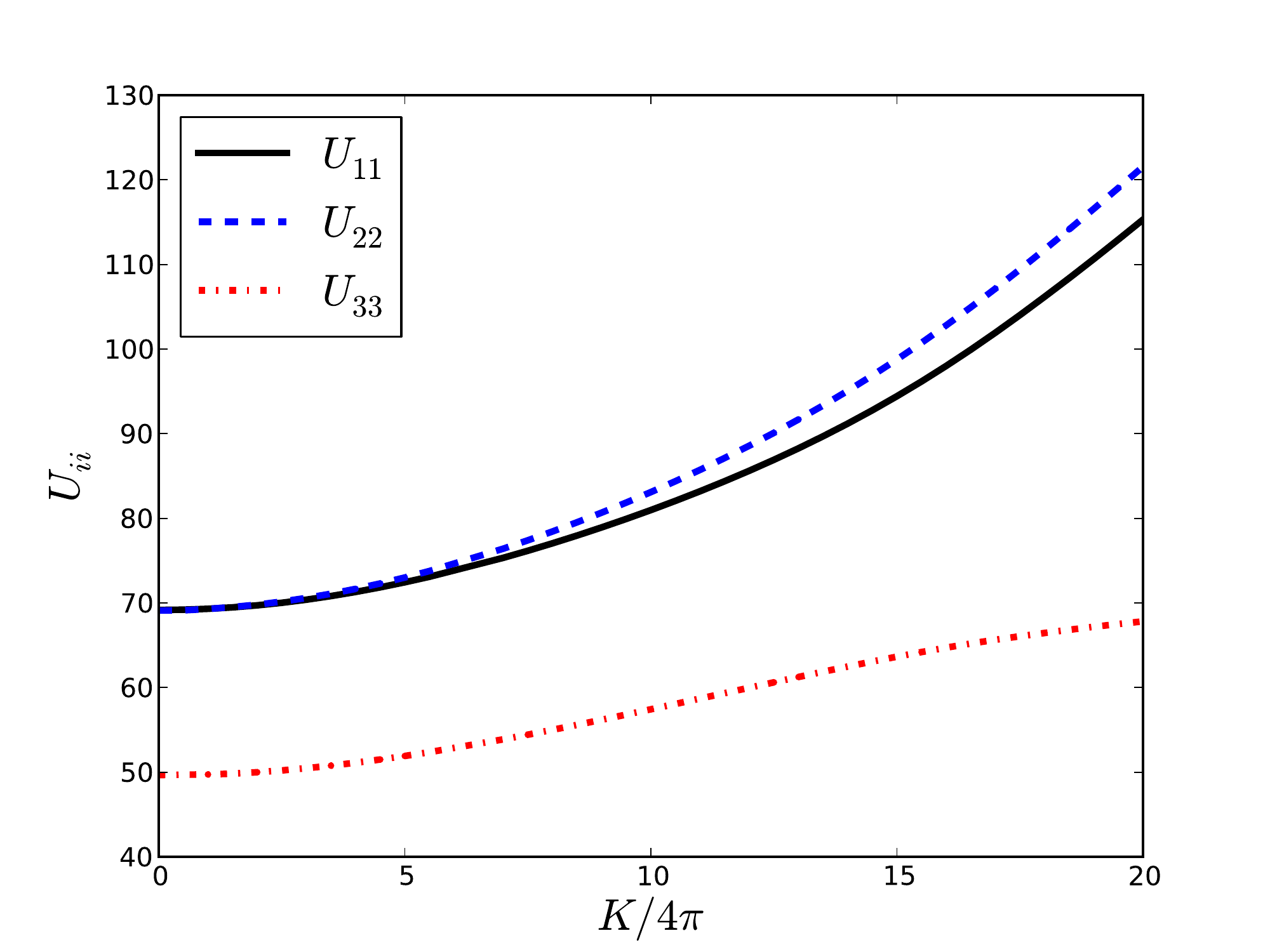}}
\caption{Diagonal elements of the isospin inertia tensor $U_{ij}$ as a function of isospin $K$ for $B=2$ Skyrme configurations with pion mass value $\mu=1.5,2$ and isospinning about $\boldsymbol{\widehat{K}}=(0,1,0)$.}
\label{B2Ky_Udiag_mass}
\end{figure}

In the following we choose the $y$ axis as our isorotation axes and set $\mu=1$. In Figs.~\ref{Fig_B2_Ky}(a) and \ref{Fig_B2_Ky}(c) we display the total energy and the moment of inertia $U_{22}$ as a function of $\omega$ up to $\omega_{\text{crit}} = 1$.  The rigid-body approximation proves to be a valid simplification for isospin values $K\le 8\times 4\pi$ ($\omega = 0.8$). However, at the critical frequency $\omega_{\text{crit}} =1$ ($K_{\text{crit}}=11.5\times 4\pi$) the  energy values turn out to be $\approx4\%$ lower than that predicted by the rigid-body approximation.  
\begin{figure}[!htb]
\centering
\includegraphics[totalheight=11.0cm]{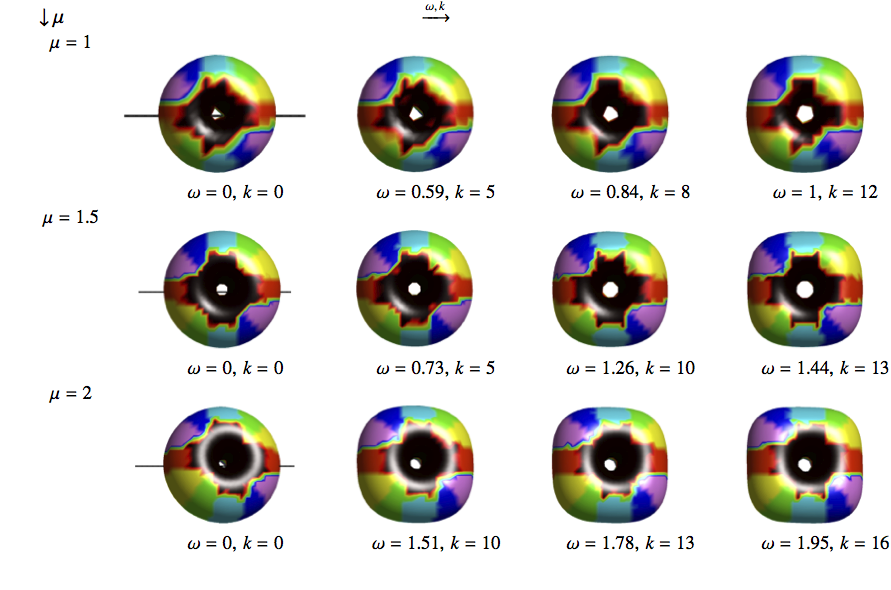}
\caption{Baryon density isosurfaces of isospinning $B=2$ Skyrmion solutions for a range of masses $\mu$ as function of isospin $K$. Here, the axis of isorotation is taken to be  $\boldsymbol{\widehat{K}}=(0,1,0)$.}
\label{Fig_B2Ky_xy_contour}
\end{figure}
The baryon density isosurfaces of isospinning charge-2 Skyrmions for a range of mass values are displayed in Fig.~\ref{Fig_B2Ky_xy_contour}. Note that we do not observe a breakup into charge-1 constituents for this mass range, instead a ``square-like'' configuration is formed. The associated moments of inertia as a function of isospin $K$ are displayed in Fig.~\ref{B2Ky_Udiag_mass}. For $\mu=1.5$ and $\mu=2$ we observe that the axial symmetry is broken at $K_{\text{SB}}=4.5\times4\pi$ ($\omega_{\text{SB}}=0.65$) and  $K_{\text{SB}}=4.0\times4\pi$ ($\omega_{\text{SB}}=0.7$), respectively.

\begin{figure}[!htb]
\subfigure[\,$B=2$]{\includegraphics[totalheight=6.cm]{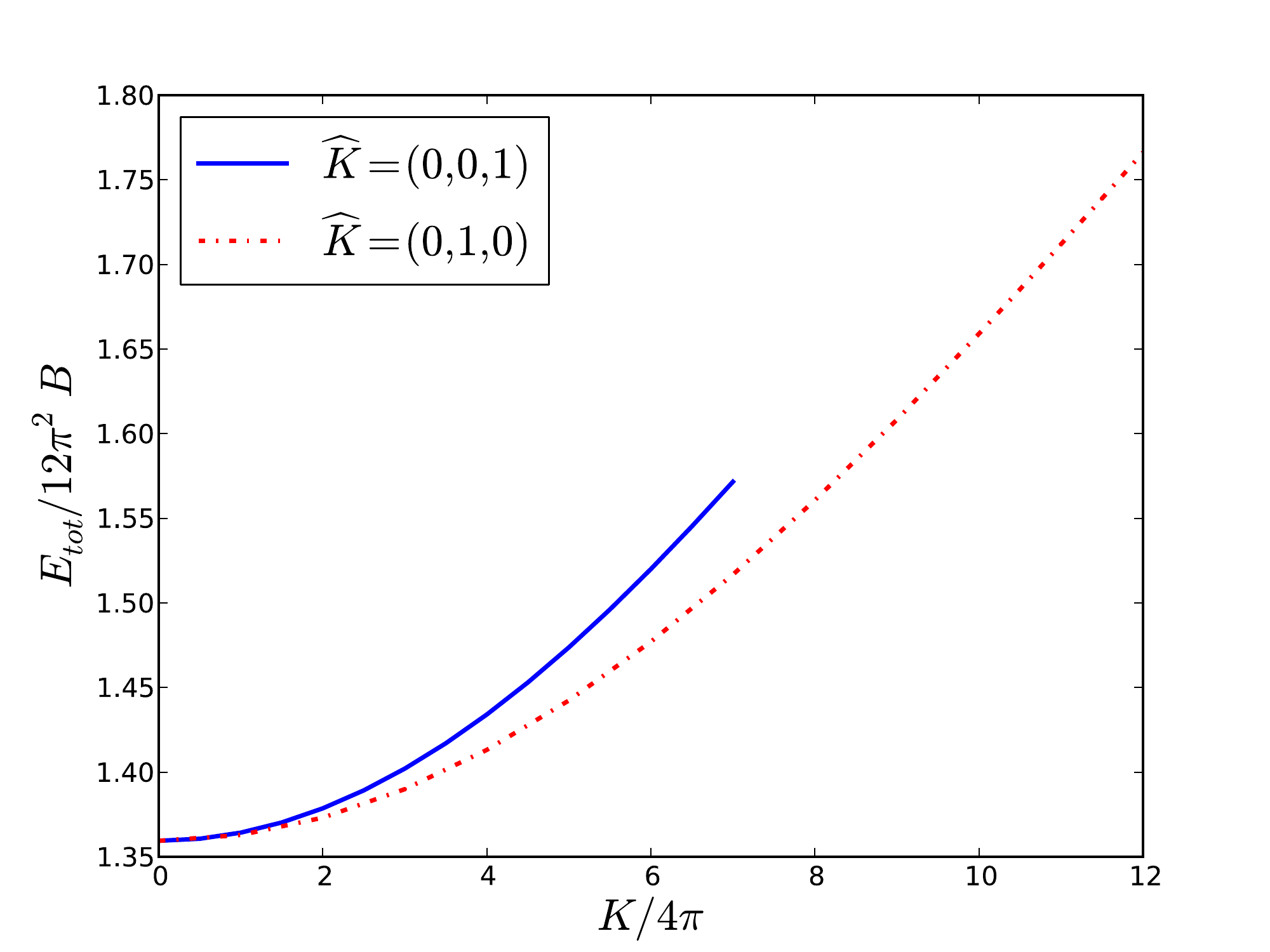}}
\subfigure[\,$B=4$]{\includegraphics[totalheight=6.cm]{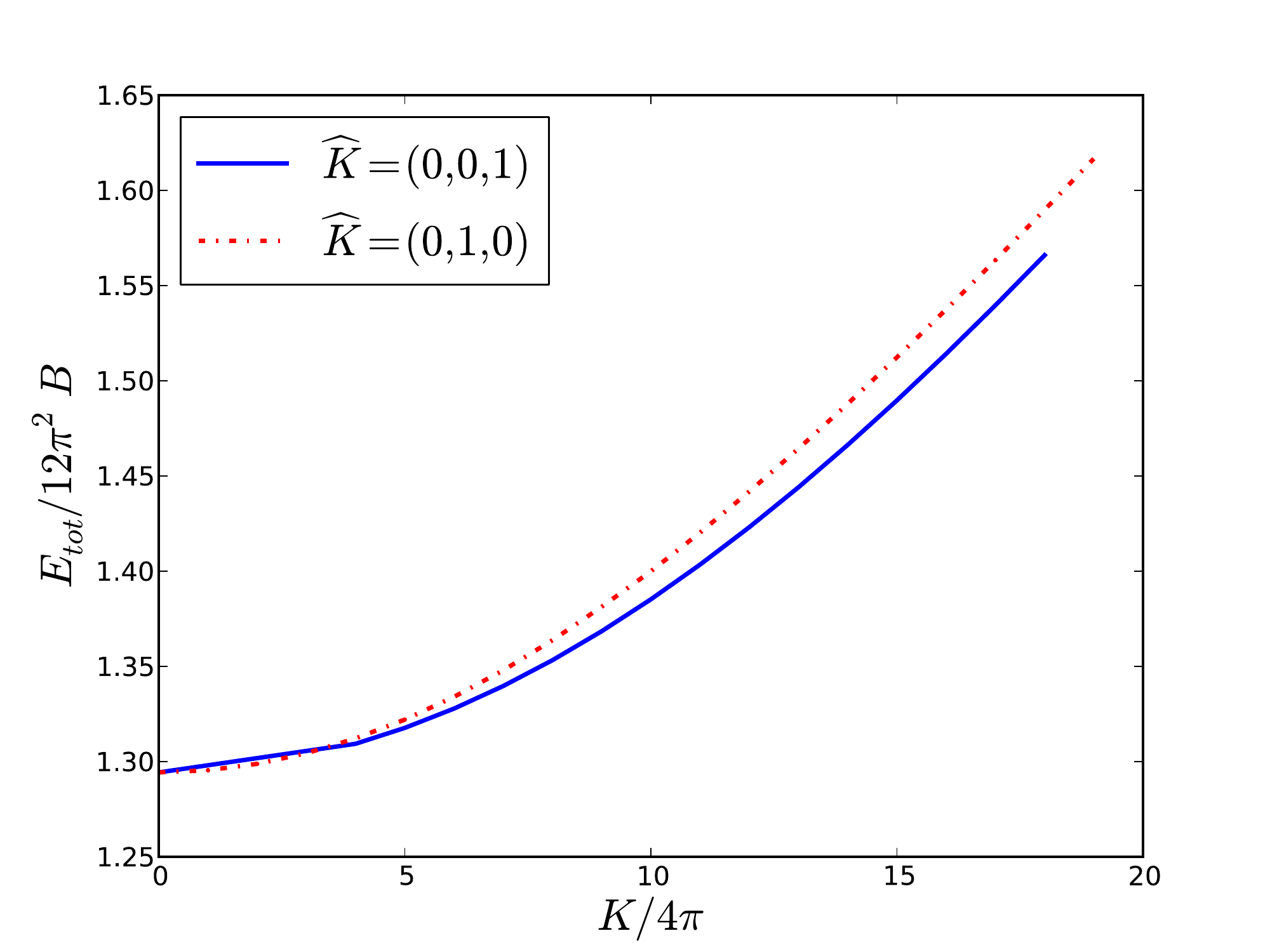}}\\
\subfigure[\,$B=8$ ($D_{6d}$ symmetry)]{\includegraphics[totalheight=6.cm]{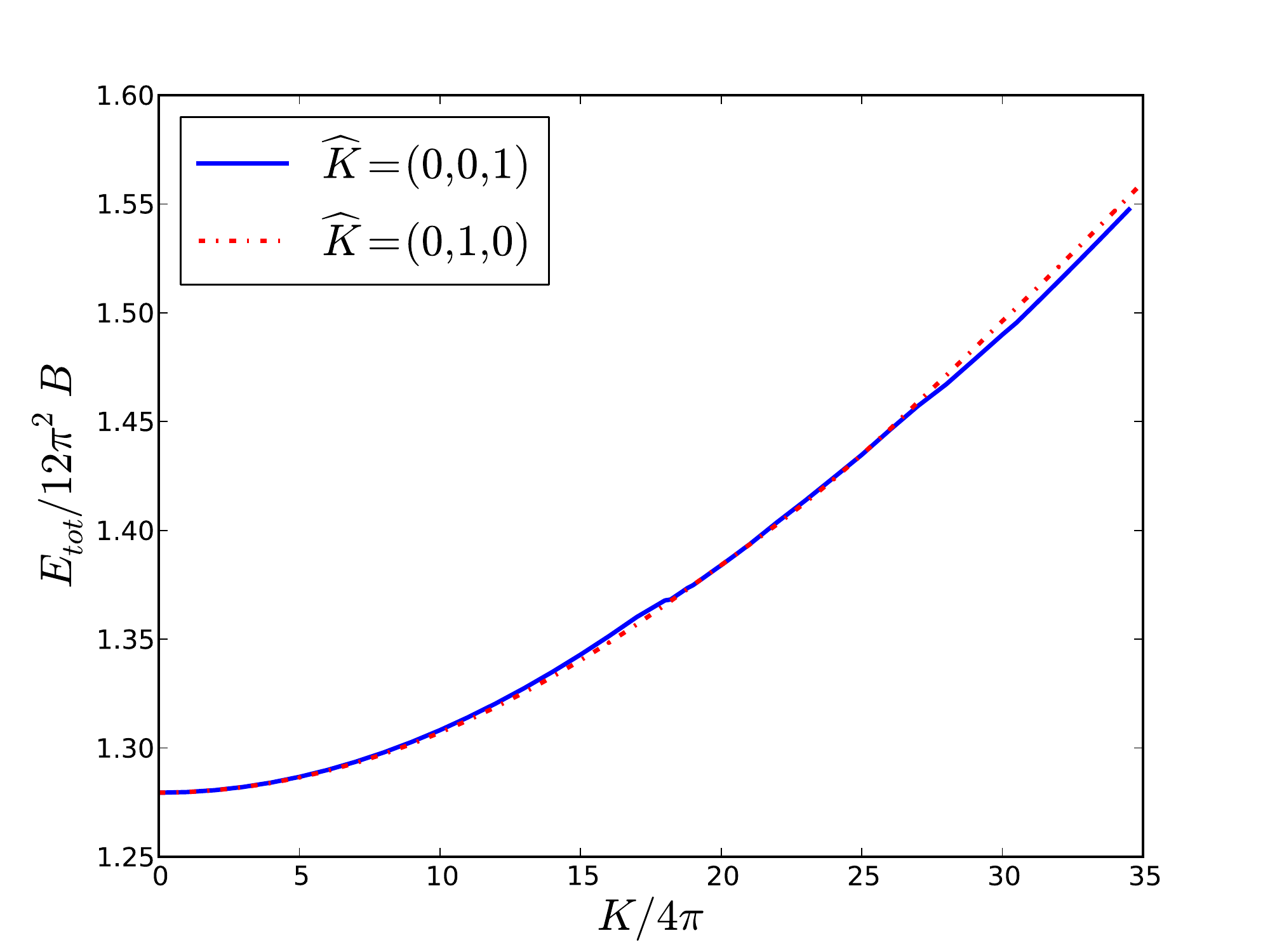}}
\subfigure[\,$B=8$ ($D_{4h}$ symmetry)]{\includegraphics[totalheight=6.cm]{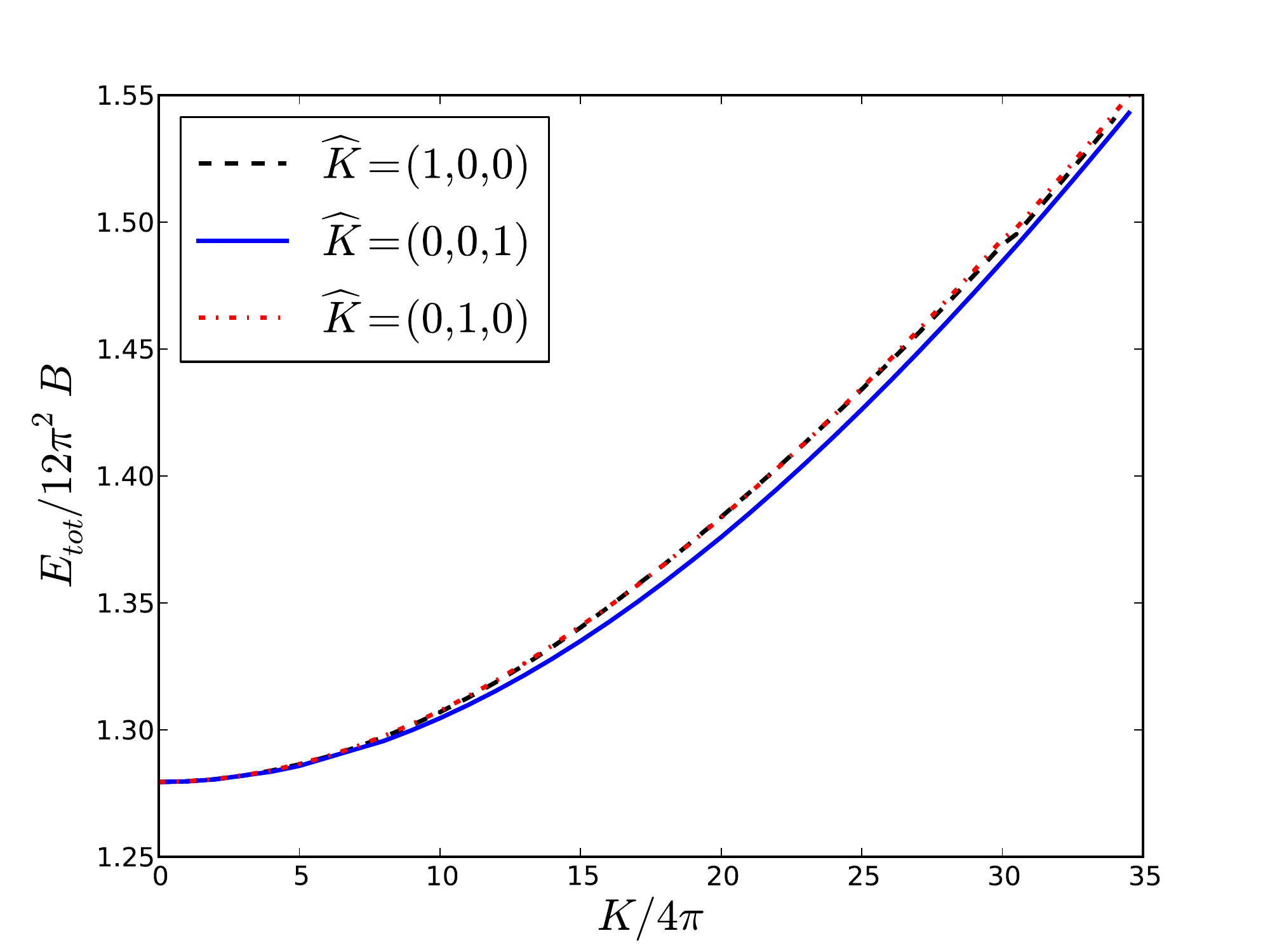}}
\caption{Energy curves $E_{\text{tot}}(K)$ for Skyrmions of baryon number $B=2,\,4,\,8$ and for different isorotation axes. Results are shown for the pion mass paramter $\mu=1$. For $B=2$ the ``square-like'' Skyrme configuration isospinning about $\boldsymbol{\widehat{K}}=(0,1,0)$ is of lowest energy, for $B=4$ the cube isospinning about $\boldsymbol{\widehat{K}}=(0,0,1)$ and for charge 8 the two cube soliton solution  with $\boldsymbol{\widehat{K}}=(0,0,1)$ has the lowest total energy. }
\label{Fig_Energy_axes}
\end{figure}

Note that for fixed isospin $K$, the $D_4$-symmetric $B=2$ configurations obtained by isospinning the $B=2$ torus about its $(0,1,0)$ axis are generally of lower energy than the solutions calculated when isospinning about  $(0,0,1)$  [see energy curves shown in Fig.~\ref{Fig_Energy_axes}(a)]. Indeed, we verified that when perturbed slightly, charge-2 Skyrmion solutions isospinning about $(0,0,1)$ can evolve into the lower energy $D_4$-symmetric solutions with isospin axis $\boldsymbol{\widehat{K}}=(0,1,0)$.

\subsubsection{$B=3$}

\begin{figure}[!htb]
\centering
\subfigure[\,Total energy vs angular frequency ]{\includegraphics[totalheight=6.cm]{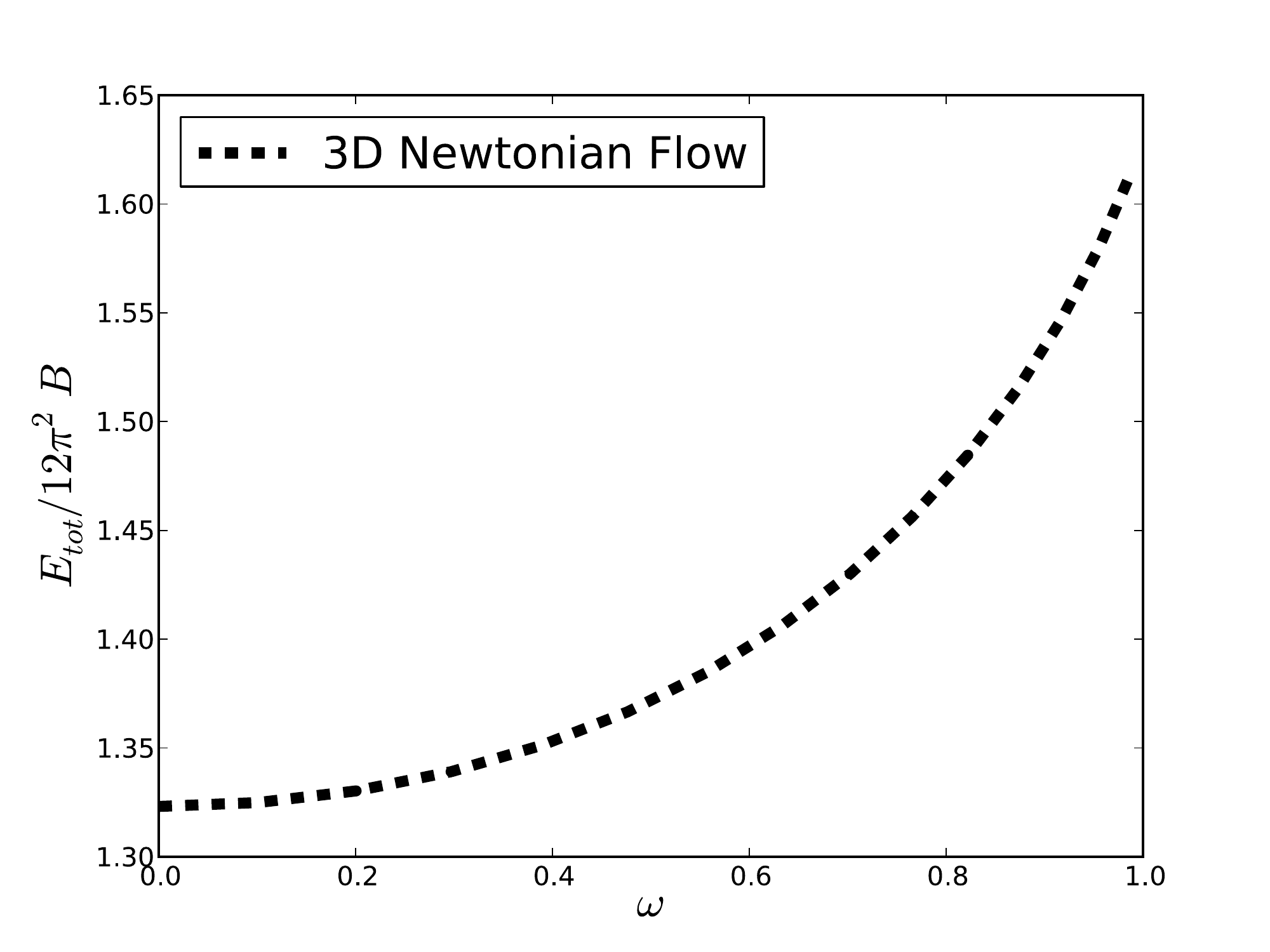}}
\subfigure[\,Mass-Spin relationship ]{\includegraphics[totalheight=6.cm]{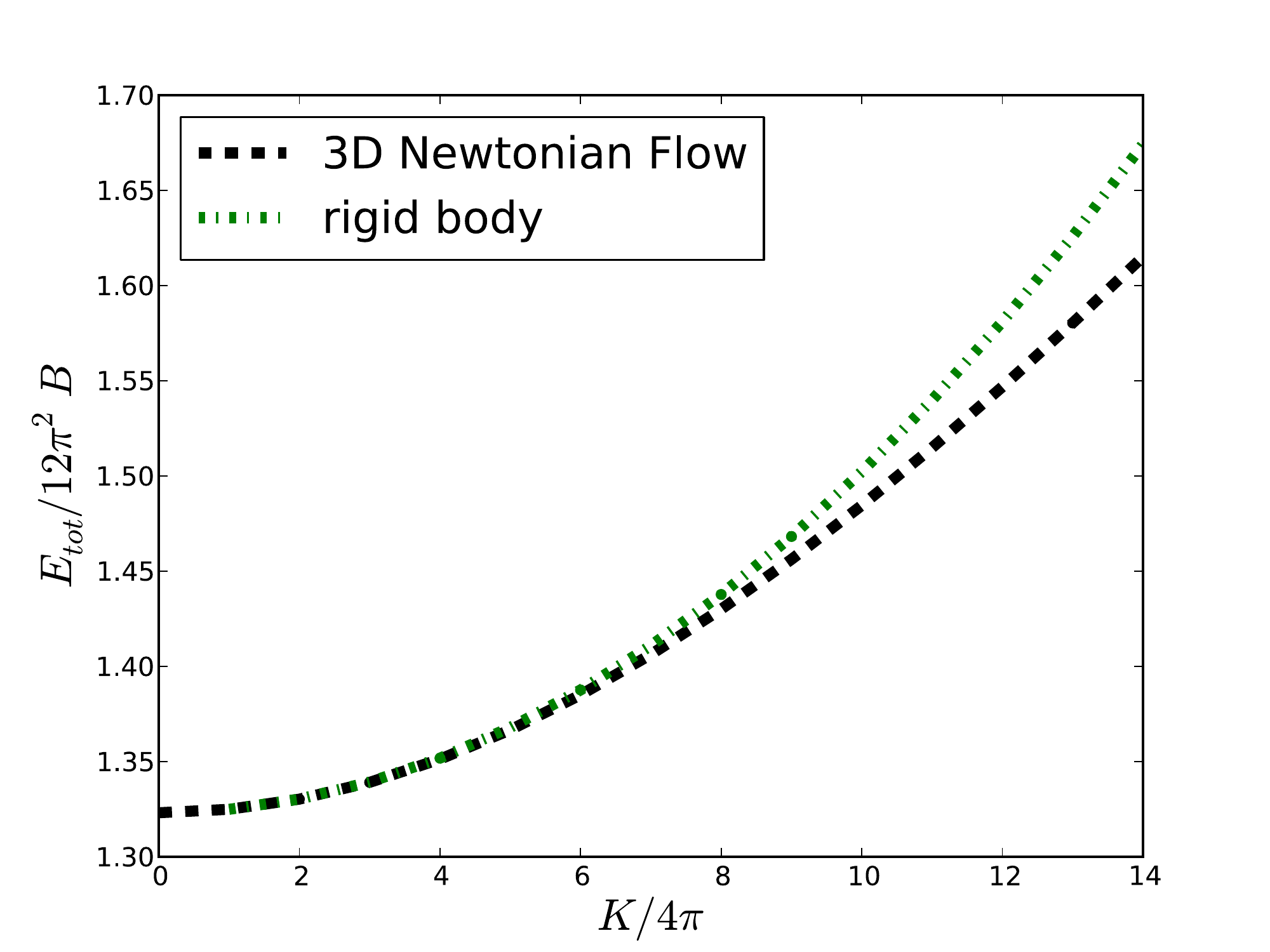}}\\
\subfigure[\,Inertia vs angular frequency]{\includegraphics[totalheight=6.cm]{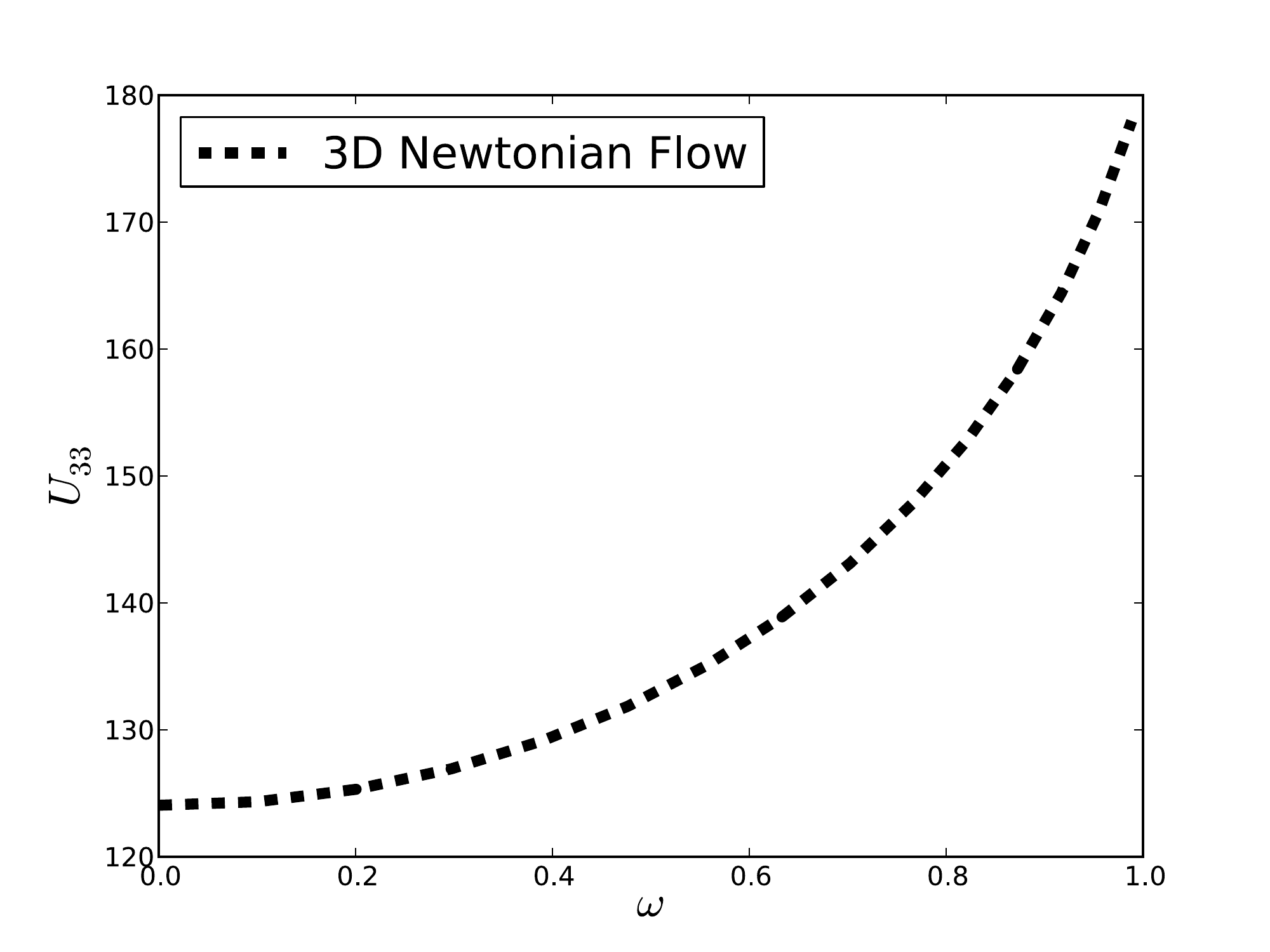}}
\subfigure[\,Inertia-Spin relationship]{\includegraphics[totalheight=6.cm]{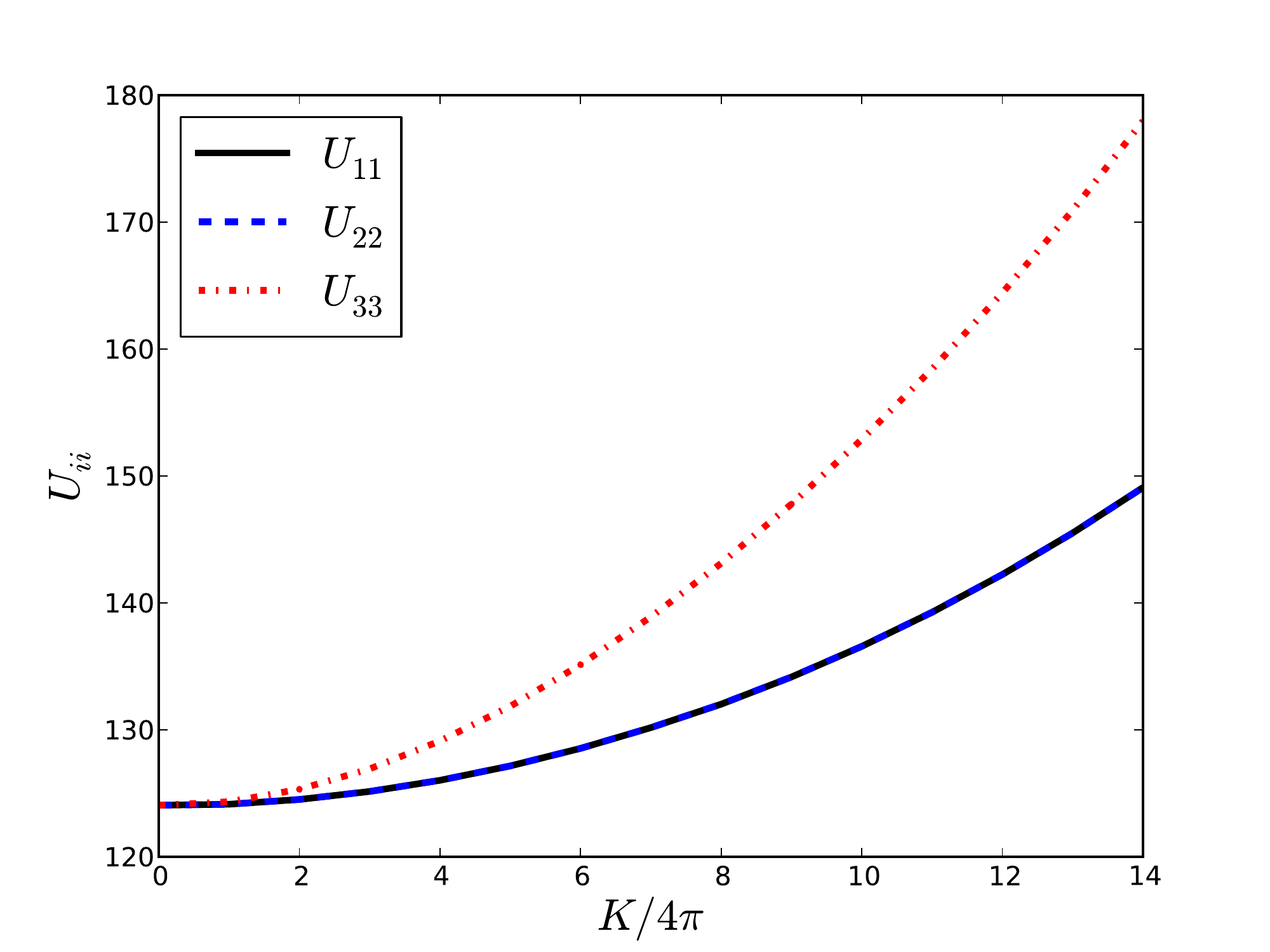}}
\caption{Isospinning $B=3$ Skyrmion for $\mu=1$. A start configuration is numerically minimized using 3D modified Newtonian flow on a $(200)^3$ grid with a lattice spacing of $\Delta x=0.1$ and a time step size $\Delta t=0.01$. We choose $\boldsymbol{\widehat{K}}=(0,0,1)$ as our isorotation axis.}
\label{Fig_B3_Kz}
\end{figure}

\begin{figure}[!htb]
\centering
\includegraphics[totalheight=11.0cm]{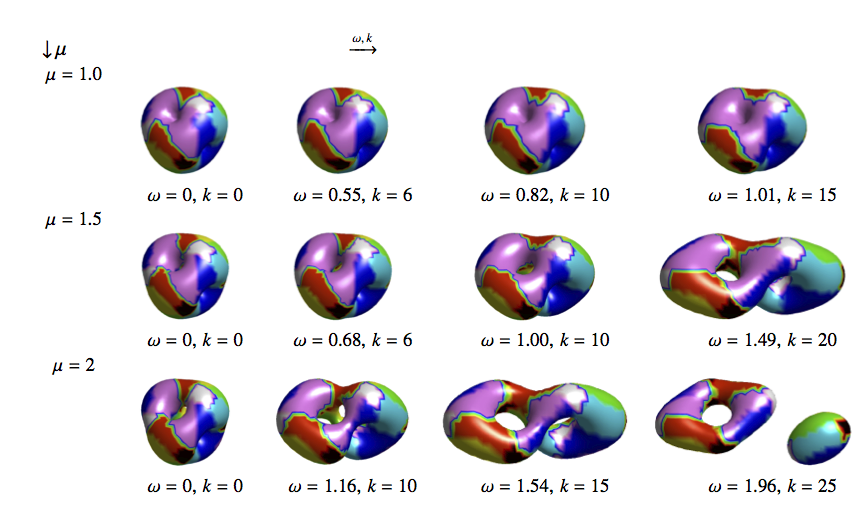}
\caption{Baryon density isosurfaces of isospinning charge-3 Skyrmion solutions for a range of mass values $\mu$. Each baryon density isosurface corresponds to the value $\mathcal{B}=0.1$. The isorotation axis is chosen to be $\boldsymbol{\widehat{K}}=(0,0,1)$.}
\label{Fig_B3_Sky_Iso}
\end{figure}

For $B=3$, we isorotate the minimal-energy tetrahedron about its $\boldsymbol{\widehat{K}}=(0,0,1)$ axis. Performing a damped field evolution for the mass value $\mu=1$ we obtain the energy and inertia dependencies on $\omega$ and $K$ displayed in Fig.~\ref{Fig_B3_Kz}. 
The corresponding baryon density isosurfaces for a range of mass values $\mu$ can be found in Fig.~\ref{Fig_B3_Sky_Iso}. 

\begin{figure}[!htb]
\subfigure[\,$\mu=1.5$]{\includegraphics[totalheight=6.cm]{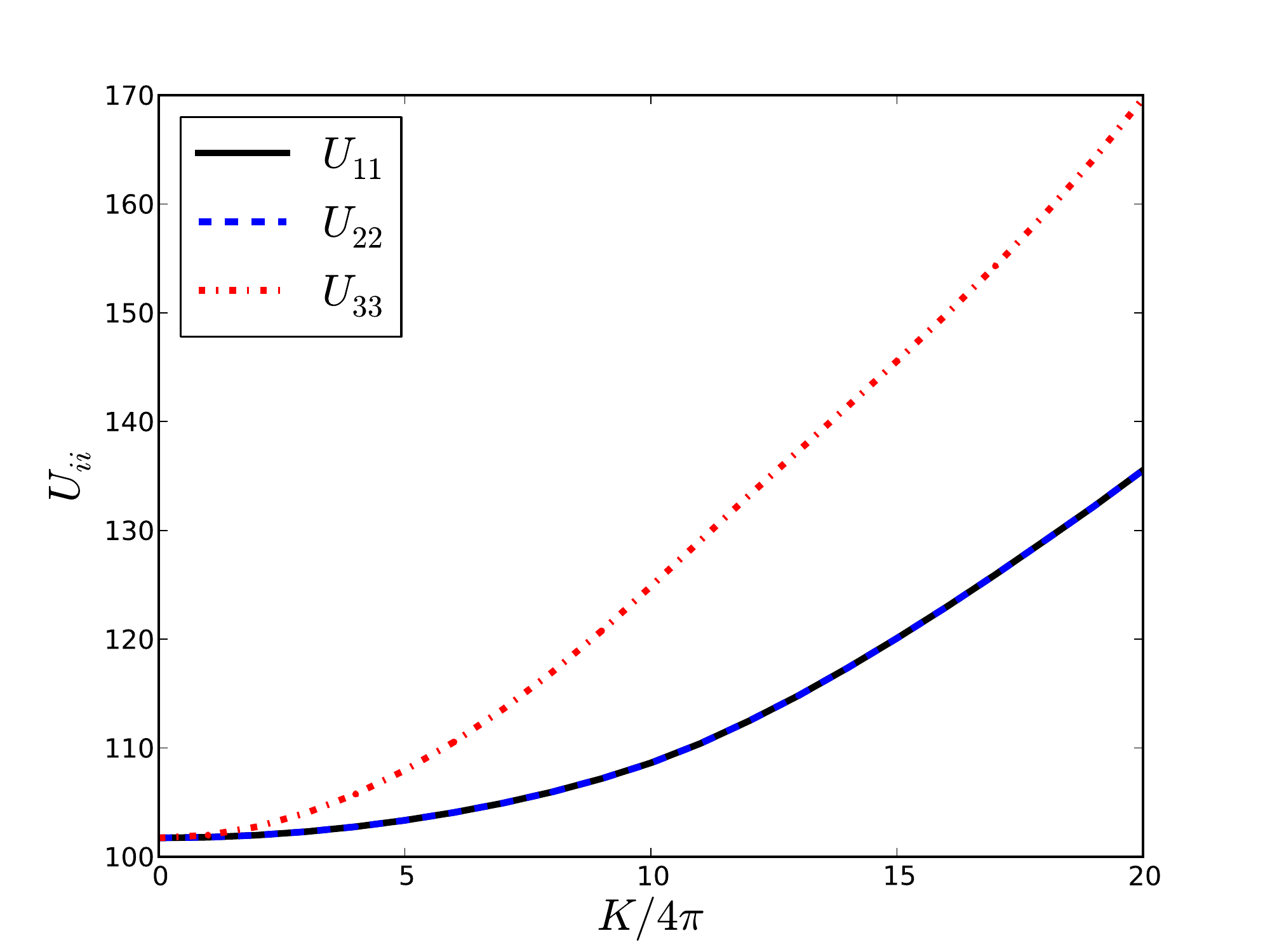}}
\subfigure[\,$\mu=2$]{\includegraphics[totalheight=6.cm]{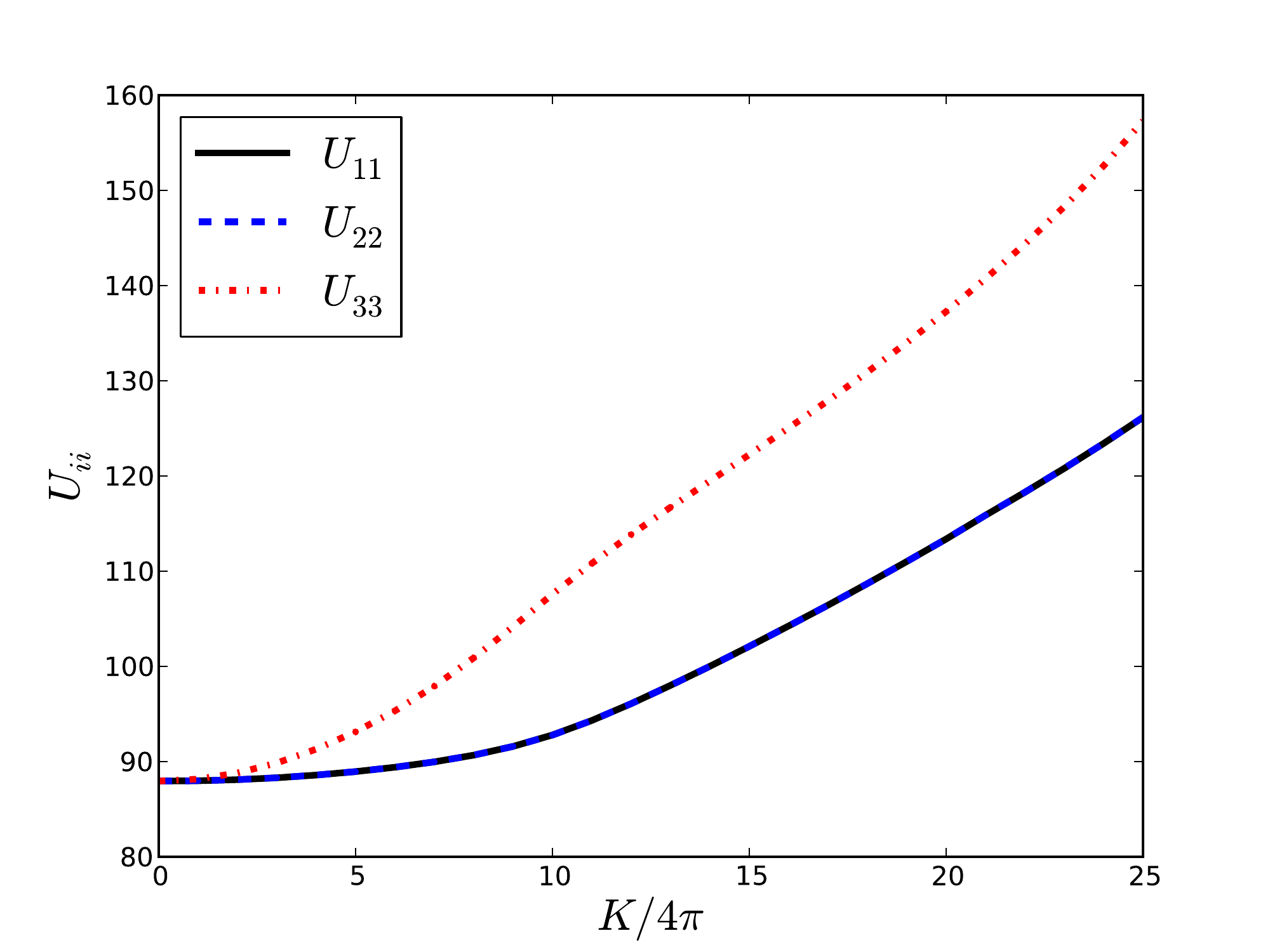}}
\caption{Diagonal elements of the isospin inertia tensor $U_{ij}$ as a function of isospin $K$ for $B=3$ Skyrme configurations with pion mass value $\mu=1.5,2$ and isospinning about the $\boldsymbol{\widehat{K}}=(0,0,1)$ axis.}
\label{B3Kz_Udiag_mass}
\end{figure}

As $\mu$ increases, the soliton's deformations due to centrifugal effects become more apparent. For $\mu=2$ the isospinning charge-3 Skyrmion solution breaks into a toroidal $B=2$ Skyrmion solution and a $B=1$ Skyrmion before reaching its upper frequency limit $\omega_\text{crit}=\mu$. Note that with increasing angular velocity $\omega$ the isospinning $B=3$ Skyrmion seems to pass through a distorted ``pretzel\,'' configuration -- a state that has previously been found to be metastable \cite{Walet:1996he,Battye:1996nt} for vanishing isospin $K$. For mass value $\mu=1.5$ the tetrahedral charge-3 Skyrmion does not break into lower-charge Skyrmions when increasing the angular frequency $\omega$. As $\omega$ increases, the charge-3 tetrahedron slowly deforms into  the ``pretzel\,''-like configuration which appears to be of lower energy than an isospinning solution with tetrahedral symmetry. Even for $\mu=1$ the tetrahedral symmetry is broken as $\omega$ increases [see inertia-spin relationship shown in Fig.~\ref{Fig_B3_Kz}(d) and baryon density isosurfaces in Fig.~\ref{Fig_B3_Sky_Iso}]. In fact, the isospinning $B=3$ Skyrme soliton (with $\mu=1$) starts to violate tetrahedral symmetry for angular frequencies $\omega>\omega_{\text{SB}}=0.06$   ($K_{\text{SB}} = 1.1\times 4\pi$). The breaking of tetrahedral symmetry occurs at  $K_{\text{SB}}= 0.8\times 4\pi$ ($\omega_{\text{SB}}= 0.1$)  and   $K_{\text{SB}}= 0.5\times 4\pi$ ($\omega_{\text{SB}}= 0.12$) for $B=3$ Skyrmions with mass $\mu=1.5$ and $\mu=2$, respectively. Furthermore, note that the $U_{11}$ and $U_{22}$ curves shown in Fig.~\ref{Fig_B3_Kz}(d) and Fig.~\ref{B3Kz_Udiag_mass} lie on top of each other.

Fig.~\ref{Fig_B3B4_rigid} shows the deviation of the deformed isospinning Skyrmion solution from its rigid rotor approximation as a function of $K$ for various mass values. Again, we note that the soliton's energy $E_{\text{tot}}(K)$ can be significantly lower than that of the rigidly isospinning Skyrmion solution. For example for $\mu=1.5$ the rigid-body approach predicts an approximately $10\%$ larger energy value close to the cutoff frequency $\omega_\text{crit}=\mu=1.5$ ($K_{\text{crit}}= 20\times 4\pi$). As already observed for $B=1$ and $B=2$ Skyrmion solutions, the accuracy of the rigid-body approximation improves for a given isospin value $K$ as $\mu$ increases (see Fig.~\ref{Fig_B3B4_rigid}).  

\subsubsection{B=4}

\begin{figure}[!htb]
\centering
\subfigure[\,Total energy vs angular frequency ]{\includegraphics[totalheight=6.cm]{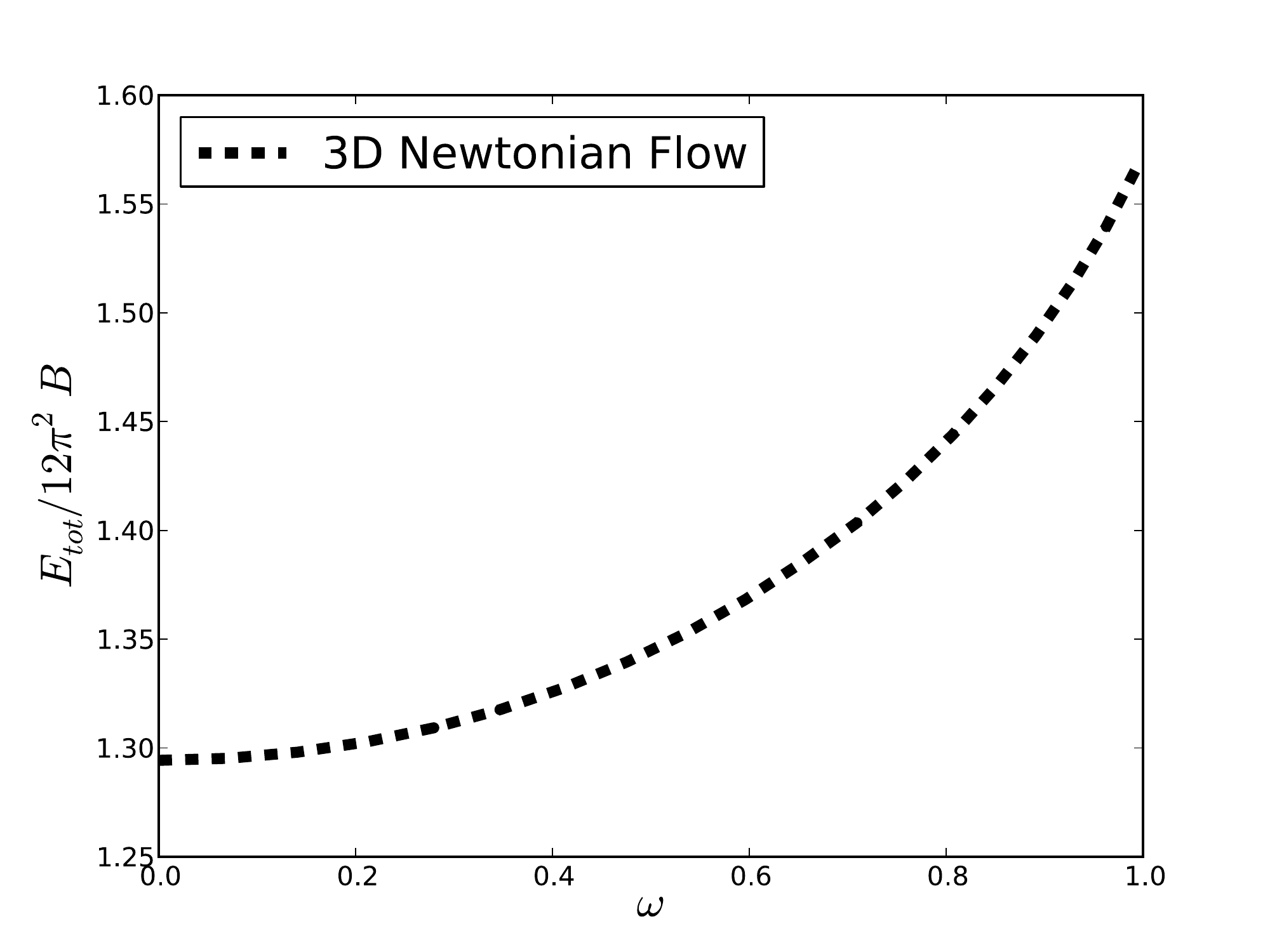}}
\subfigure[\,Mass-Spin relationship ]{\includegraphics[totalheight=6.cm]{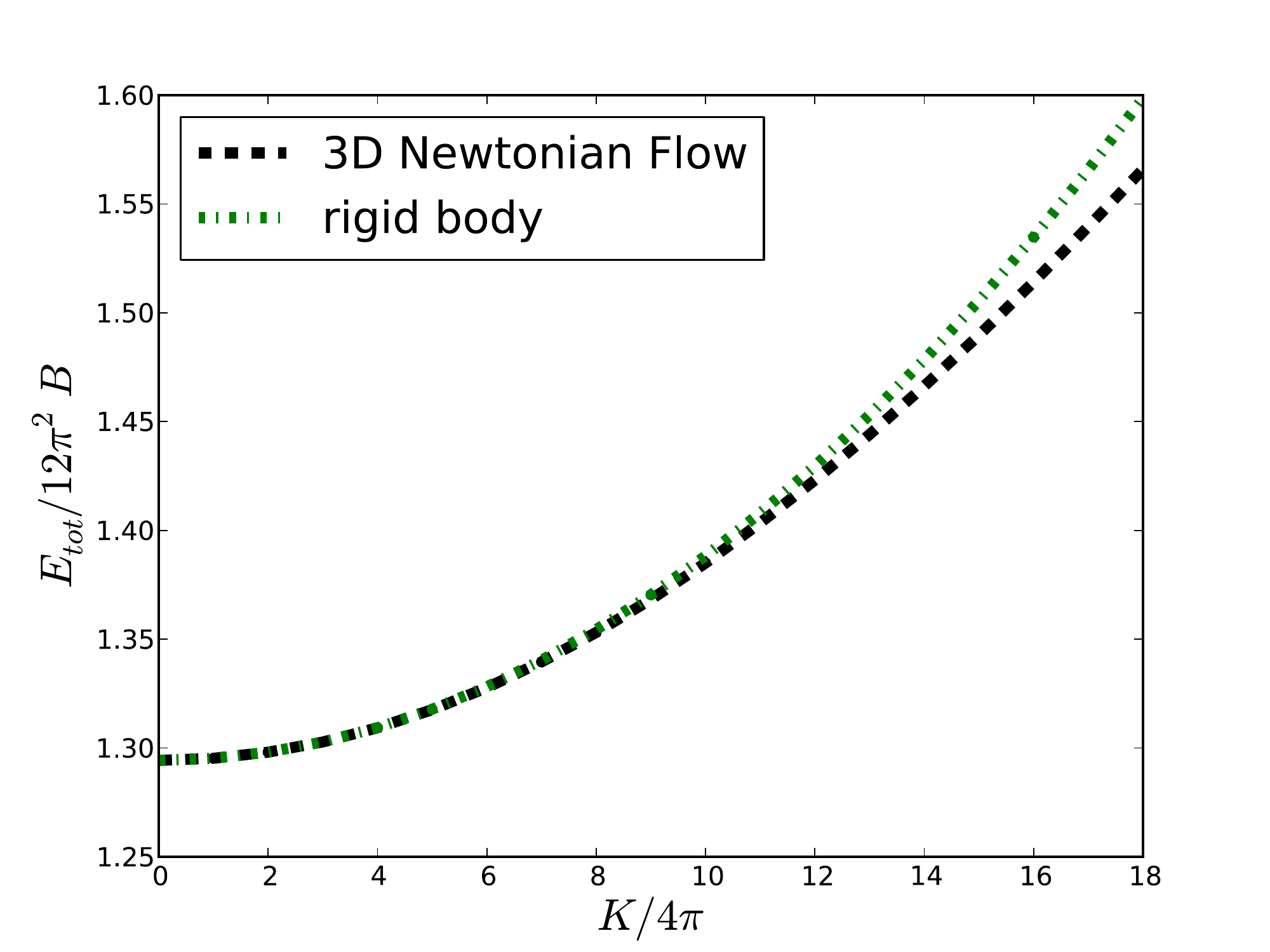}}\\
\subfigure[\,Inertia vs angular frequency]{\includegraphics[totalheight=6.cm]{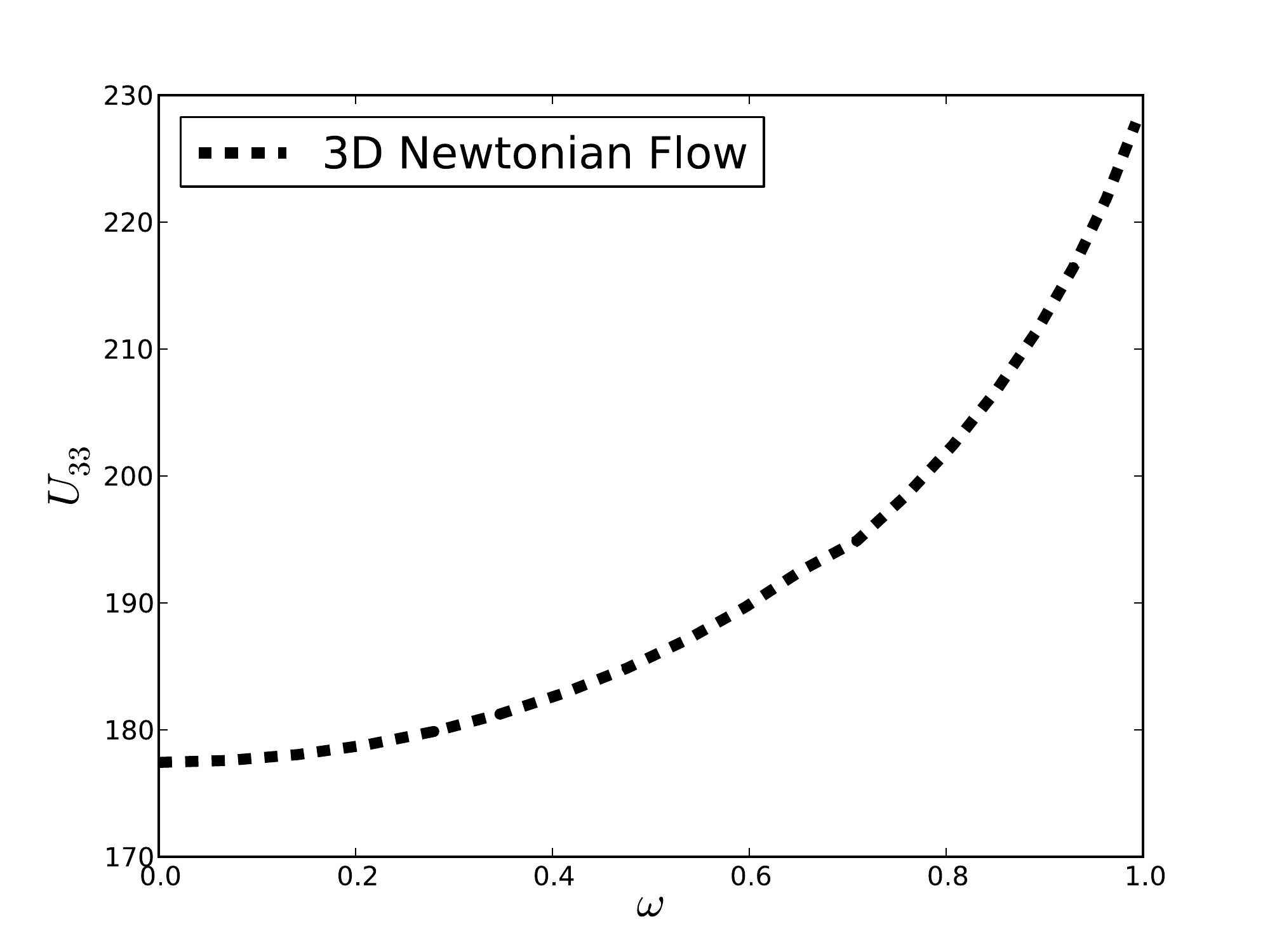}}
\subfigure[\,Inertia-Spin relationship]{\includegraphics[totalheight=6.cm]{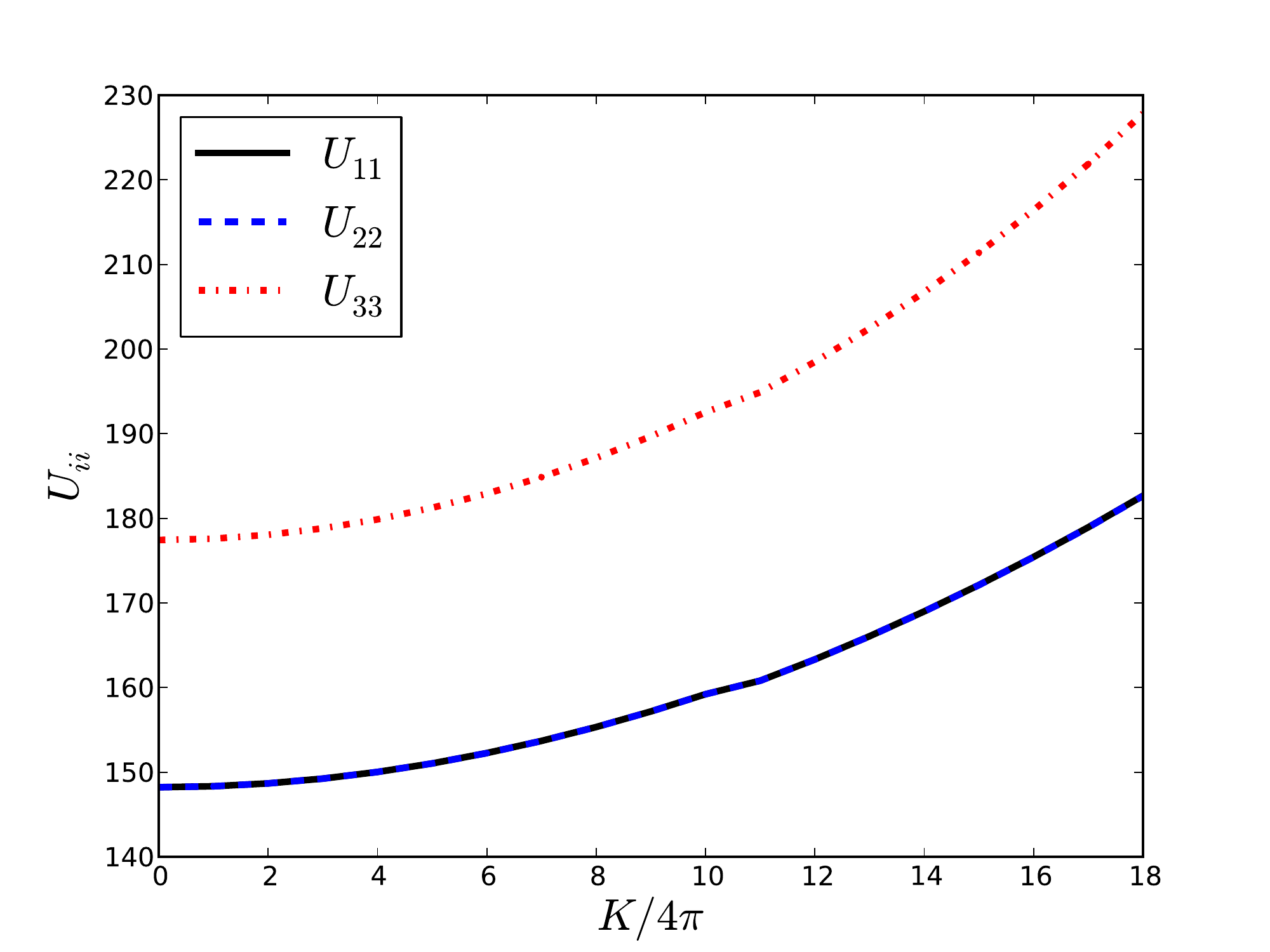}}
\caption{Isospinning $B=4$ Skyrmion. $(\mu=1)$. A start configuration is numerically minimized using 3D modified Newtonian flow on a $(200)^3$ grid with a lattice spacing of $\Delta x=0.1$ and a time step size $\Delta t=0.01$. We choose $\boldsymbol{\widehat{K}}=(0,0,1)$ as our isorotation axis.}
\label{Fig_B4_Kz}
\end{figure}

\begin{figure}[!htb]
\centering
\includegraphics[totalheight=11.0cm]{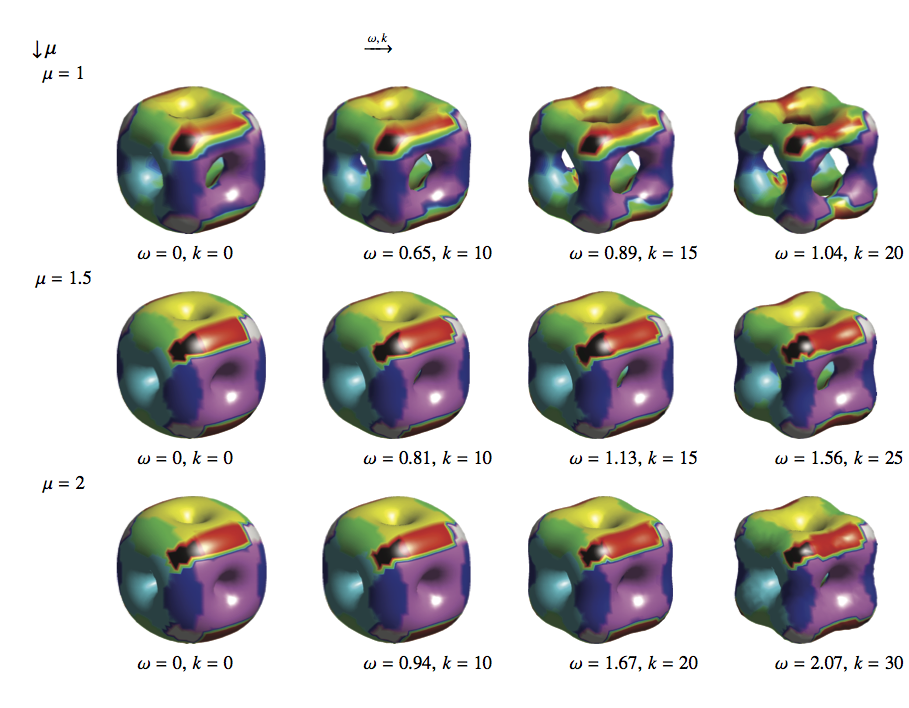}
\caption{We display the baryon density isosurfaces (not to scale) of isospinning, octahedrally symmetric charge-4 Skyrmion solutions for a range of mass values. The isorotation axis is chosen to be $\boldsymbol{\widehat{K}}=(0,0,1)$.}
\label{Fig_B4_Sky_Iso}
\end{figure}

\begin{figure}[!htb]
\centering
\subfigure[\,$B=3$]{\includegraphics[totalheight=6.cm]{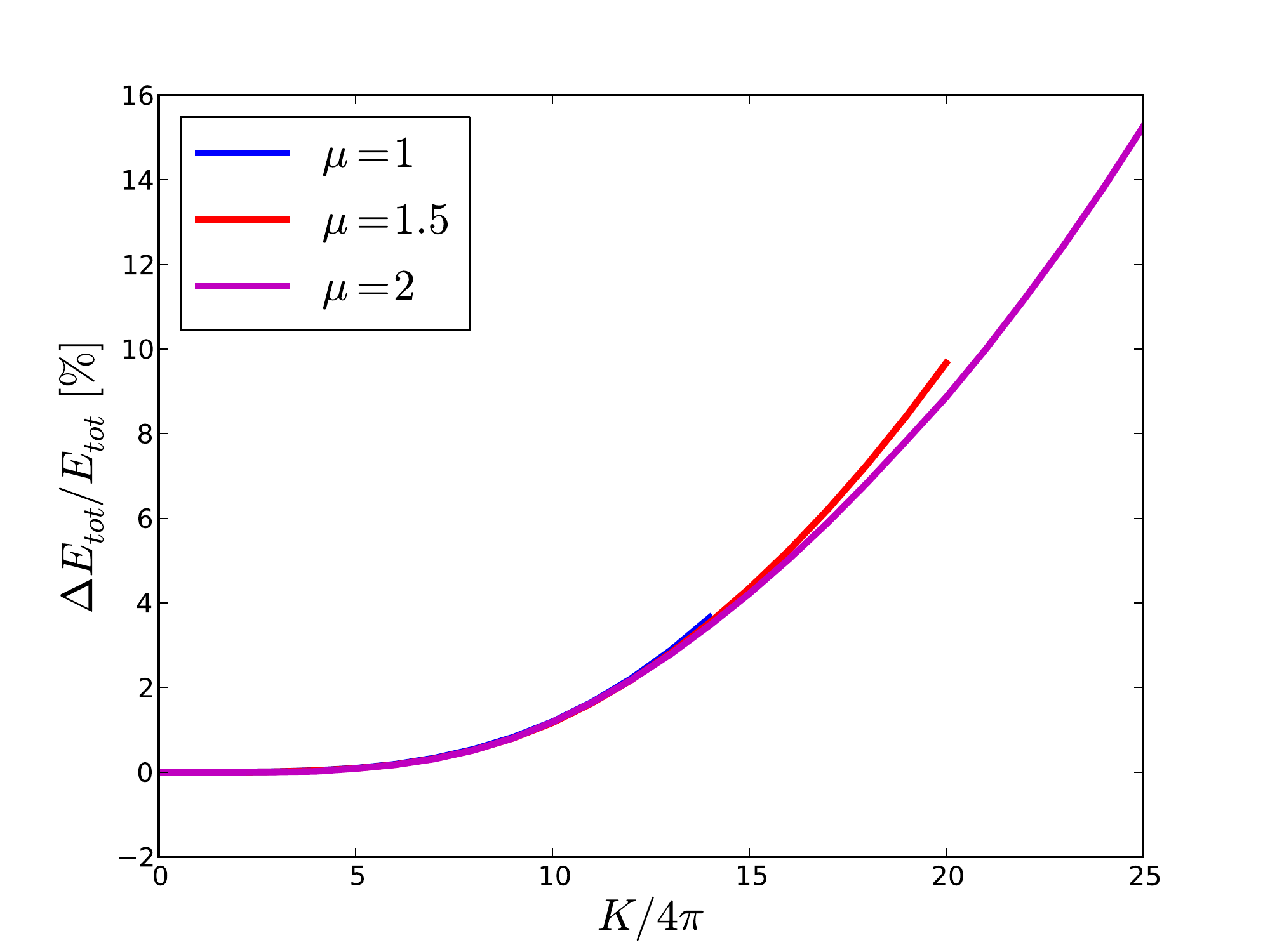}}
\subfigure[\,$B=4$]{\includegraphics[totalheight=6.cm]{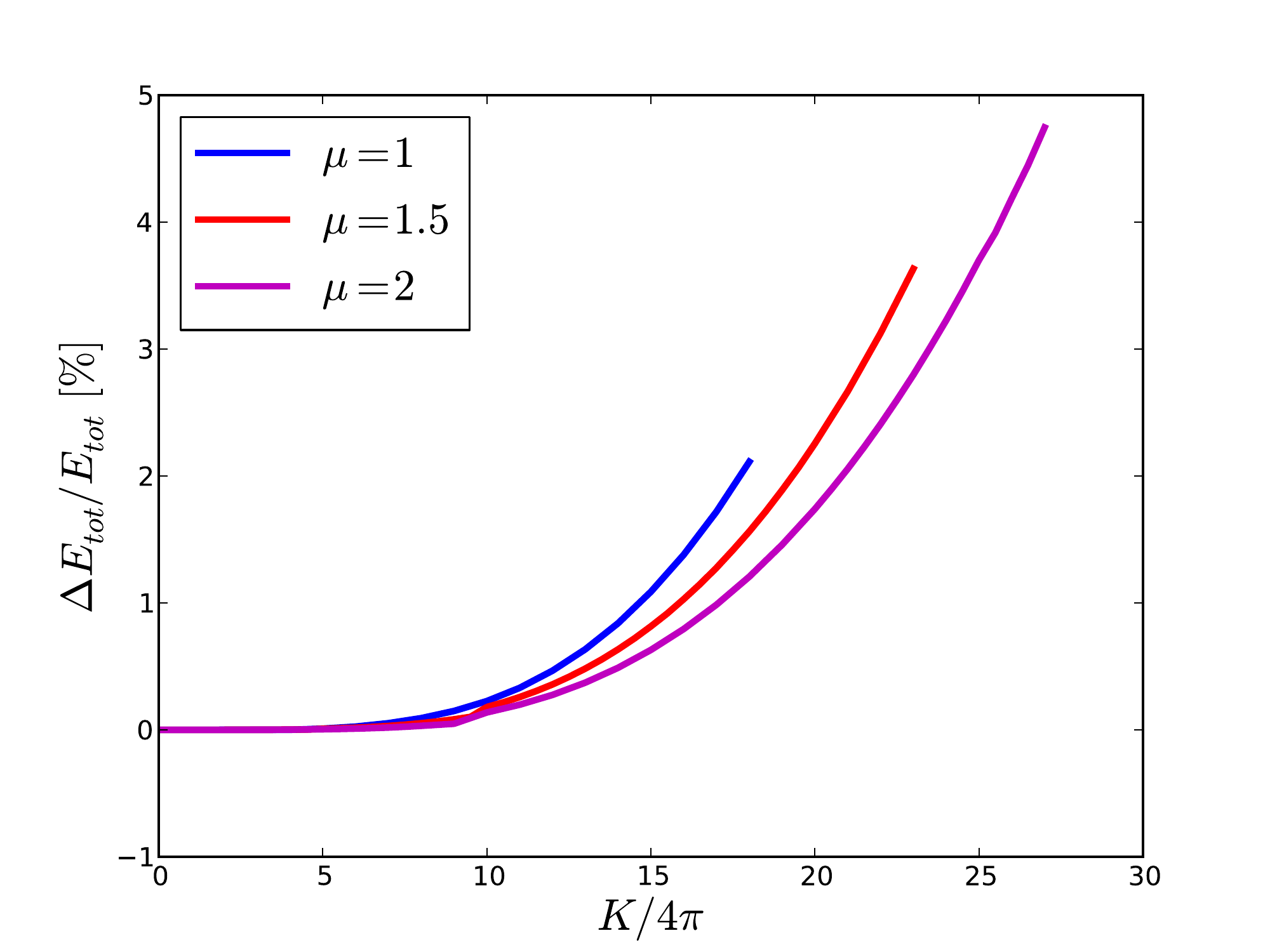}}
\caption{The deviation $\Delta E_{\text{tot}}/E_\text{tot}=\left(E_\text{Rigid}-E_{\text{tot}}\right)$ from the rigid body approximation for charge-3 and charge-4 Skyrmions as a function of
 isospin $K$ for various rescaled mass values $\mu$. In each case the isorotation axis is chosen to be $\boldsymbol{\widehat{K}}=(0,0,1)$.}
\label{Fig_B3B4_rigid}
\end{figure}

 We do not observe any breaking of the octahedral symmetry for the minimal-energy $B=4$ Skyrmion solution when isospinning the configuration around the $\boldsymbol{\widehat{K}}=(0,0,1)$ axis (see the baryon density isosurfaces shown in Fig.~\ref{Fig_B4_Sky_Iso} as function of $K$ for various mass values $\mu$). Plotting the elements of the isospin, spin and mixed  inertia tensor as a function of $K$ confirms that the isospinning charge-4 Skyrmion preserves $O_h$ symmetry up to the maximal angular frequency $\omega=\mu$. In particular, the isospin inertia tensor $U_{ij}$ is diagonal and satisfies $U_{11}=U_{22}\ne U_{33}$ for all allowed values of $K$ [compare Fig.~\ref{Fig_B4_Kz}(d) for mass value $\mu=1$]. Although, the Skyrmion's shape turns out to be largely independent of the angular frequency $\omega$, the Skyrmion's size increases monotonically with $\omega$. In Fig.~\ref{Fig_B3B4_rigid} we compare our numerical energy values for arbitrarily deforming Skyrmion solutions with those obtained when assuming a rigidly rotating, cubically symmetric $B=4$ Skyrmion solution. Evidently, the accuracy of the rigid-body approximation improves with increasing soliton mass and baryon number $B$. For example, for $\mu=2$ and $B=4$ the energy values predicted by the rigid-body formula are at the critical frequency ($\omega_{\text{crit}}=\mu$) roughly $5\%$ larger.  For comparison, for charge-1 Skyrmions  (with mass value $\mu=2$ and $\omega=\omega_{\text{crit}}$) the rigid-body approximation gives energy values which are $15\%$ larger than those for the deforming soliton solutions.

\begin{figure}[!htb]
\centering
\includegraphics[totalheight=11.0cm]{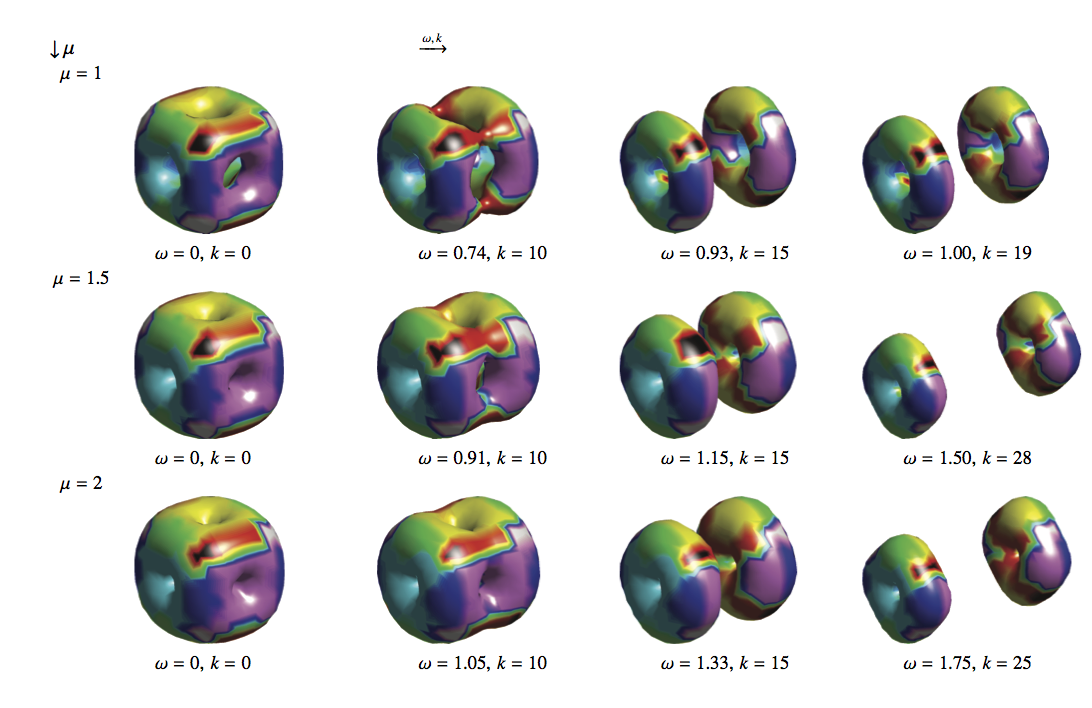}
\caption{We display the baryon density isosurfaces (not to scale) of isospinning, octahedrally symmetric charge-4 Skyrmion solutions for a range of mass values $\mu$. Each baryon density isosurface corresponds to the value $\mathcal{B}=0.2$. The isorotation axis is chosen to be $\boldsymbol{\widehat{K}}=(0,1,0)$. }
\label{Fig_B4_Sky_Iso_Ly}
\end{figure}

\begin{figure}[!htb]
\subfigure[\,$\mu=1$]{\includegraphics[totalheight=4.cm]{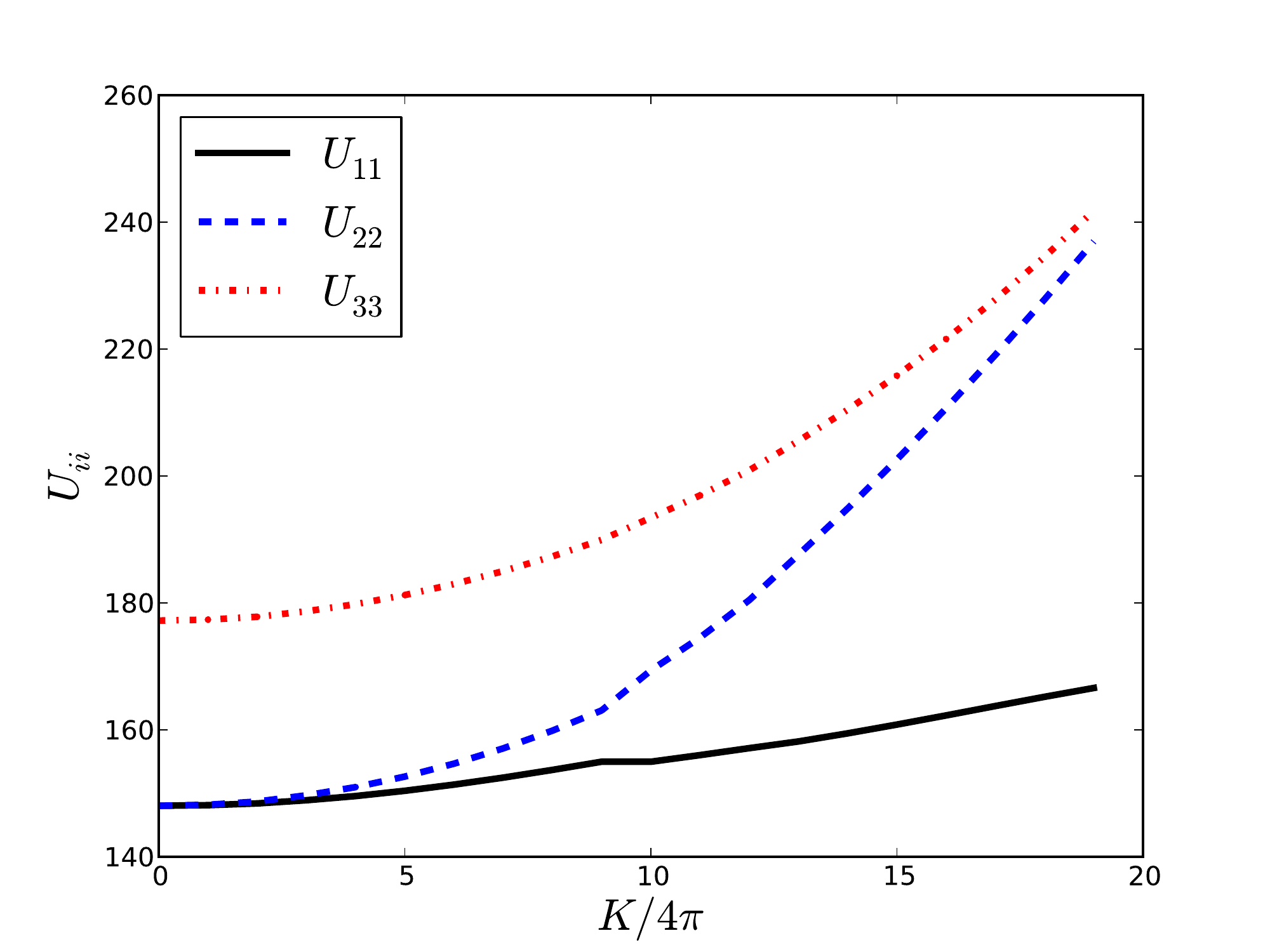}}
\subfigure[\,$\mu=1.5$]{\includegraphics[totalheight=4.cm]{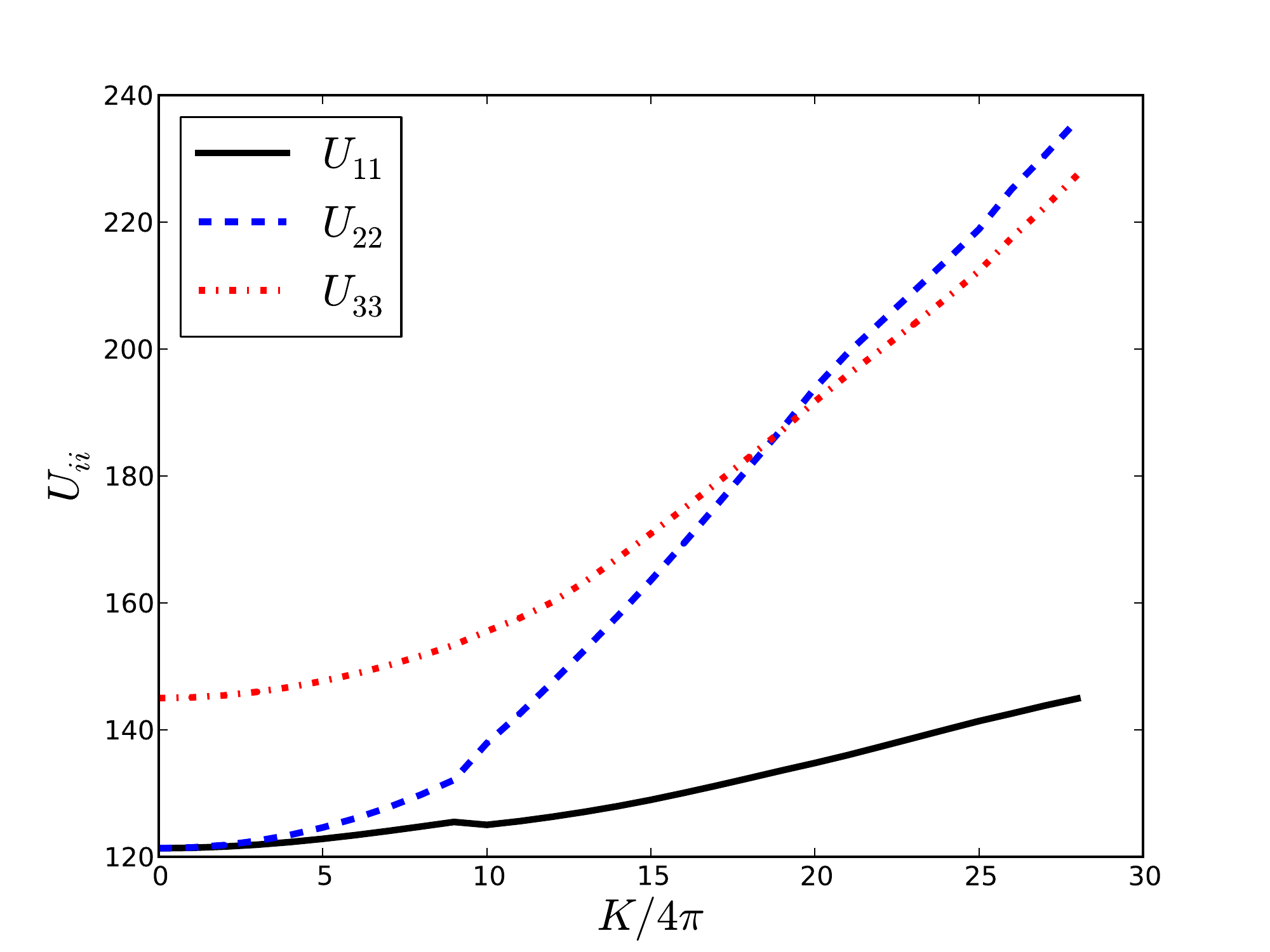}}
\subfigure[\,$\mu=2$]{\includegraphics[totalheight=4.cm]{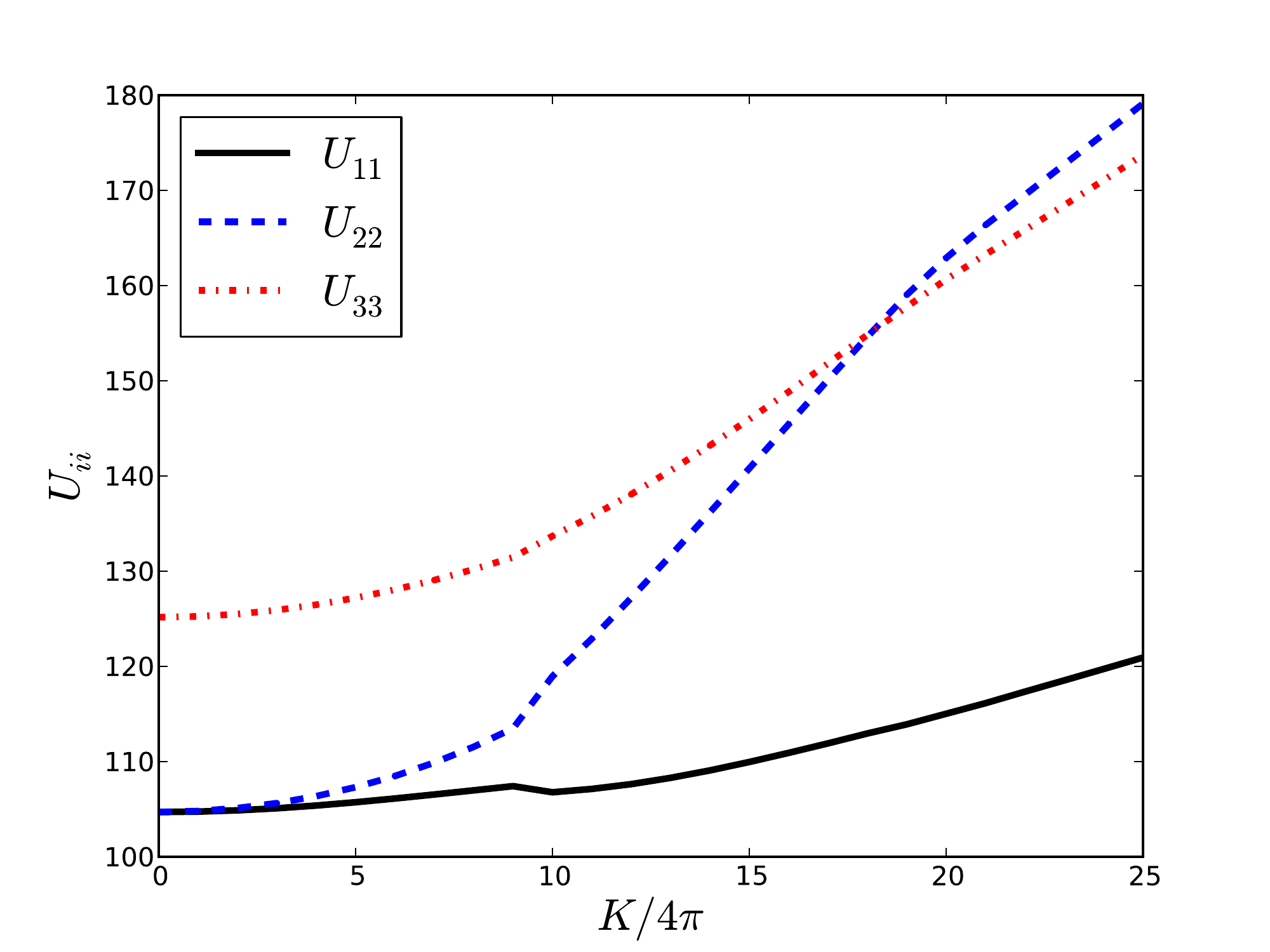}}
\caption{Diagonal elements of the isospin inertia tensor $U_{ij}$ as a function of isospin $K$ for $B=4$ Skyrme configurations with pion mass value $\mu=1,1.5,2$ and isospinning about the $\boldsymbol{\widehat{K}}=(0,1,0)$-axis.}
\label{B4Ky_Udiag_mass}
\end{figure}

\begin{figure}[!htb]
\centering
\includegraphics[totalheight=8.0cm]{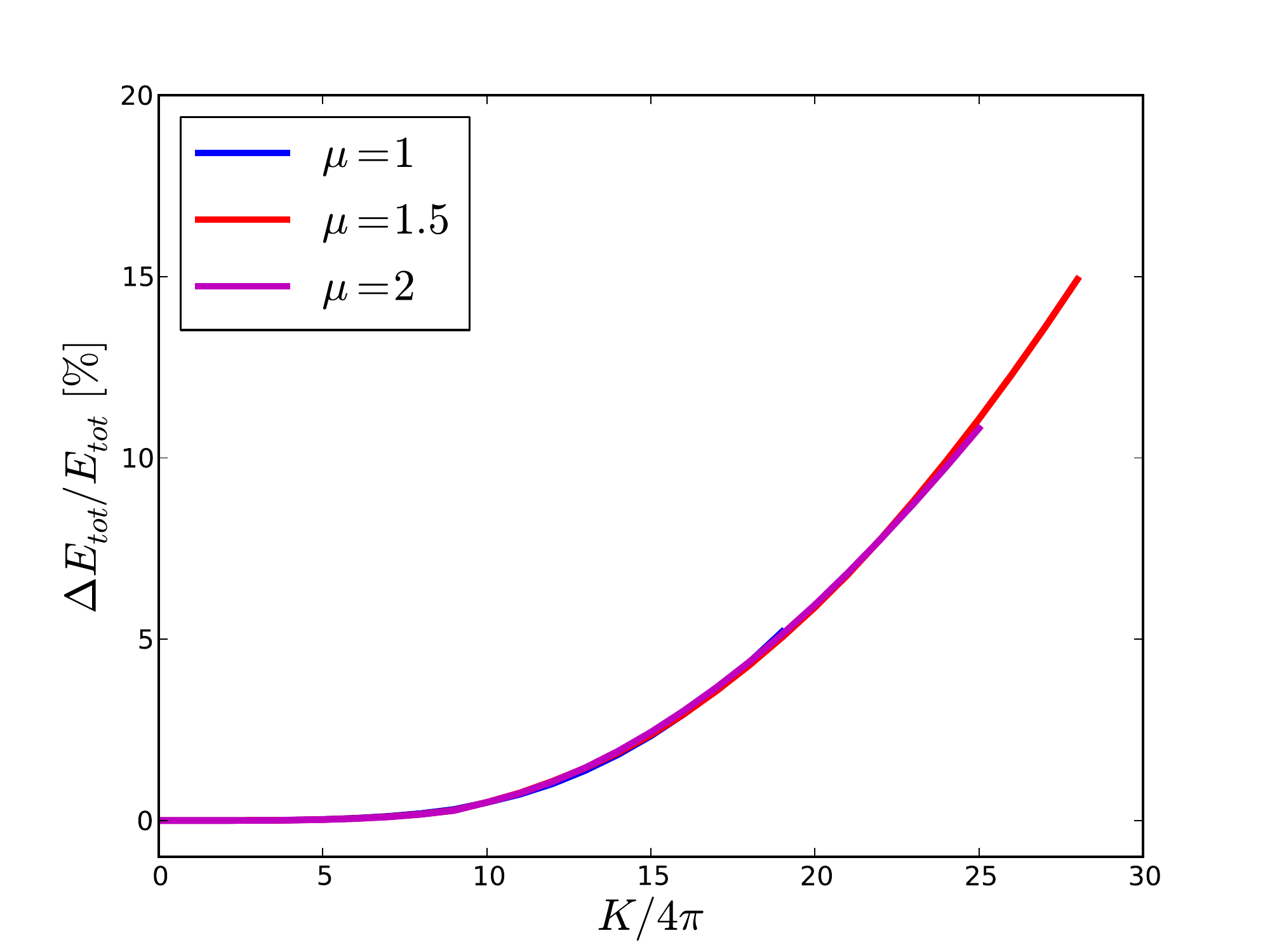}
\caption{The deviation $\Delta E_{\text{tot}}/E_\text{tot}=\left(E_\text{Rigid}-E_{\text{tot}}\right)$ from the rigid body approximation for charge-4 Skyrmions as a function of isospin $K$ for various rescaled mass values $\mu$. The isorotation axis is chosen to be $\boldsymbol{\widehat{K}}=(0,1,0)$.}
\label{Fig_B4y_rigid}
\end{figure}

When choosing $\boldsymbol{\widehat{K}}=(1,0,0)$ or $\boldsymbol{\widehat{K}}=(0,1,0)$ as our isorotation axis, we observe that with increasing angular frequency $\omega$ the octahedrally symmetric charge-4 Skyrmion solution becomes unstable to break up into a pair of toroidal $B=2$ Skyrmions. In Fig.~\ref{Fig_B4_Sky_Iso_Ly} we display the baryon density isosurfaces for isospinning $B=4$ Skyrmion solutions with rescaled pion masses $\mu=1,1.5$ and $2$. The breaking of the cubic symmetry is found to occur at $K_{\text{SB}}= 3.58 \times 4\pi$ ($\omega_{\text{SB}} = 0.18$) for $\mu=1$, $K_{\text{SB}}= 3.32\times 4\pi$ ($\omega_{\text{SB}} = 0.28$) for $\mu=1.5$ and $K_{\text{SB}}= 2.46\times 4\pi$ ($\omega_{\text{SB}}= 0.31$) for $\mu=2$ (see the nonzero elements of the isospin inertia tensor plotted in Fig.~\ref{B4Ky_Udiag_mass} as a function of $K$). 
As shown in Fig.~\ref{Fig_B4y_rigid}, for pion mass values up to 2 the energy values predicted by the rigid-body formula can be up to 15\% higher than those obtained without imposing any spatial symmetries on the isospinning Skyrme configurations.

Similar to the $B=2$ case, we find that the Skyrmion configuration of lowest energy for given isospin $K$ is the solution for which the constituents stay closer together, that is the $B=4$ cube isospinning about  $\boldsymbol{\widehat{K}}=(0,0,1)$. The charge-4 solutions which split into two $D_4$-symmetric charge-2 tori are of higher energy for all classically allowed isospin values $K$  [see energy curves in \ref{Fig_Energy_axes}(b)].

\subsection{Higher Charge Skyrmions: $B=8$}

\begin{figure}[!htb]
\centering
\subfigure[\,$\boldsymbol{\widehat{K}}=(0,1,0)$]{\includegraphics[totalheight=6.cm]{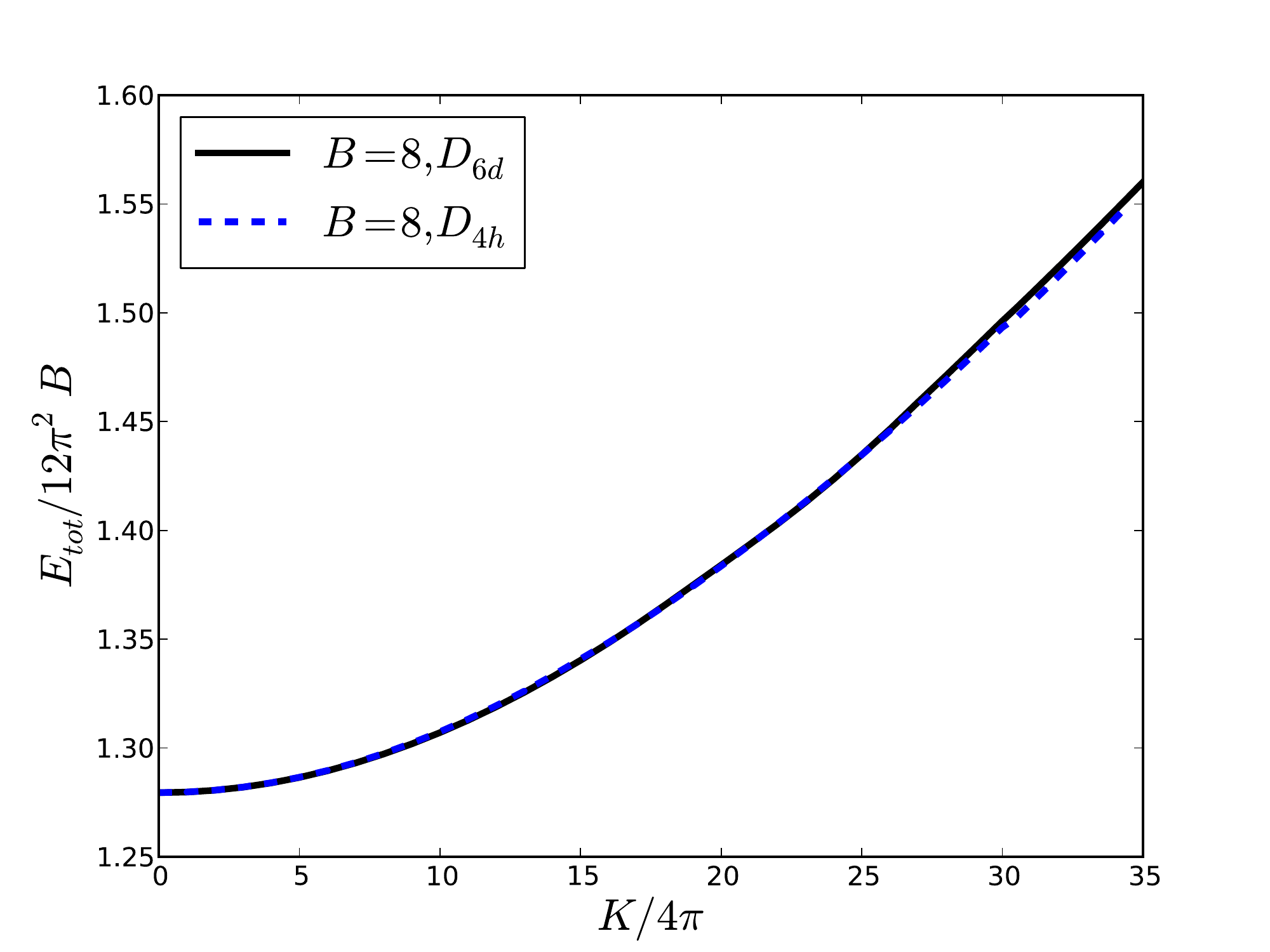}}
\subfigure[\,$\boldsymbol{\widehat{K}}=(0,0,1)$]{\includegraphics[totalheight=6.cm]{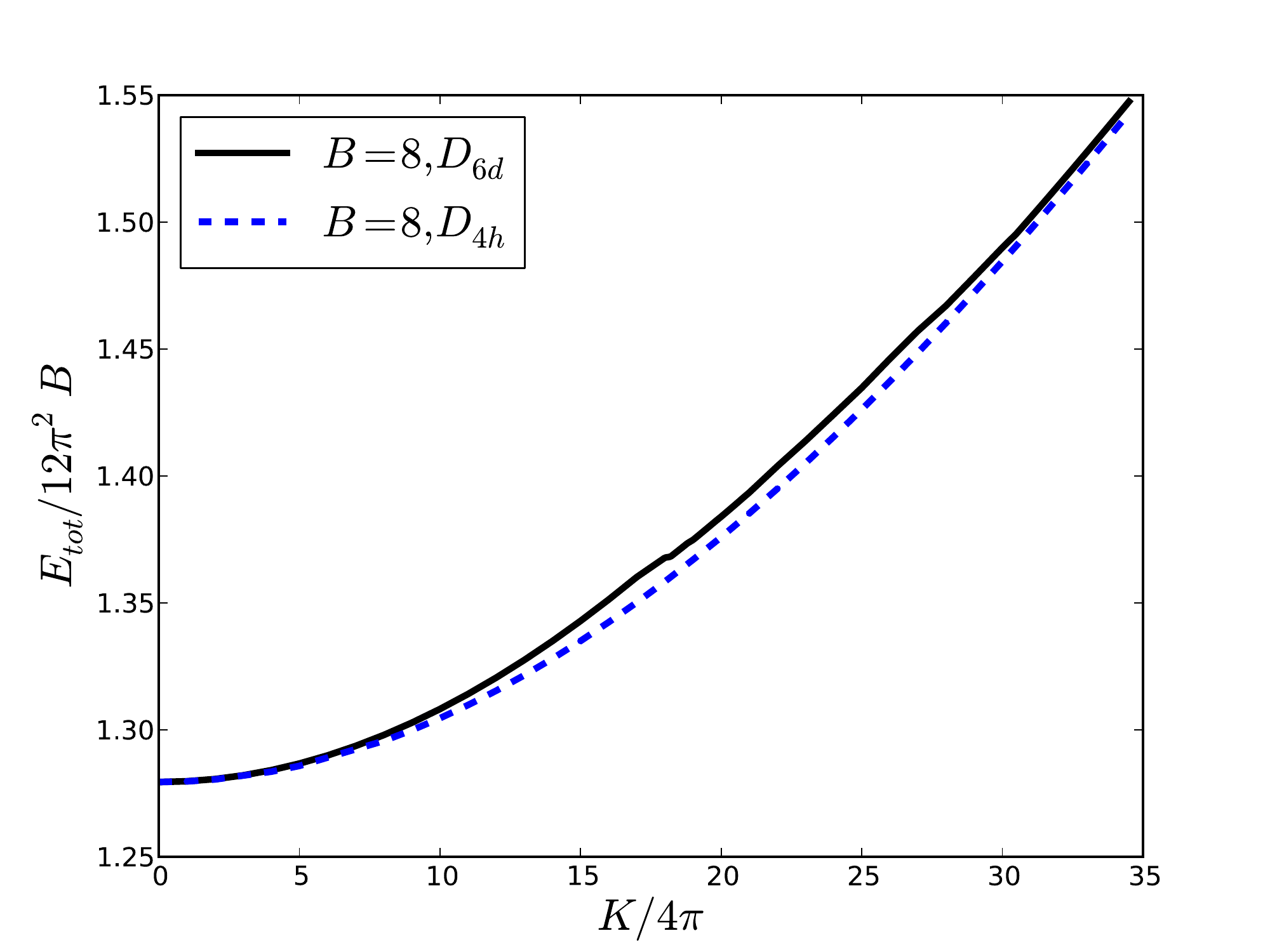}}
\caption{Total energy $E_{\text{tot}}$  as a function of isospin $K$ for $D_{4h}$- and $D_{6d}$- symmetric $B=8$ Skyrme configurations with pion mass value $\mu=1$. The isorotation axes are chosen as indicated above.}
\label{Fig_EB8}
\end{figure}

\begin{figure}[!htb]
\centering
\includegraphics[totalheight=9.0cm]{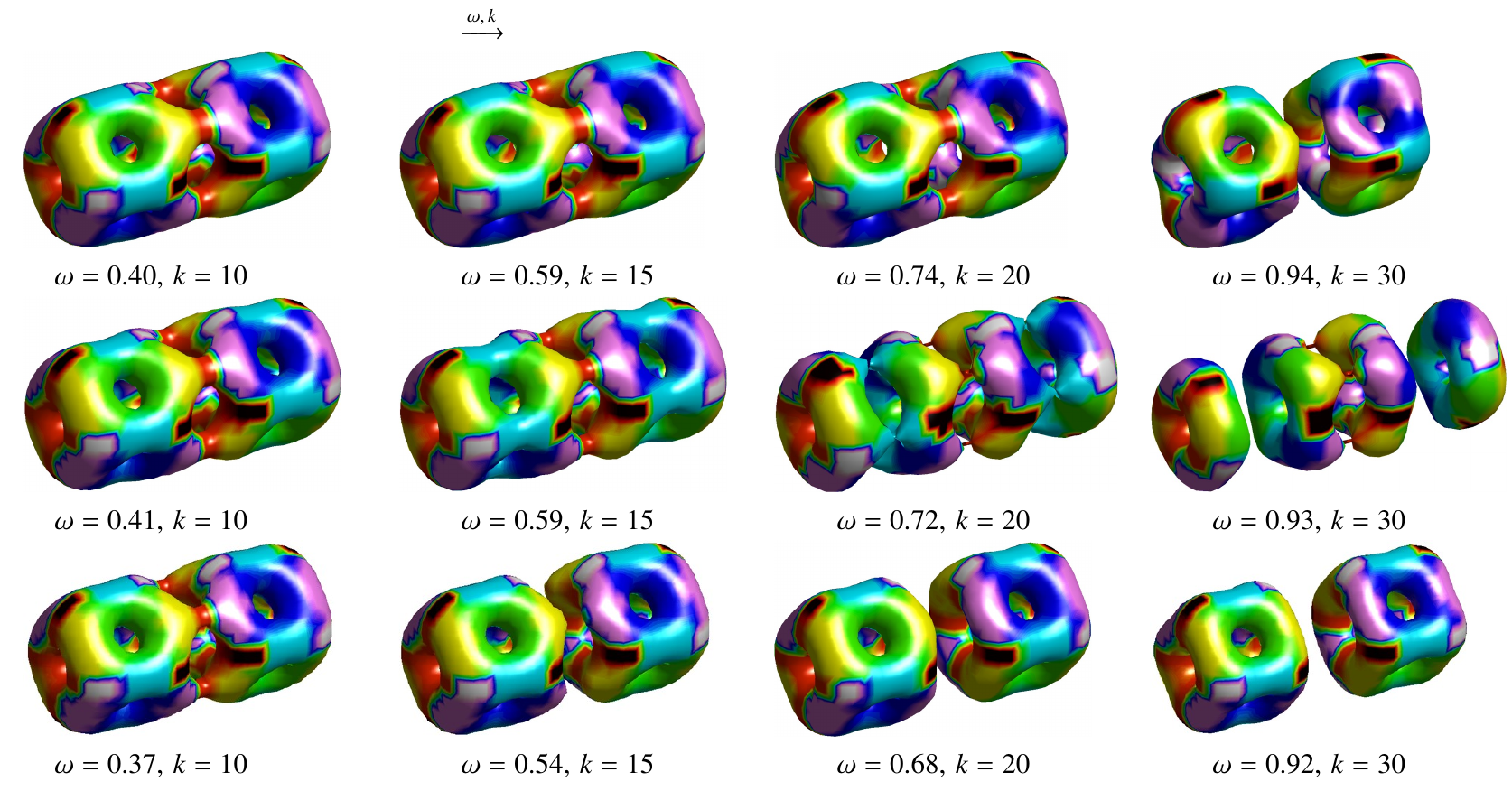}
\caption{Baryon density isosurfaces of isospinning $D_{4h}$-symmetric  charge-$8$ Skyrmion solutions for mass parameter $\mu=1$ as a function of isospin $K$ and angular frequency $\omega$. The isorotation axis is chosen to be $\boldsymbol{\widehat{K}}=(1,0,0)$ (first row), $\boldsymbol{\widehat{K}}=(0,1,0)$ (second row) and $\boldsymbol{\widehat{K}}=(0,0,1)$ (third row). Each baryon density isosurface corresponds to the value $\mathcal{B}=0.15$.}
\label{Fig_B8_Sky_Iso_D4h}
\end{figure}

\begin{figure}[!htb]
\subfigure[\,$\boldsymbol{\widehat{K}}=(1,0,0)$]{\includegraphics[totalheight=4.cm]{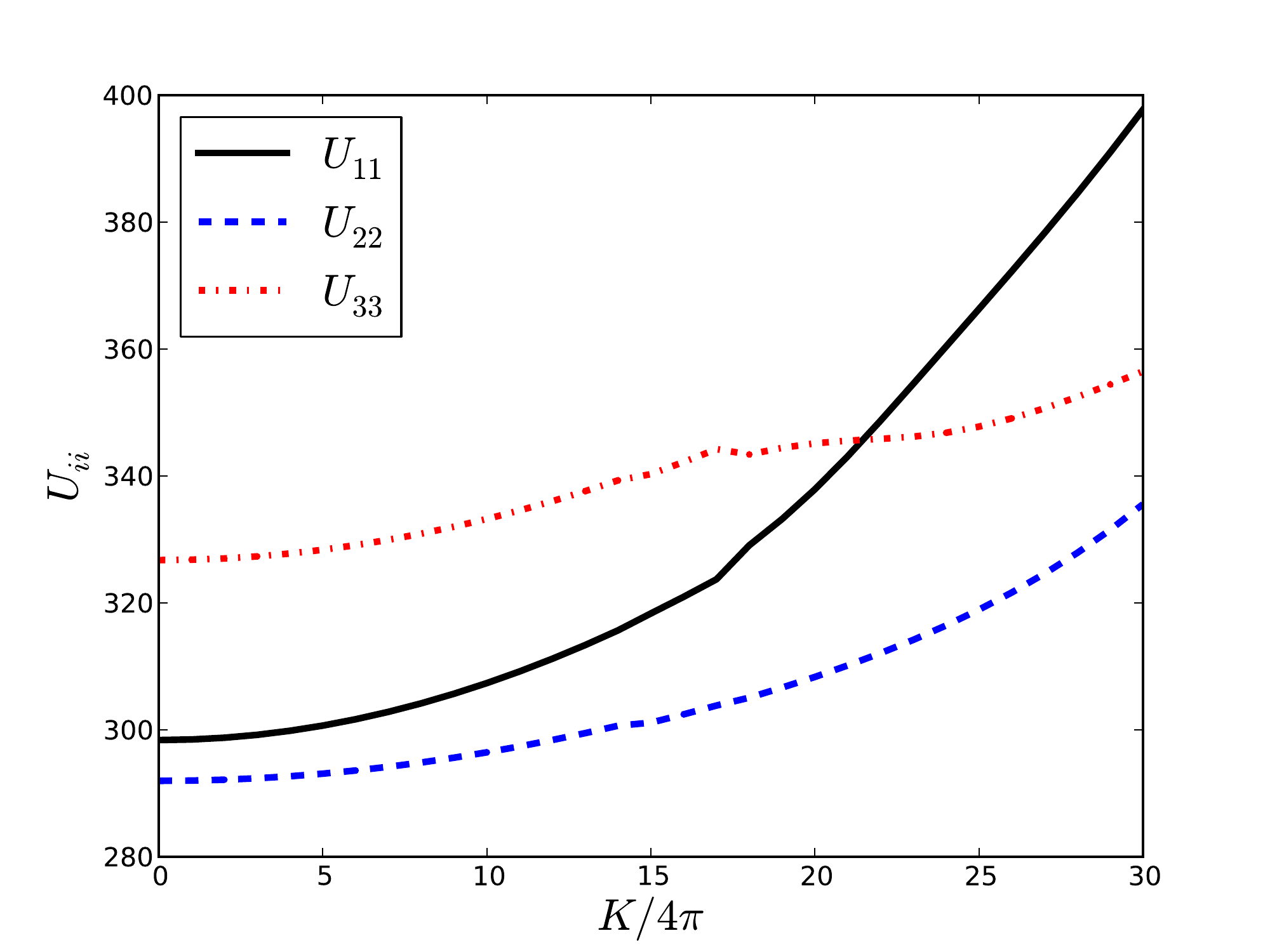}}
\subfigure[\,$\boldsymbol{\widehat{K}}=(0,1,0)$]{\includegraphics[totalheight=4.cm]{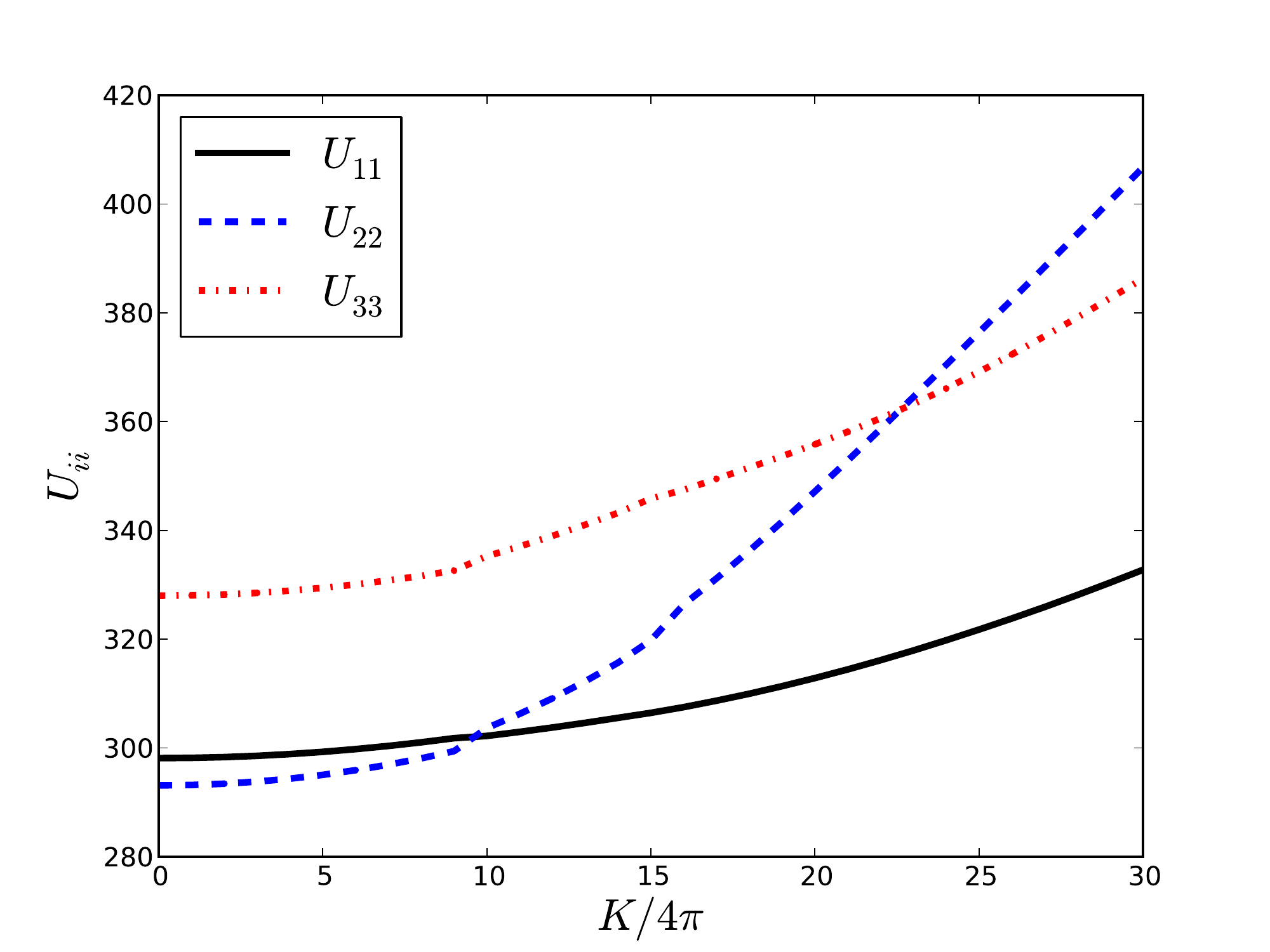}}
\subfigure[\,$\boldsymbol{\widehat{K}}=(0,0,1)$]{\includegraphics[totalheight=4.cm]{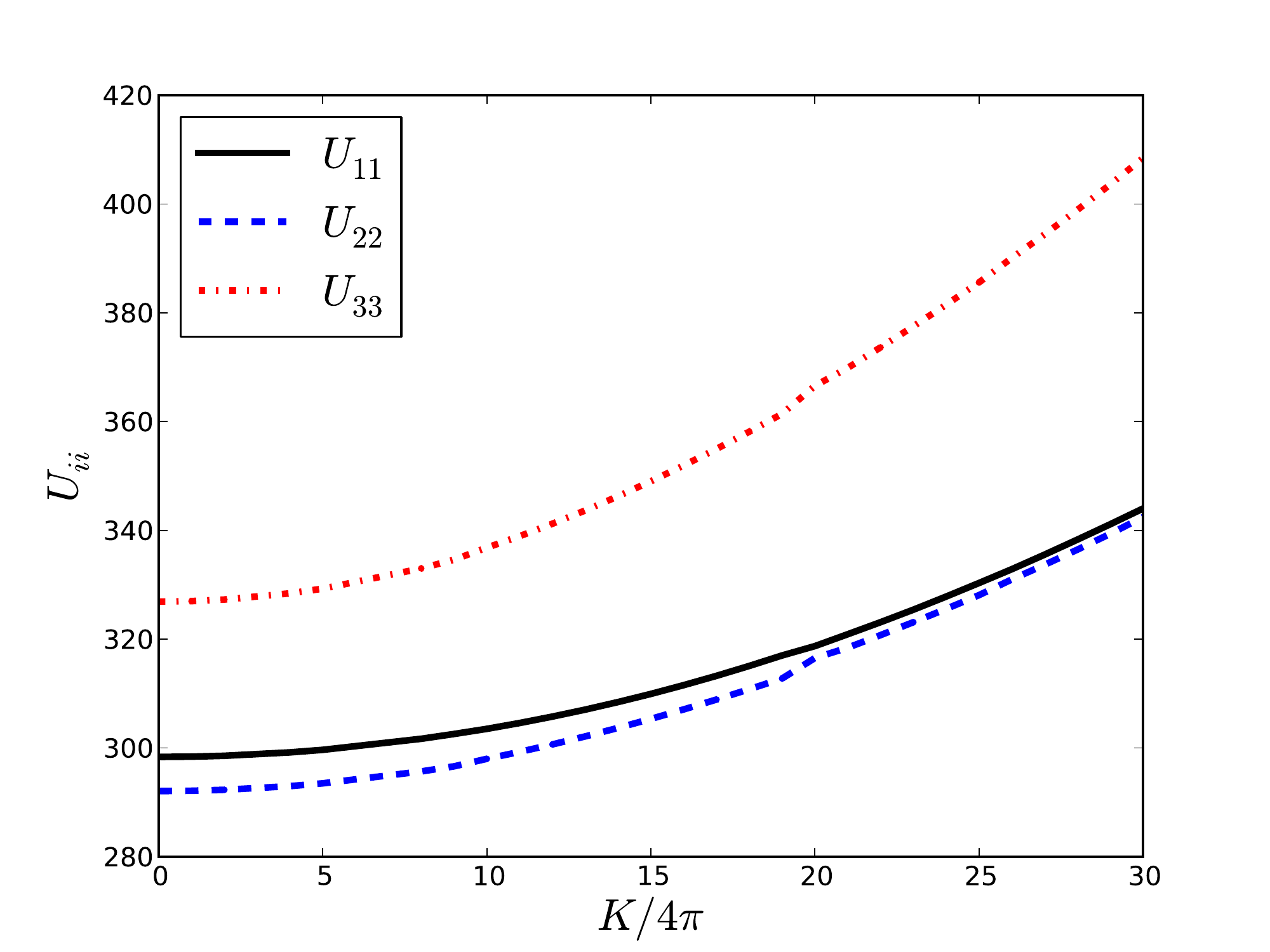}}
\caption{Diagonal elements of the isospin inertia tensor $U_{ij}$ as a function of isospin $K$ for $D_{4h}$-symmetric $B=8$ Skyrme configurations with pion mass value $\mu=1$. The isorotation axes are chosen as indicated above.}
\label{B8D4h_Udiag_mass}
\end{figure}

\begin{figure}[!htb]
\centering
\includegraphics[totalheight=8.0cm]{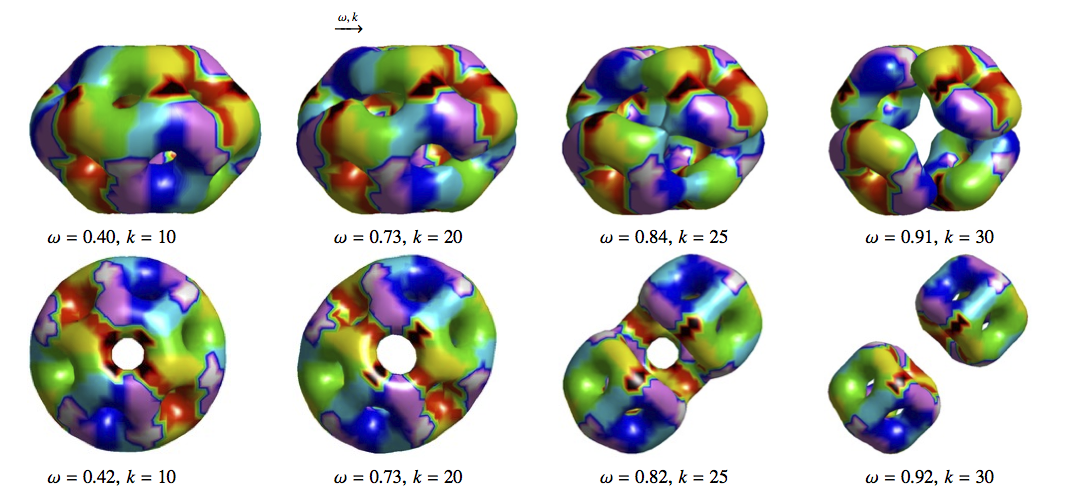}
\caption{Baryon density isosurfaces of isospinning $D_{6d}$-symmetric  $B=8$ Skyrmion solutions for $\mu=1$ as a function of isospin $K$ and angular frequency $\omega$. The isorotation axis is chosen to be $\boldsymbol{\widehat{K}}=(0,1,0)$ (first row) and $\boldsymbol{\widehat{K}}=(0,0,1)$ (second row). Each baryon density isosurface corresponds to the value $\mathcal{B}=0.15$.}
\label{Fig_B8_Sky_Iso_D6d}
\end{figure}

\begin{figure}[!htb]
\subfigure[\,$\boldsymbol{\widehat{K}}=(0,1,0)$]{\includegraphics[totalheight=6.cm]{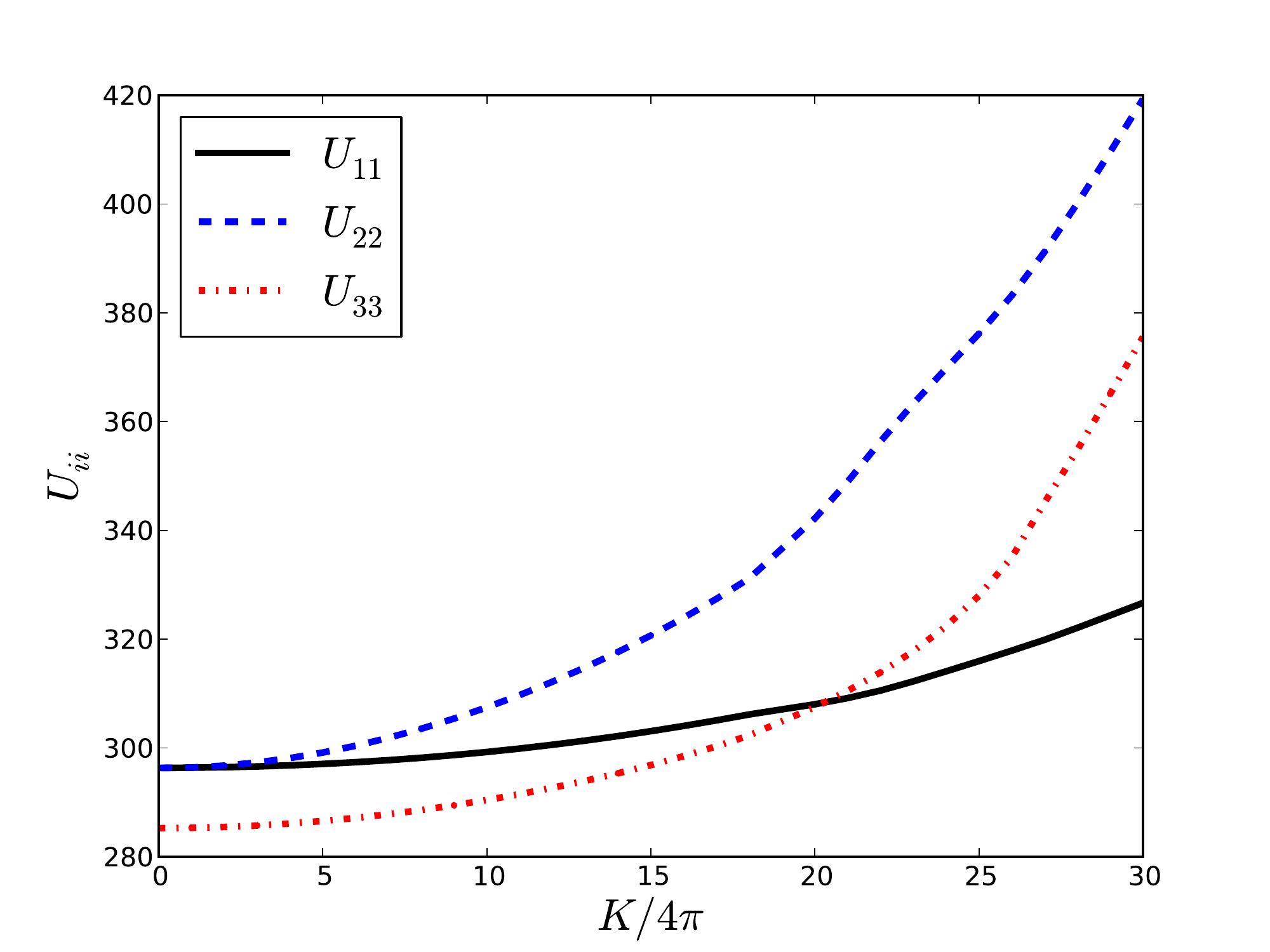}}
\subfigure[\,$\boldsymbol{\widehat{K}}=(0,0,1)$]{\includegraphics[totalheight=6.cm]{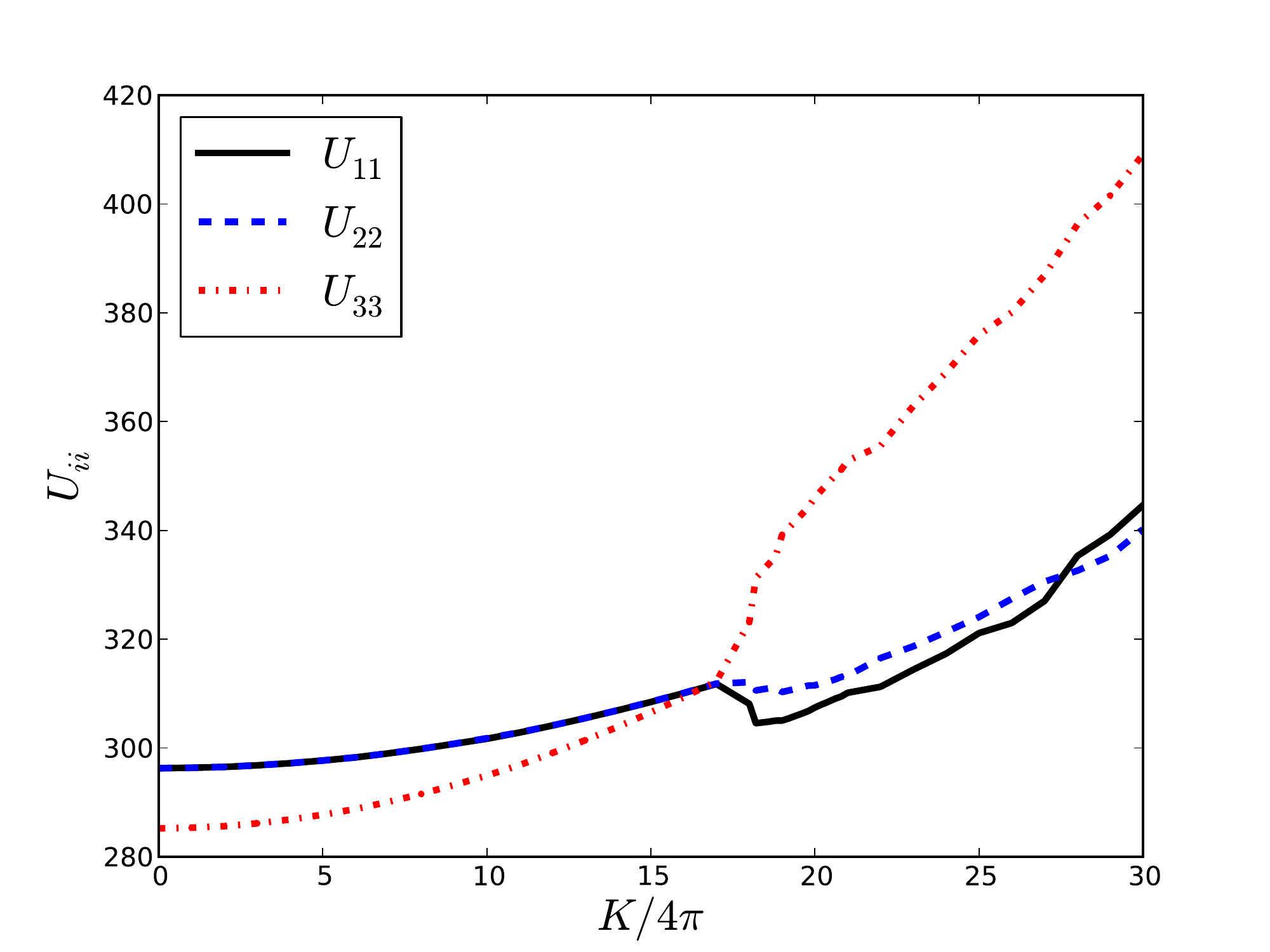}}
\caption{Diagonal elements of the isospin inertia tensor $U_{ij}$ as a function of isospin $K$ for $D_{6d}$-symmetric $B=8$ Skyrme configurations with pion mass value $\mu=1$. The isorotation axes are chosen as indicated above.}
\label{B8D6d_Udiag_mass}
\end{figure}

For baryon number $B=8$ we investigate the effect of isospin on $D_{4h}$- and $D_{6d}$-symmetric Skyrme configurations (with rescaled pion mass value $\mu$ set to 1).  For $B=8$ Skyrme solitons with approximate $D_{4h}$ symmetry we find that when isospinning about their $(0,1,0)$ axis  there exists a breakup frequency at which the isospinning solution splits into four $B=2$ tori (illustrated by the baryon density isosurfaces presented in Fig.~\ref{Fig_B8_Sky_Iso_D4h}). Similarly, we observe that the $D_{6d}$ configuration when isospinning about its $(0,1,0)$ axis breaks apart into four $B=2$ tori, but aligned in a different way (see baryon density isosurfaces presented in Fig.~\ref{Fig_B8_Sky_Iso_D6d}). When choosing $\boldsymbol{\widehat{K}}=(0,0,1)$ as isorotation axis,  $D_{4h}$- and $D_{6d}$-symmetric Skyrmion solutions -- are found to break up into two well-separated $B=4$ clusters (see Fig.~\ref{Fig_B8_Sky_Iso_D4h} and Fig.~\ref{Fig_B8_Sky_Iso_D6d}, respectively). Finally, for $B=8$ Skyrmions with approximate $D_{4h}$ symmetry another possible isorotation axis choice is given by  $\boldsymbol{\widehat{K}}=(1,0,0)$. Again, as the isospin $K$ increases, we observe a breakup into two $B=4$ cubic Skyrme solutions (see Fig.~\ref{Fig_B8_Sky_Iso_D4h}). The associated diagonal elements of the isospin inertia tensor $U_{ij}$ can be found as a function of $K$ in Fig.~\ref{B8D4h_Udiag_mass} and Fig.~\ref{B8D6d_Udiag_mass}, respectively. In Fig.~\ref{Fig_EB8} we compare the energy curves for $D_{4h}$-  and $D_{6d}$-symmetric Skyrmions isospinning about $\boldsymbol{\widehat{K}}=(0,1,0)$ and $\boldsymbol{\widehat{K}}=(0,0,1)$, respectively. For $\boldsymbol{\widehat{K}}=(0,1,0)$ $D_{4h}$- and $D_{6d}$-symmetric solutions remain energy degenerate within the limits of our numerical accuracy as the angular velocity increases. However, when choosing the isorotation axis $\boldsymbol{\widehat{K}}=(0,0,1)$ the $D_{4h}$ solution appears to be of lower energy for fixed, nonzero isospin value $K$.

For $D_{6d}$-symmetric $B=8$ solutions, we cannot decide within the limits of our numerical accuracy which isospin axis results in the lowest energy configuration for fixed isospin [compare energy curves in Fig.~\ref{Fig_Energy_axes}(c)].  For $D_{4h}$-symmetric $B=8$ solutions, the two cube soliton solution  with $\boldsymbol{\widehat{K}}=(0,0,1)$ has the lowest total energy [see Fig.~\ref{Fig_Energy_axes}(d)]. If slightly perturbed, the solution composed of four aligned $B=2$ tori with $\boldsymbol{\widehat{K}}=(0,1,0)$ can evolve into two cubes isospinning about  $\boldsymbol{\widehat{K}}=(0,0,1)$. 

\subsection{Critical Angular Frequencies}

It has been observed \cite{Harland:2013uk,Battye:2013xf,Battye:2013tka,Halavanau:2013vsa} that isospinning soliton solutions in models of the Skyrme family suffer from two different types  of instabilities: One is related to the deformed metric in the pseudoenergy functional $F_\omega\left(\boldsymbol{\phi}\right)$ becoming singular at $\omega_1=1$ \cite{Harland:2013uk} and the other to the Hamiltonian no longer being bounded from below at $\omega_2$. The first critical frequency $\omega_1$ is \emph{independent} of the concrete potential choice, whereas the second critical frequency $\omega_2$ \emph{crucially depends} on the particular choice of the potential term.

\begin{figure}[!htb]
\centering
\subfigure{\includegraphics[totalheight=4.0cm]{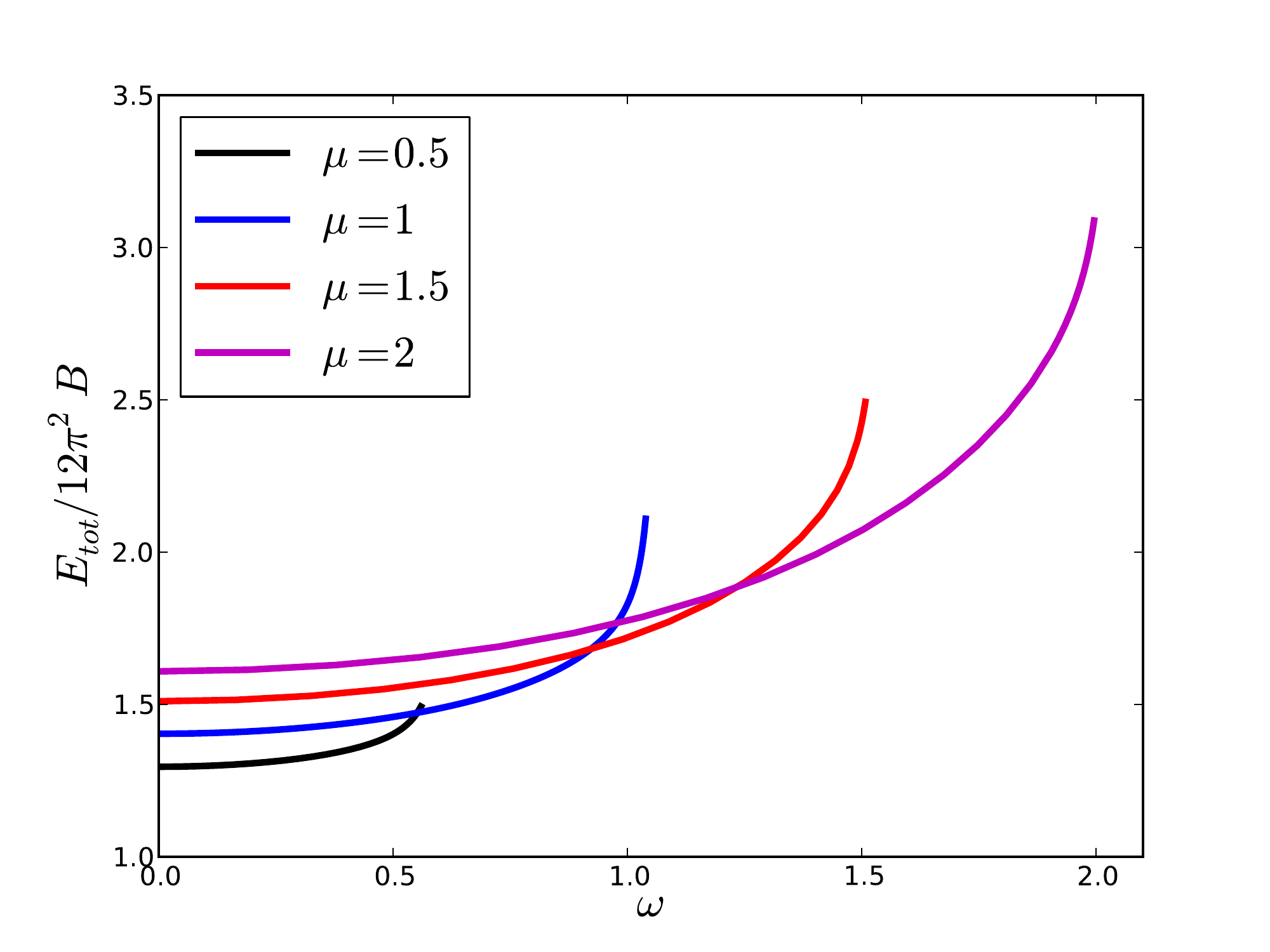}}
\subfigure{\includegraphics[totalheight=4.0cm]{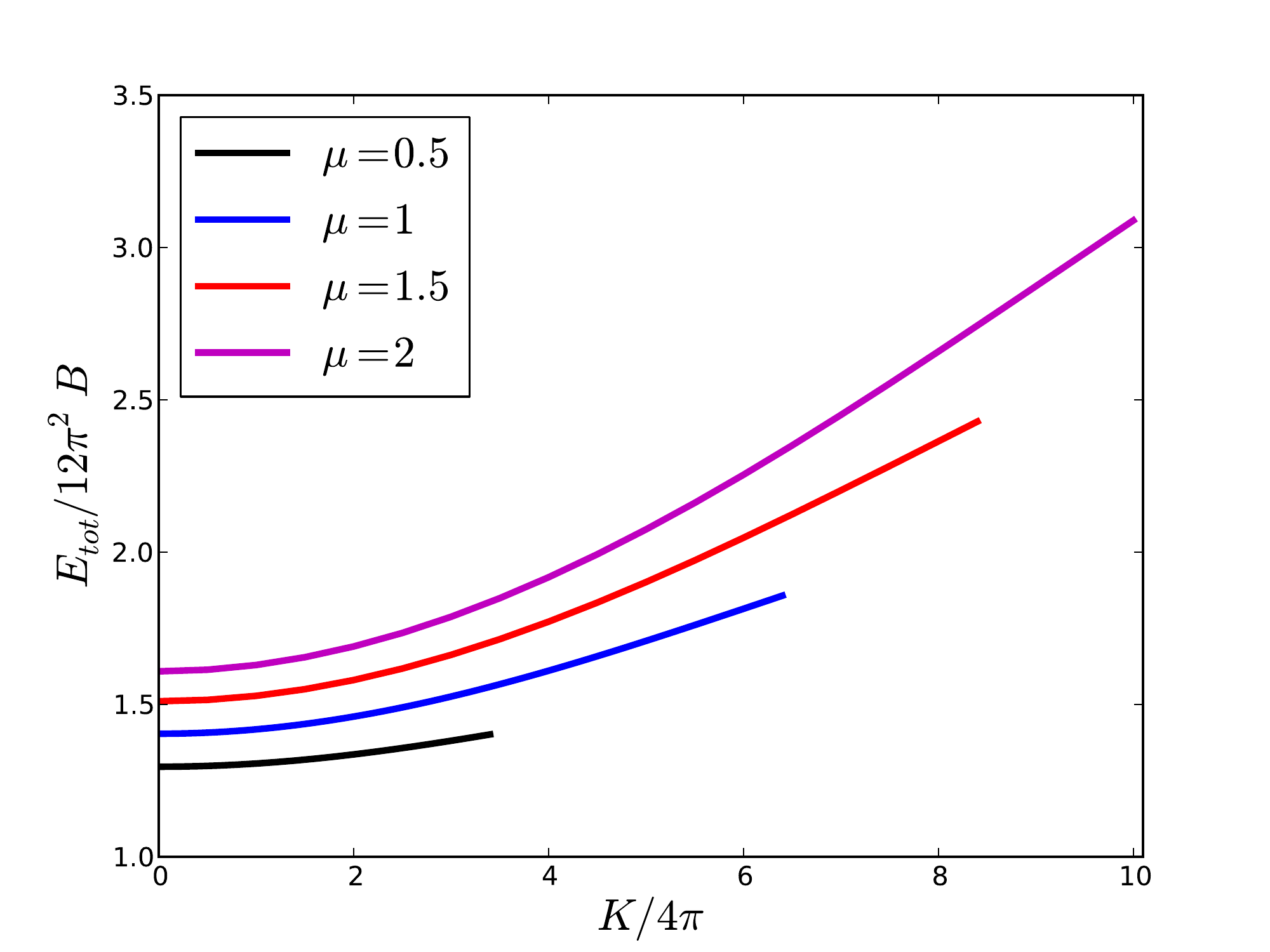}}
\subfigure{\includegraphics[totalheight=4.0cm]{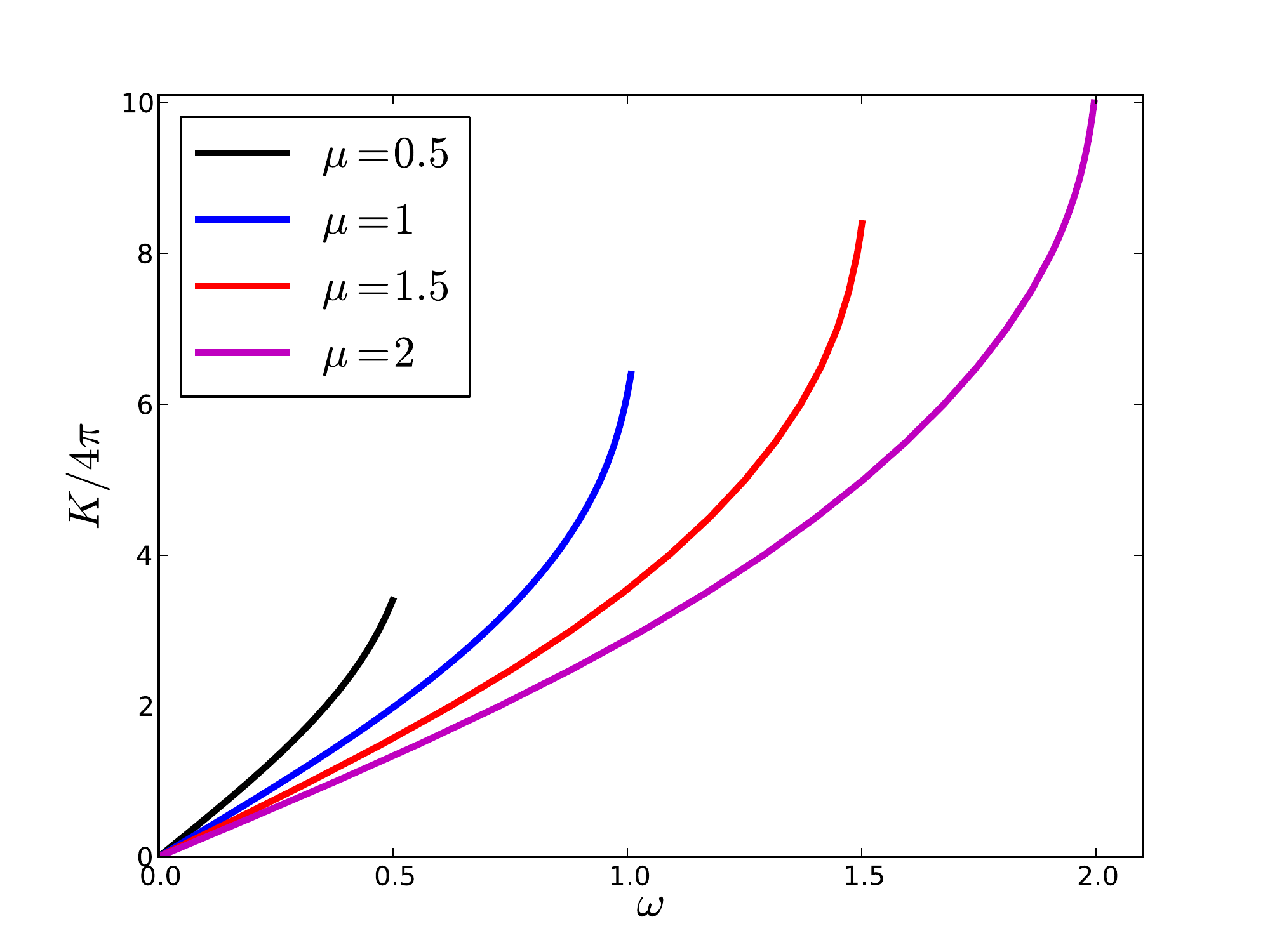}}
\caption{Isospinning $B=1$ Skyrme solitons in the Skyrme model with the potential term $V=2\mu^2(1-\sigma)$ for a range of mass values $\mu$. We plot the total energy $E_{\text{tot}}$ and isospin $K$  as function of angular frequency $\omega$ and the total energy as a function of isospin $K$ at $\mu=0.5,1,1.5,2$. The $z$-axis is chosen to be the axis of isorotation. Our 3D-relaxation calculations are performed on a $(100)^3$ grid with grid spacing $\Delta x=0.2$. }
\label{Fig_B1_Sky_mass}
\end{figure}

Note that we \emph{do not} observe the same pattern of critical frequencies in the full Skyrme model (\ref{Lag_SU2}). In Fig.~\ref{Fig_B1_Sky_mass} we plot energy $E_{\text{tot}}$ and isospin $K$ of the $B=1$ Skyrmion with the standard potential term $V=2\mu^2\left(1-\sigma\right)$ as functions of the angular frequency $\omega$ for a range of values of the mass parameter $\mu$. In addition, the energy is shown as a function of angular momentum $K$. Contrary to observations in the baby Skyrme model \cite{Battye:2013tka} and in the Skyrme-Faddeev model \cite{Battye:2013xf} energy minimizer can be found up to $\omega = \omega_{\text{crit}}$ when the minimisation problem becomes ill defined. In particular, there \emph{does not} exist a critical behavior at $\omega=\omega_1$. 

\begin{figure}[!htb]
\centering
\subfigure{\includegraphics[totalheight=4.0cm]{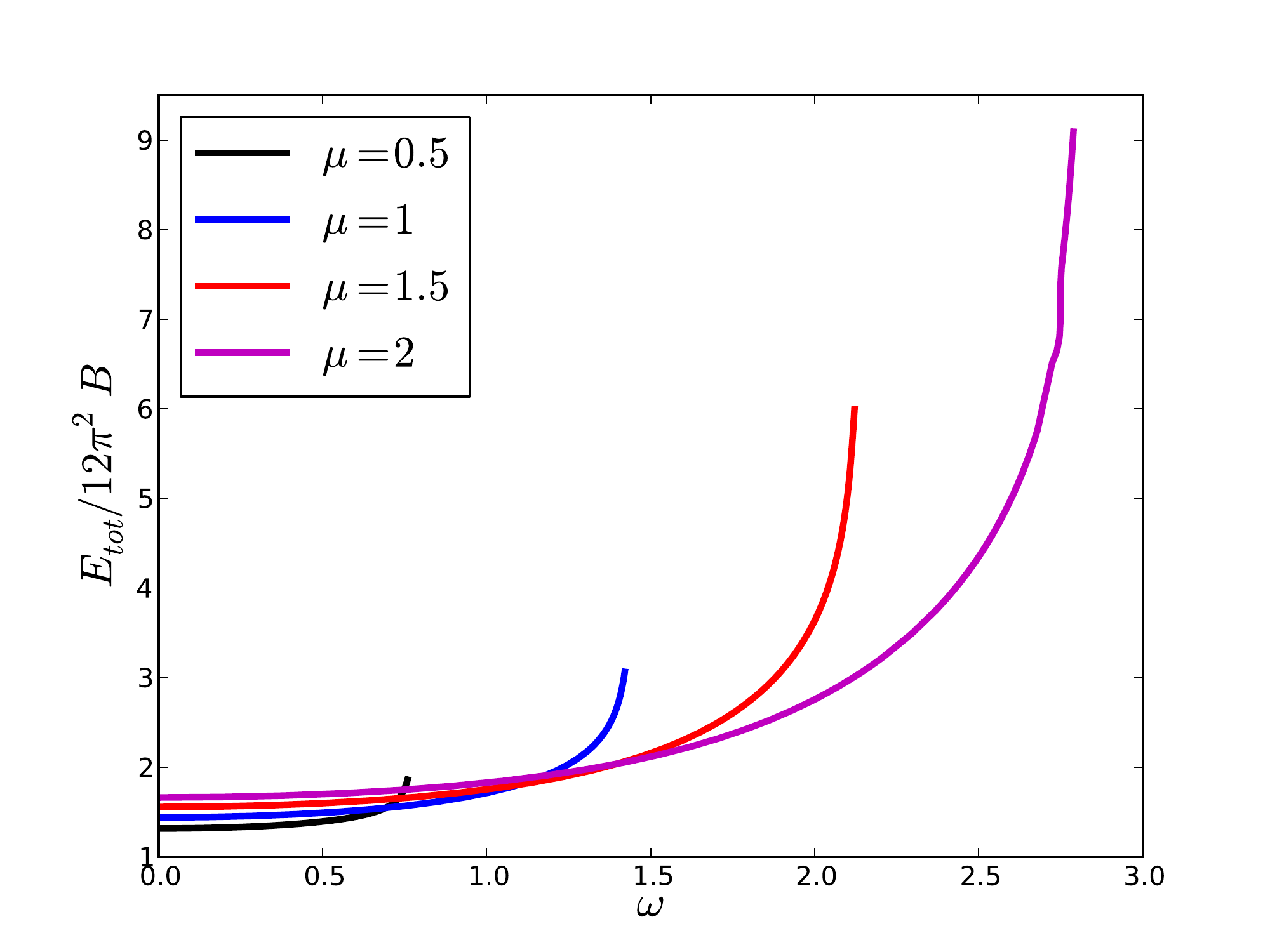}}
\subfigure{\includegraphics[totalheight=4.0cm]{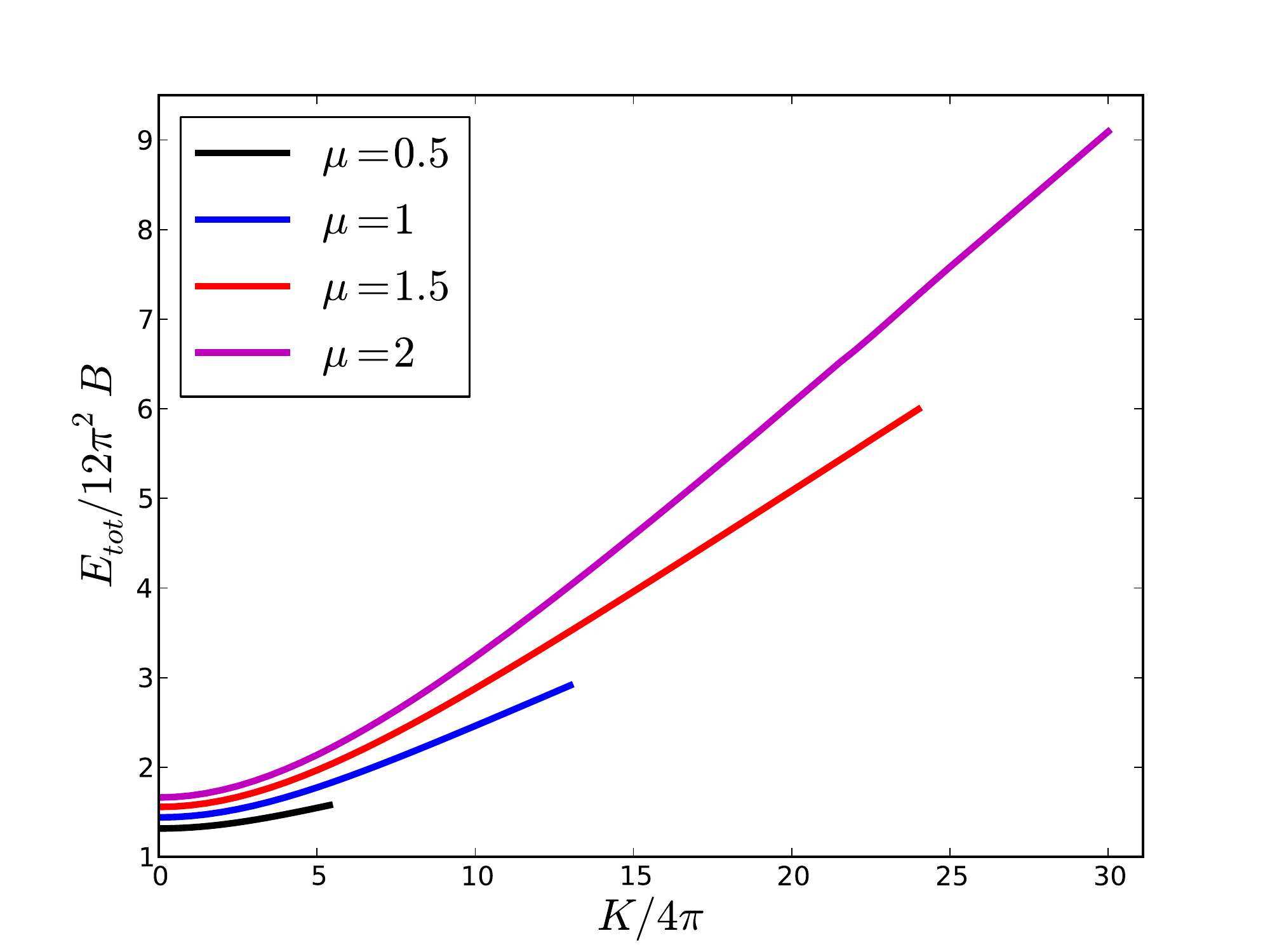}}
\subfigure{\includegraphics[totalheight=4.0cm]{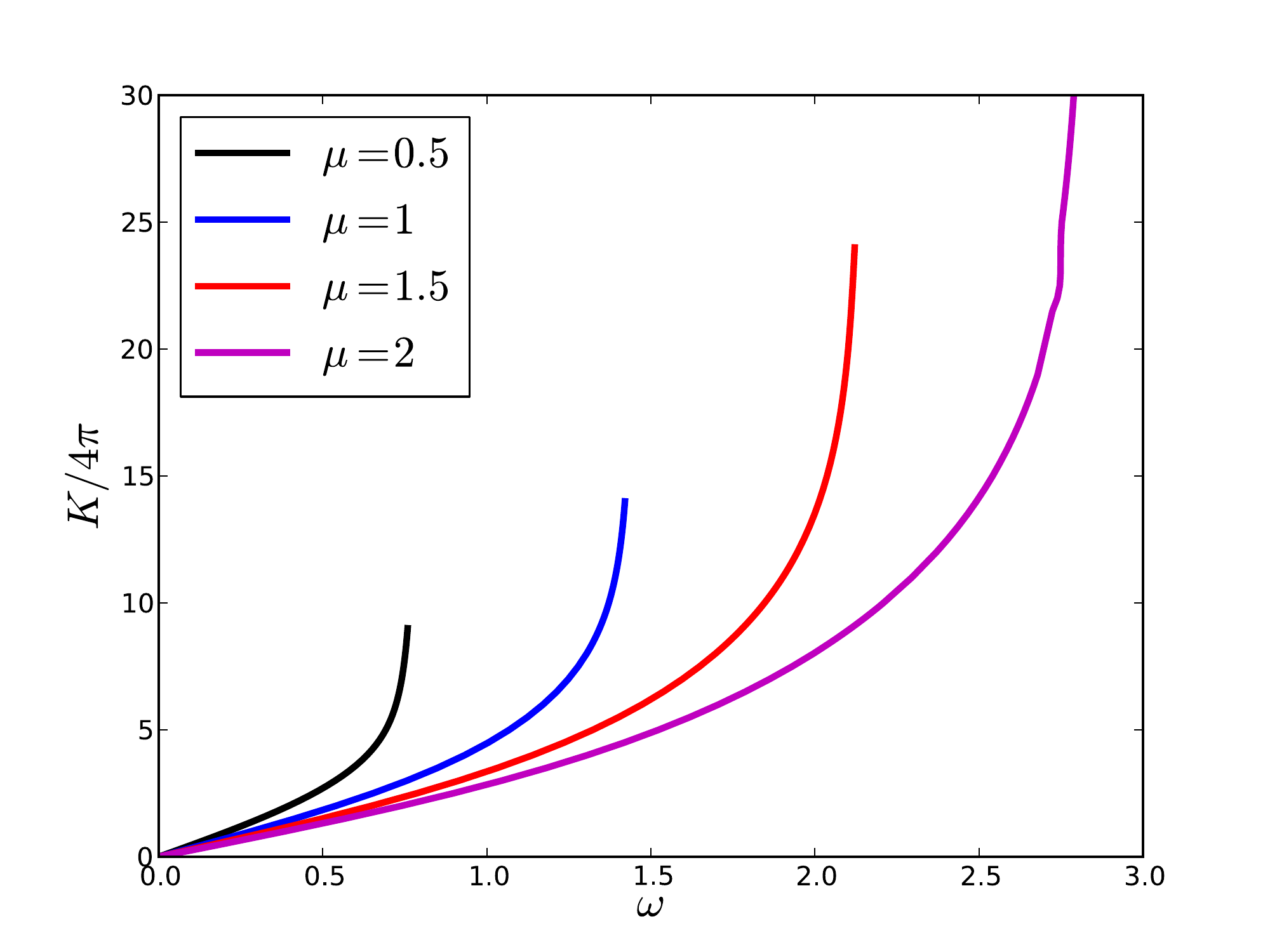}}
\caption{Isospinning charge-1 soliton solutions in the Skyrme model with the potential term $V=2\mu^2(1-\sigma^2)$  for a range of mass values $\mu$.  We display the total energy $E_{\text{tot}}$ and isospin $K$  as function of angular frequency $\omega$ and isospin $K$ at $\mu=0.5,1,1.5,2$. The isorotation axis is chosen to be $\boldsymbol{\widehat{K}}=(0,0,1)$. We perform the 3D relaxation calculations on a $(100)^3$ grid with grid spacing $\Delta x=0.2$.}
\label{Fig_B1_Sky_mass_New}
\end{figure}

When choosing the double vacuum potential $V=2\mu^2(1-\sigma^2)$ in (\ref{Sky_mass}), we obtain for the isospinning $B=1$ Skyrmion solution the energy and isospin curves  shown in Fig.~\ref{Fig_B1_Sky_mass_New}. Again, we observe that Skyrmions cannot spin with angular frequencies $\omega > \sqrt{2}\mu $ -- the meson mass of the model -- since they become unstable to the emission of radiation.

\subsection{Mean Charge Radii}\label{Sec_Skyrme_Prop}

\begin{figure}[!htb]
\subfigure{\includegraphics[totalheight=6.0cm]{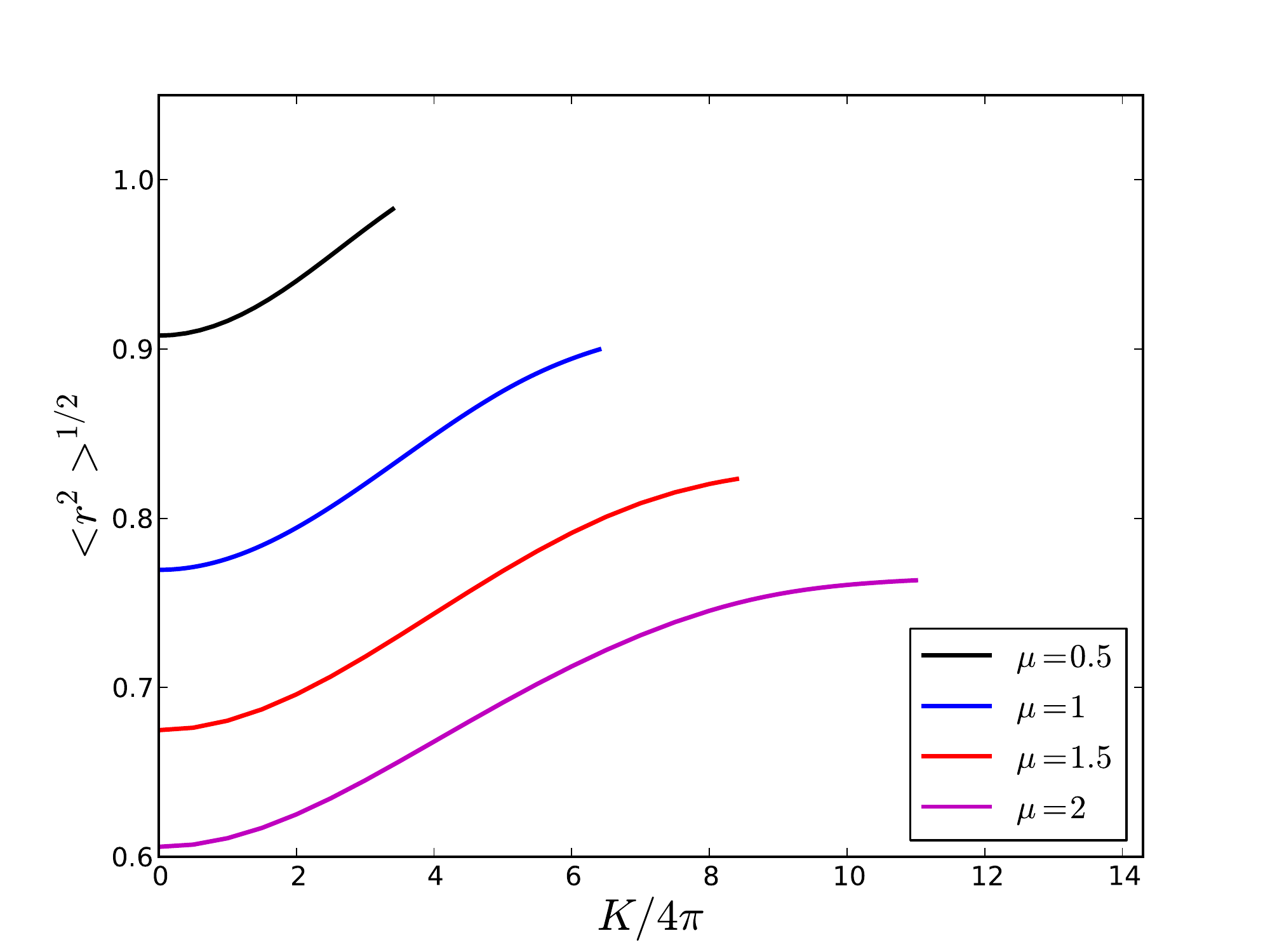}}
\subfigure{\includegraphics[totalheight=6.0cm]{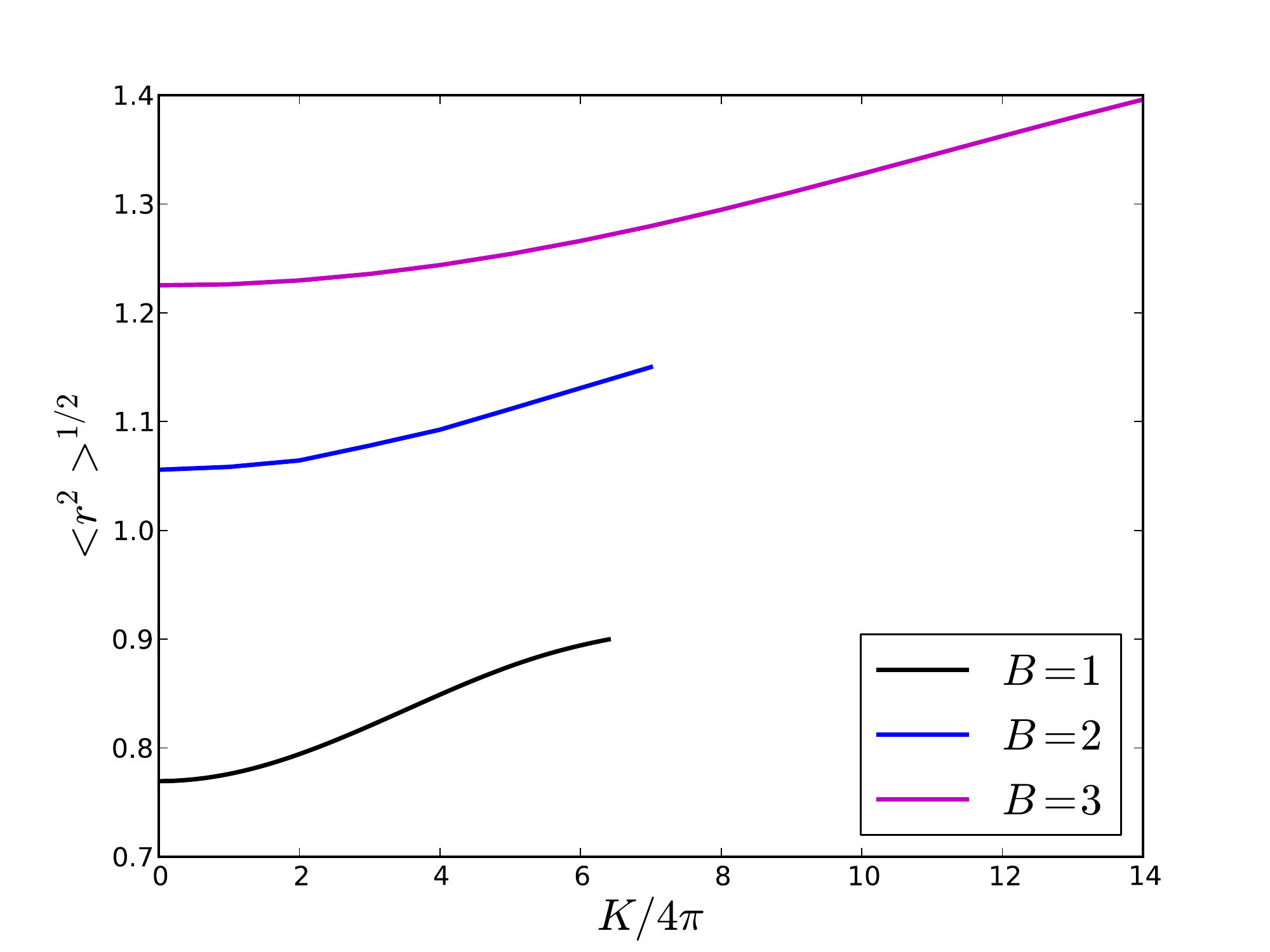}}
\caption{Mean charge radii $<r^2\!>^{1/2}$ (\ref{Sky_mean_rad}) for Skyrmion solutions of topological charge $1\le B\le 3$ as a function of isospin $K$. These calculations have been performed with the conventional potential term $V=2\mu^2\left(1-\sigma\right)$ for isospinning $B=1$ Skyrmions with mass value $\mu=0.5,1,1.5,2$ (left) and for isospinning $B=1,2,3$ Skyrmion solutions with $\mu=1$ (right).}
\label{Fig_Sky_mean_charge}
\end{figure}

In Fig.~\ref{Fig_Sky_mean_charge} we present as a function of isospin $K$ the mean charge radii of $B=1-3$ soliton solutions in the Skyrme model (\ref{Lag_SU2}) with the rescaled mass $\mu=1$. In addition, we show for $B=1$ the mean charge radii for a range of mass values. We define the mean charge radius of a Skyrmion solution  as the square root of the second moment of the topological charge density $\mathcal{B}(\boldsymbol{x})$ in (\ref{Sky_bary_dens}) 
\begin{equation}
<r^2>=\frac{\displaystyle\int\,r^2\,\mathcal{B}(\boldsymbol{x})\,\text{d}^3x}{\displaystyle\int \,\mathcal{B}(\boldsymbol{x})\,\text{d}^3x}\,.
\label{Sky_mean_rad}
\end{equation}

Similar to our observations of isospinning soliton solutions in the standard baby Skyrme model \cite{Battye:2013tka}, we note that the changes in the Skyrmion's shape are reflected by the changes in slopes of the mean charge radius curves in Fig.~\ref{Fig_Sky_mean_charge}.
For example for $B=1$ and mass value $\mu=1$ we find that the radius $<r^2\!>^{1/2}$ grows with the isospin and that there exists an inflection point.
 For $B=2$ and $B=3$ we observe that the inflection point occurs at higher isospin values. These inflection points are in reasonable agreement with the maximal isospin values stated in Sec.~\ref{Sec_low} up to which the rigid-body approximation is a good simplification. 
Furthermore, as $\mu$ increases the inflection point of the  mean charge radius $<r^2\!>^{1/2}$ occurs at increasingly higher isospin values. This confirms that the rigid-body approximation becomes more accurate as $B$ or $\mu$ increases.

\section{Spin induced from Isospin }\label{Sec_spin_induced}

\begin{figure}[!htb]
\subfigure[\,$B=1$, $\boldsymbol{\widehat{K}}=(0,0,1)$]{\includegraphics[totalheight=6.cm]{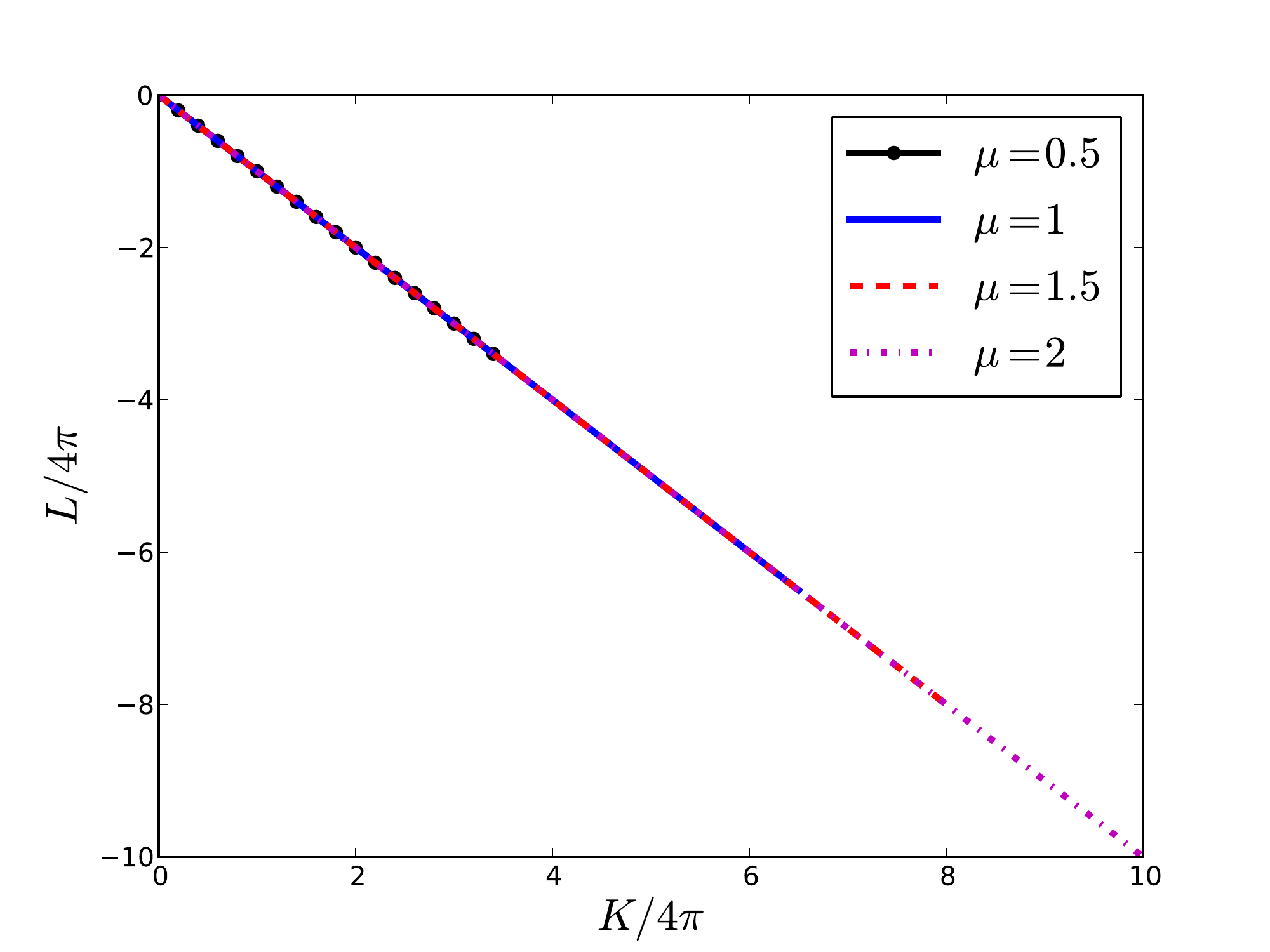}}
\subfigure[\,$B=2$, $\boldsymbol{\widehat{K}}=(0,0,1)$]{\includegraphics[totalheight=6.cm]{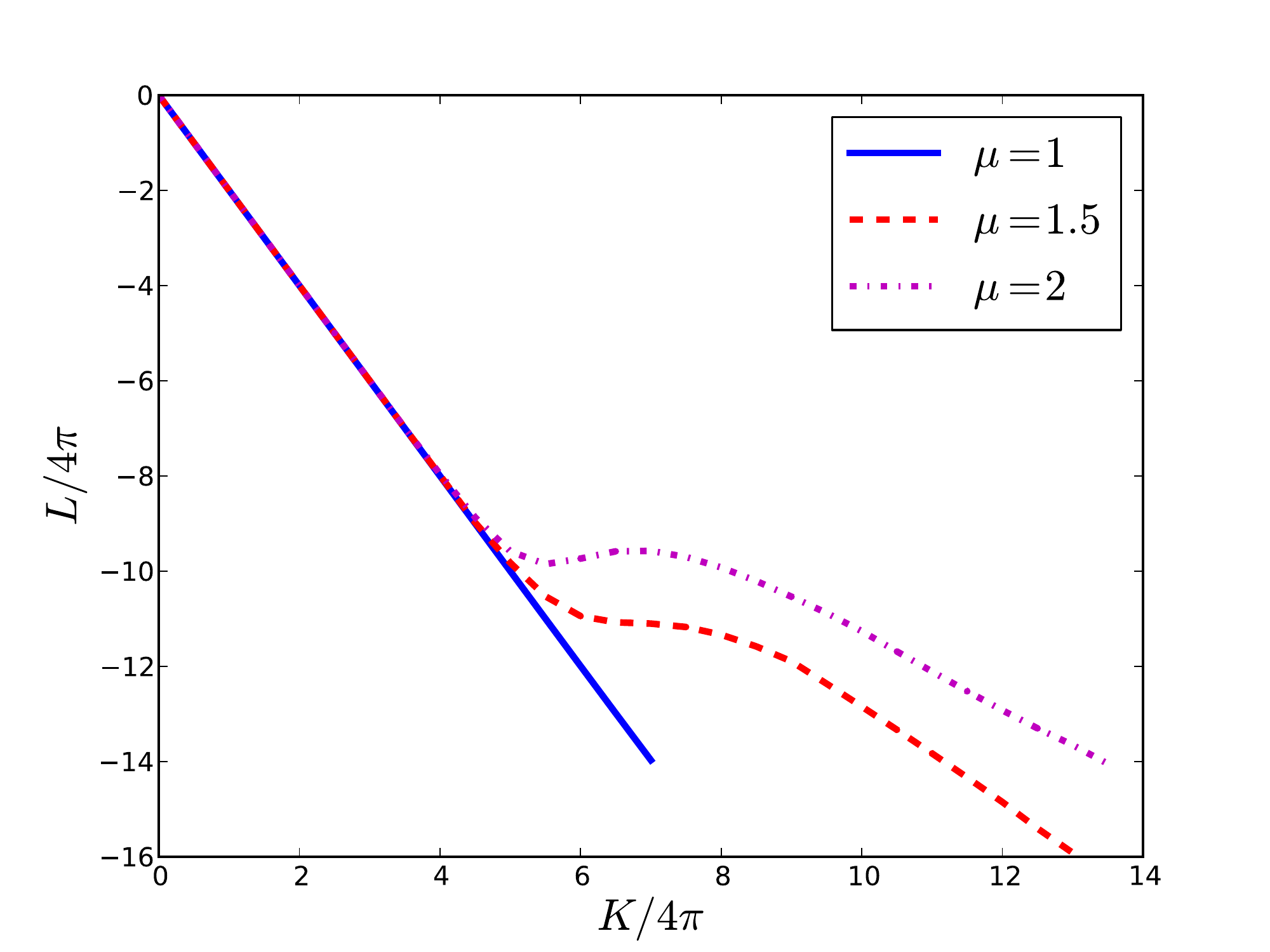}}\\
\subfigure[\,$B=3$, $\boldsymbol{\widehat{K}}=(0,0,1)$]{\includegraphics[totalheight=6.cm]{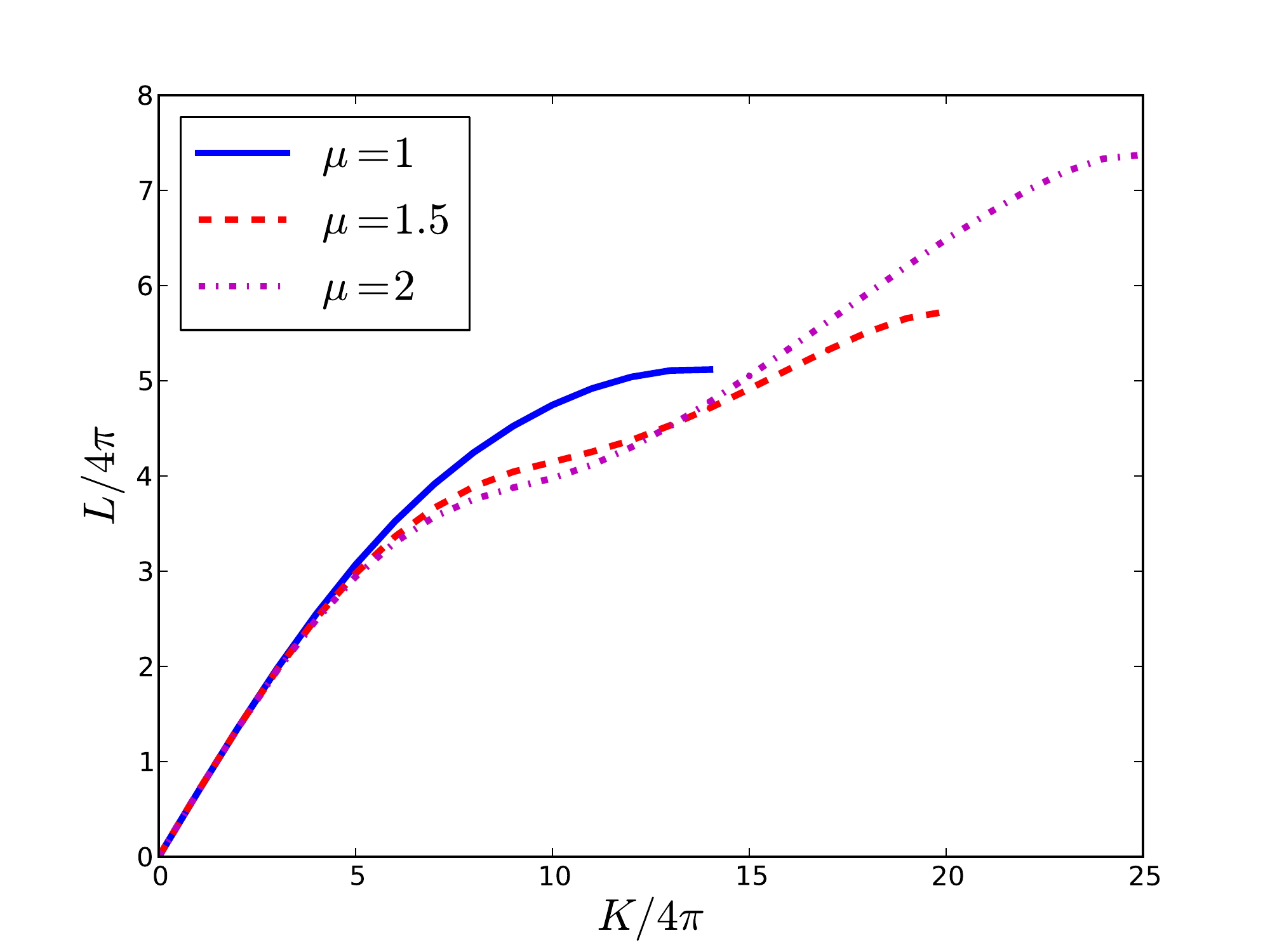}}
\subfigure[\,$B=8$ ($D_{6d}$ symmetry), $\boldsymbol{\widehat{K}}=(0,0,1)$]{\includegraphics[totalheight=6.cm]{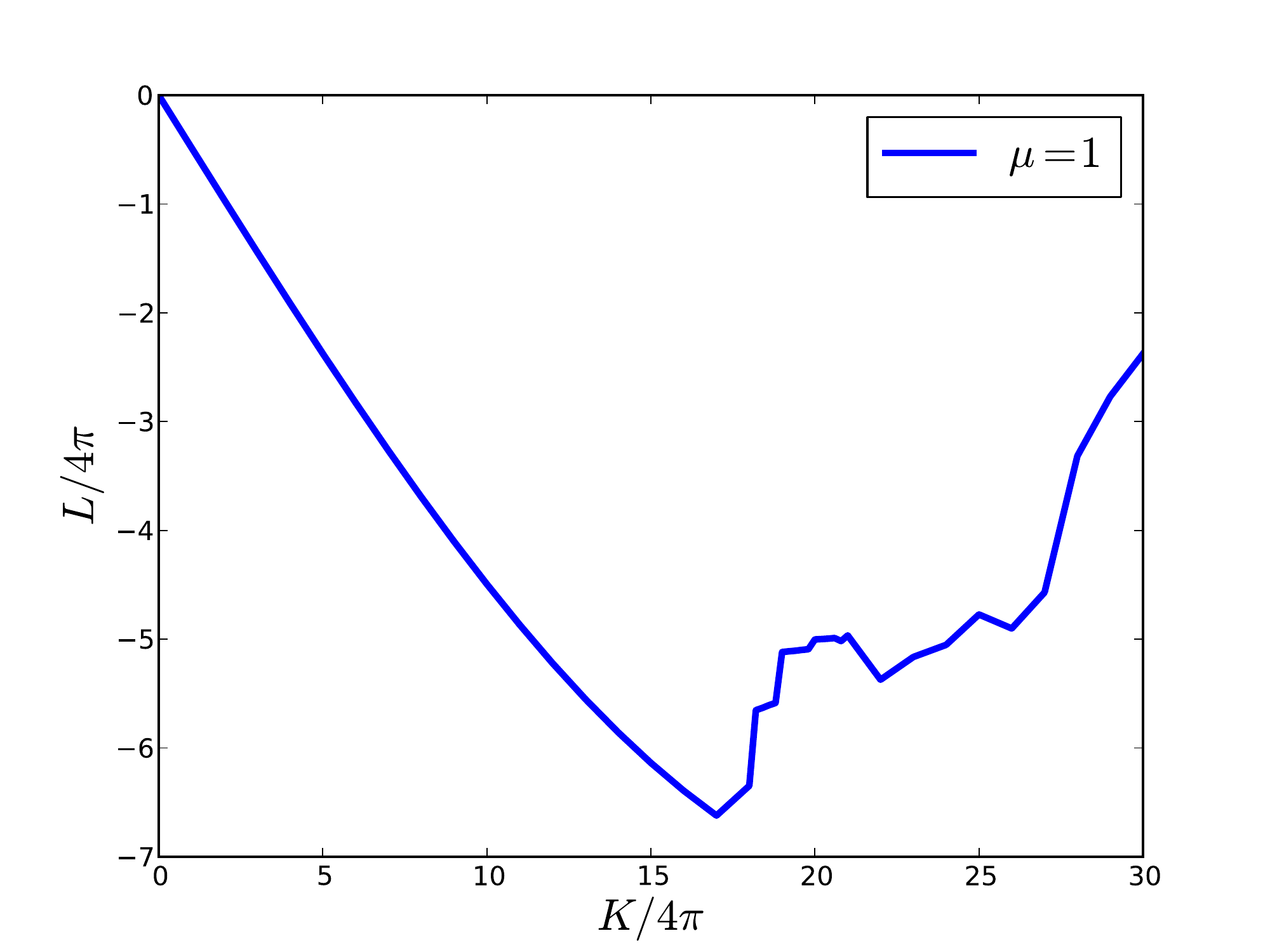}}
\caption{Spin $L$ as a function of isospin $K$ for Skyrmion solutions of topological charge $B=1-3,8$ and with pion mass $\mu$. The isospin axes are chosen as indicated. Here, we only display isospinning Skyrme configurations with nonzero $L$. }
\label{B_Spin}
\end{figure}

If $W_{ij}$ is nonzero, then Skyrme configurations will obtain \emph{classical} spin when isospin is added. For $B=1$ we display in Fig.~\ref{B_Spin}(a) the acquired spin $L$ as a function of isospin $K$ for a range of pion masses $\mu$ when isospinning about $\boldsymbol{\widehat{K}}=(0,0,1)$. Since the isospinning charge-1 Skyrmion preserves axial symmetry, $L$ is found to grow linearly with $K$. Numerically, the slope is found to be $-0.99$ and agrees well with the expected one $L/K=-W_{33}/U_{33}=-1$ for an axially symmetric charge-1 configuration with $W_{33}=U_{33}$. Stationary, isospinning Skyrme configurations can only be constructed up to the critical isospin $K_{\text{crit}}=U_{33}\omega_{\text{crit}}$ with $\omega_{\text{crit}}=\mu$. Consequently, the functions $L(K)$ terminate at the points $L_{\text{crit}}=-K_{\text{crit}}$. Spin $L$ and isospin $K$ have the same magnitude and are of opposite sign. This agrees with the  Finkelstein-Rubinstein constraints that are commonly imposed when quantizing $B=1$ Skyrmions \cite{Krusch:2002by} . 

For $B=2$, we investigated isospinning Skyrmion solutions with isospin axes $\boldsymbol{\widehat{K}}=(0,1,0)$ and $\boldsymbol{\widehat{K}}=(0,0,1)$. Isospinning around $\boldsymbol{\widehat{K}}=(0,1,0)$ leads to a novel configuration with $D_4$ symmetry  (see Fig.~\ref{Fig_B2Ky_xy_contour}). For fixed isospin, this configuration has lower energy than isorotation around $\boldsymbol{\widehat{K}}=(0,0,1)$, see Fig.~\ref{Fig_Energy_axes}. For this $D_4$-symmetric  configuration, we verify numerically that the spin $L=-W_{22}\omega$ vanishes for all classically allowed isospin values $K$.  Therefore, this configuration may become important for calculating excited states of the Deuteron with nonzero isospin \cite{Manton:2011mi}.

However, if we choose $\boldsymbol{\widehat{K}}=(0,0,1)$ as isospin axis, the $B=2$ Skyrmion gains spin $L=-W_{33} K$ as $K$ increases, see Fig.~\ref{B_Spin}(b).  For $\mu=1$ the  axial symmetry remains unbroken and hence $L$ depends linearly on $K$. We confirm that the numerically calculated slope $-1.97$ agrees well with the expected one $L/K=-W_{33}/U_{33}=-2$ as $W_{33}=2U_{33}$ for an axially symmetric   charge-2 configuration. For larger mass values $\mu$  isospinning around the
$\boldsymbol{\widehat{K}}=(0,0,1)$ leads to the breakup into two $B=1$ Skyrmions orientated in the attractive channel. The axial symmetry is broken at $K_{\text{SB}}$, and the $B=2$ Skyrmion solution starts to split into two $B=1$ Skyrmions.  As the isospin $K$ increases further $L$ is approximately $-K$. In this regime the isospinning configuration is well described by two separated, axially symmetric deformed $B=1$ Skyrmions.  This is consistent with head-on scattering of two spinning $B=1$ Skyrmions in the attractive channel \cite{Foster:2014vca} where the configuration of closest approach is not the torus but a configuration of two separated Skyrmions. The attractive channel has also been discussed in Ref.~\cite{Leese:1994hb} when quantising the Deuteron. This degree of freedom was essential for comparing the spatial probability distribution of the deuteron with experimental values \cite{Braaten:1988cc,Forest:1996kp,Manton:2011mi}.

In summary, for $B=2$ we observe that Skyrme configurations with nonzero $W$ and hence nonzero spin $L$ show centrifugal effects and separate out whereas states with $W=0$ tend to stay more compact. Isospinning the charge 2 Skyrmion about its $(0,0,1)$ axis results in the breakup into two well-separated charge-1 Skyrmions, whereas isospinning about $(0,1,0)$ yields compact $D_4$-symmetric $B=2$ configurations of lower energy.

For $B=3$ we display in Fig.~\ref{B_Spin}(c) the $L(K)$ graphs for pion masses $\mu=1,1.5,2$. We observe that as long as the tetrahedral symmetry remains unbroken the spin $L$ inreases linearly with $K$. Breaking of the tetrahedral symmetry results in a lower $W$ and hence a lower increase in $L$ for higher $K$ values.    

For $B=4$ there are two different isospin axes: $\boldsymbol{\widehat{K}}=(0,0,1)$ and $\boldsymbol{\widehat{K}}=(0,1,0)$.  For 
$\boldsymbol{\widehat{K}}=(0,0,1)$ we find that the octahedral symmetry remains unbroken (see baryon density isosurfaces shown in Fig.~\ref{Fig_B4_Sky_Iso_Ly}). The mixed inertia tensor $W$ and hence $L$ vanish for all classically allowed isospin values $K$.  When isospinning the cubic $B=4$ Skyrmion solution about $\boldsymbol{\widehat{K}}=(0,1,0)$ we observe breakup into two $D_4$-symmetric charge-2 Skyrmions (see Fig.~\ref{Fig_B4_Sky_Iso_Ly}). Again, the mixed inertia tensor $W$ and hence spin $L$ is found to vanish for all classically allowed $K$ values.

For $B=8$ it is interesting to observe that all configurations are breaking up into constituents, either into two $B=4$ parts or into four $B=2$ parts. Note that physically $B=8$ may describe beryllium $^8{\text{Be}}$ which is unstable to splitting up into two $\alpha$ particles, see e.g. Ref.~\cite{Krane:1987}. For $D_{6d}$-symmetric Skyrmion solutions we find that when isospinning about $\boldsymbol{\widehat{K}}=(0,0,1)$ there exist a critical isospin value at which the soliton solution splits up into two $B=4$ cubes. This breakup process is reflected in the $L(K)$ graph shown in Fig.~\ref{B_Spin}(d). For $K\le 17\times 4\pi$ ($L\ge -6.6 \times 4 \pi$) the spin $|L|$ grows linearly with $K$ and  the isospinning Skyrme configuration preserves its dihedral symmetry. For higher isospin values $K$ the dihedral symmetry is broken and the isospinning solution starts to break apart into two $B=4$ cubes of zero total spin $L$. Figure~\ref{B_Spin}(d) shows how $L(K)$ decreases as $K$ increases beyond  $K=17\times 4\pi$.  When choosing $\boldsymbol{\widehat{K}}=(0,1,0)$ as isorotation axis, the $D_{6d}$ isospinning symmetric Skyrmion solutions breaks up into four $B=2$ tori. In this case, the total spin $L$ is found to be zero (within the limits of our numerical accuracy) for all classically allowed isospin values. For $B=8$ Skyrme solitons with approximate $D_{4h}$ symmetry we find that all the isospinning solutions investigated in this article (see baryon density isosurfaces in Fig.~\ref{Fig_B8_Sky_Iso_D4h})  possess zero total spin for all values of $K$.

\begin{table}[!htb]
\caption{We list the critical isospin values $K_{\text{crit}}$ up to which stable isospinning Skyrmion solutions exist. Here, we calibrate the Skyrme model using the standard parameter set $F_\pi=108\,\text{MeV}$ and $e=4.84$ \cite{Adkins:1983ya,Adkins:1983hy}. Recall that in these units $\hbar$ takes the value $46.8$. For $K_{\text{crit}}$, we list the energy per baryon $M_B/B$, the isorotational energy per baryon $E_{\text{Iso}}/B$ and the total energy per baryon $E_{\text{tot}}/B$. }
\begin{tabularx}{\textwidth}{cXXXXXXXX}
\hline\hline
 $B$& $G$ &$\mu$ & $m_\pi$ [MeV] &$\boldsymbol{\widehat{K}}$ & $K_{\text{crit}}\,[\hbar]$  &$M_B/B$ [MeV] &$E_{\text{Iso}}/B$ [MeV]  &$E_{\text{tot}}/B$ [MeV]  \\
\hline
1& $O(3)$            &  0.5 & 130.7 & $(0,0,1)$ & 0.9   & 868.7    & 59.7   &928.4 \\
&             &  1 &261.4   & $(0,0,1)$ &  1.6 &   1008.8     &  230.0   & 1238.8\\
&             &  1.5 & 392.0& $(0,0,1)$ & 2.3&   1180.6     & 452.9 &  1633.5\\
&             &  2 &  522.7 & $(0,0,1)$ &  2.7&    1349.0     &    698.4 &2047.4 \\
2& $D_{\infty h}$  & 1 &261.4& $(0,1,0)$&    3.2      &    955.2   & 211.7 &  1166.9  \\
&  & 1.5 & 392.0& $(0,1,0)$&    4.0       &   1091.5 &  393.2    &  1484.7   \\
&  & 2 &522.7 & $(0,1,0)$       &   4.7       &    1235.7   &   616.4    &   1852.1\\
  &    &0.5&130.7 & $(0,0,1)$& 1.1&       824.9  &   34.8 &   859.7\\
  &    &1&261.4& $(0,0,1)$&           1.9 &    914.7   &   123.7& 1038.4 \\
  &    &1.5&392.0& $(0,0,1)$&           3.5 &  1074.2  &  340.5    &  1414.7  \\
  &    &2&522.7& $(0,0,1)$&       4.8 &     1280.7 &    631.0   &   1911.7 \\
3& $T_d$     &   1       &261.4& $(0,0,1)$ &       4.0&      912.8  &     178.0 &  1090.8  \\
&      &   1.5       &392.0& $(0,0,1)$ &  5.6    & 1060.9   &    368.2 &     1429.1  \\
&   &   2      &522.7& $(0,0,1)$ &     6.6 &    1207.7     &   583.4&   1791.1 \\
4& $O_h$      &  1      & 261.4& $(0,1,0)$&  4.7    &  878.0    &  156.6  &   1034.6 \\
&      &  1.5      & 392.0& $(0,1,0)$ & 7.5& 1063.8     &  365.7  &   1429.5 \\
&       &  2      &522.7 & $(0,1,0)$&9.1 &    1219.1&597.9 &  1817.0\\
&        &   1    &261.4 & $(0,0,1)$&       4.7 &    878.0   &    156.6 &   1034.6  \\
&      &   1.5    & 392.0& $(0,0,1)$&  6.3     &    990.4      & 309.1&   1299.5\\
&        &   2    & 522.7& $(0,0,1)$&   7.3    &   1106.7   &  482.1&  1588.8\\
8 &$D_{6d}$    &   1   &261.4&   $(0,1,0)$ &  9.4       &   880.4   &  150.3 &   1030.7\\
 &         &1 &261.4 &  $(0,0,1)$ &   9.2   & 870.7  &   151.7 & 1022.4\\
  &$D_{4h}$  &  1       & 261.4&  $(1,0,0)$ &  9.1     &   868.2    &  150.0 &1018.2 \\
 &         & 1& 261.4&  $(0,1,0)$ &   9.2    &872.4    &     151.6 &   1024.0 \\
 &         & 1&  261.4& $(0,0,1)$ &   9.1      &    866.5 &  148.5 &  1015.0 \\\hline\hline
\end{tabularx}
\label{Tab_Kcrit_Skyrme}
\end{table}

\section{Conclusion}\label{Sec_Skyrme_Con}

In this article we have performed fully three-dimensional numerical relaxations of isospinning soliton solutions with topological charges $B=1-4,8$ in the Skyrme model with the conventional mass term included and without imposing any assumptions about the soliton's spatial symmetries. Our calculations show that the qualitative shape of isospinning Skyrmion solutions can differ drastically from the ones of the static ($\omega=0$) solitons. The deformations become increasingly pronounced as the mass value $\mu$ increases. Briefly summarized, we distinguish the following types of behavior:

\begin{enumerate}[(i)]
\item \emph{Breakup into lower-charge Skyrmions:} Isospinning Skyrmion solutions can split into lower- charge Skyrmions at some critical breakup frequency value. Examples are the breakup (for $\mu$ sufficiently large) of the $B=2$ solution into two $B=1$ Skyrmions when isospinning about $\boldsymbol{\widehat{K}}=(0,0,1)$; the breakup of the charge-3 Skyrmion into a $B=1$ hedgehog and a $B=2$ torus; the breakup of the $D_{4h}$- and $D_{6d}$-symmetric $B=8$ Skyrme configurations into four $B=2$ tori when isospinning about $\boldsymbol{\widehat{K}}=(0,1,0)$ and the breakup of isospinning $D_{4h}$ and $D_{6d}$ Skyrmions into charge-4 subunits. These breakup processes do not occur as pronounced as observed for isospinning soliton solutions in the $(2+1)$-dimensional version of the Skyrme model \cite{Battye:2013tka,Halavanau:2013vsa}.
\item \emph{Formation of new solution types:}  Isospinning Skyrmion solutions can deform into configurations that do not exist at vanishing $\omega$ or are only metastable at $\omega=0$. An example is the tetrahedral $B=3$ Skyrmion (with $\mu=1.5$) which evolves with increasing $\omega$ into a ``pretzel\,''-like configuration -- a state that is only metastable at $\omega=0$ \cite{Walet:1996he,Battye:1996nt}.  
 \item \emph{Lifting of energy degeneracies:} Adding isospin can remove energy degeneracies. For example isospinning $D_{6d}$ and $D_{4h}$ symmetric Skyrme solitons about their $\boldsymbol{\widehat{K}}=(0,0,1)$ axes results for the configuration with approximate $D_{4h}$ symmetry in a lower energy value than found for isospinning $D_{6d}$ solitons, thereby removing the degeneracy.
\item \emph{Spin generated from Isospin:} 
If $W_{ij}$ is nonzero, then Skyrme configurations will obtain classical spin when isospin is added. For example for $B=1$ this gives states with spin and isospin opposite, as required by the Finkelstein-Rubinstein constraints \cite{Krusch:2002by}. For $B=2$ we observe that Skyrme configurations with nonzero mixed inertia tensor $W_{ij}$ show centrifugal effects and separate out whereas states with $W_{ij}=0$ tend to stay more compact. Isospinning around 
$\boldsymbol{\widehat{K}}=(0,0,1)$ leads to the breakup into two $B=1$ Skyrmions orientated in the
attractive channel and with $L_{\text{crit}}$ given by approximately $-K_{\text{crit}}$. Isospinning around $\boldsymbol{\widehat{K}}=(0,1,0)$ leads to a novel configuration with $D_4$ symmetry of vanishing total spin for all classically allowed isospin values $K$. 
\end{enumerate}

Finally, we also investigated numerically the critical behavior of isospinning Skyrmion solutions. Contrary to previous numerical studies on isospinning soliton configurations in the baby Skyrme \cite{Battye:2013tka,Halavanau:2013vsa} and Skyrme-Faddeev model \cite{Harland:2013uk,Battye:2013xf} we found numerically only one type of instability being present. Skyrmion solutions can isospin up to a critical angular frequency $\omega_{\text{crit}}$ that is given by the meson mass of the model. For $\omega>\omega_{\text{crit}}$ Skyrmion solutions become unstable to pion radiation. Recall that at $\omega_{\text{crit}}$ the values of the energy and angular momenta are finite, and therefore, the corresponding angular momenta $K_{\text{crit}}$ (and $L_{\text{crit}}$) are also finite, see Ref.~\cite{Battye:2005nx}. The situation is different for baby Skyrmions where energy and moment of inertia diverge at $\omega_{\text{crit}}$ \cite{Piette:1994mh}.

In Table~\ref{Tab_Kcrit_Skyrme} we give a classical bound on how fast Skyrme configurations of baryon number $B=1-4,8$ for given pion mass $\mu$ and isorotation axis $\boldsymbol{\widehat{K}}$ are allowed to isospin. For simplicity, we took the following approach: We fixed the Skyrme parameters $e$ and $F_\pi$ as in Refs. \cite{Adkins:1983ya,Adkins:1983hy} and calculated the induced pion mass via $m_\pi=2\mu /eF_\pi$. The critical isospin values $K_{\text{crit}}$ up to which stationary solutions exist are given in $\hbar$ units, where  $\hbar=46.8$ for the standard parameter set $F_\pi=108\,\text{MeV}$ and $e=4.84$. The associated energy values are obtained by multiplying our numerical energy values by the energy scale $F_\pi/4e= 5.58$ MeV. Note that our findings are in qualitative agreement with earlier literature \cite{Battye:2005nx}. In particular, we find that the nucleon ($E^N_{\text{tot}}=939$ MeV) and delta ($E^\Delta_{\text{tot}}=1232$ MeV) cannot be reproduced if the pion mass mass value is set to its experimental value $m_\pi=138$ MeV ($\mu=0.526$) and the standard values for the Skyrme parameters $e$ and $F_\pi$ are used. However, if the pion mass is taken to be larger than its experimental value, then we can reproduce nucleon and delta masses. Note that the calibration of the Skyrme model is still an open problem \cite{Battye:2005nx,Manton:2006tq}.

The types of deformations observed in this article have been largely ignored in previous work \cite{Manko:2007pr,Battye:2009ad} on modeling nuclei by quantized Skyrmion solutions and are exactly the ones we would like to take into account when quantizing the Skyrme model. Spin and isospin quantum numbers of ground states and excited states have so far almost exclusively been calculated within the rigid body approach \cite{Irwin:1998bs,Krusch:2002by,Krusch:2005iq,Lau:2014sva}, that is by neglecting any deformations and symmetry changes due to centrifugal effects. Our numerical full field simulations clearly demonstrate the limitations of this simplification. The symmetries of isospinning soliton solutions can change drastically and the solitons are found to be of substantially lower energies than predicted by the rigid-body approach. This work offers interesting new insights into the classical behavior of Skyrmions and gives an indication of which effects have to be taken into account when quantising Skyrmions.

\section*{Acknowledgements}
This work was undertaken on the COSMOS Shared Memory system at DAMTP, University of Cambridge operated on behalf of the STFC DiRAC HPC Facility. This equipment is funded by BIS National E-infrastructure capital grant ST/J005673/1 and STFC grants ST/J001341/1, ST/H008586/1, ST/K00333X/1. Many thanks to David Foster, Nick Manton, Yasha Shnir, Paul Sutcliffe and Niels Walet for useful discussions. This work was financially supported by the UK Engineering and Physical Science Research Council (grant number EP/I034491/1).

\appendix

\section{Skyrmion Inertia Tensor}\label{App_Inertia}

For completeness, we explicitly list in Tables~\ref{Tab_3}-\ref{Tab_7} static energies $M_B$, mean charge radii $<r^2\!>^{1/2}$ and all diagonal elements of the spin and mixed inertia tensors together with the off-diagonal inertia tensor elements for static Skyrmion solutions with $1\le B\le 4,\,8$ and rescaled mass parameter $\mu=1,1.5$ and 2. Note that the static energies and isospin inertia tensor diagonal elements  for $\mu=1$ have already been given in Table \ref{Tab_Stat_sky_low}.
\subsection{$\mu=1$}

\begin{table}[H]
\caption{We list the off-diagonal elements of the isospin $(U_{ij})$, spin $(V_{ij})$ and mixed $(W_{ij})$ inertia tensors, the mean charge radii  $<r^2\!>^{1/2}$ and the symmetries $G$ of the Skyrme solitons. The mass parameter $\mu$ is chosen to be 1.}
\centering 
\begin{tabularx}{\textwidth}{cXXXXXXXXXXc}
\hline\hline
 $B$& $G$  & $U_{12}$ &$U_{13}$ &$U_{23}$  & $V_{12}$ & $V_{13}$ & $V_{23}$&$W_{12}$ & $W_{13}$ & $W_{23}$&$<r^2\!>^{1/2}$\\
\hline
1& $O(3)$ &  0.0 &   0.0 &    0.0 &   0.0 &   0.0 &   0.0 &    0.0 &    0.0 &    0.0 &0.777\\
2& $D_{\infty h}$  & 0.0 & 0.0 &  0.0 &  0.6 &   0.2 &   0.2 &  0.0 &  0.0 & 0.0&1.055\\
3& $T_d$ & 0.0 & 0.0 &  0.0 &  0.2  &  0.2 &  0.2 & 0.0 &  0.0 &0.0&1.225\\
4& $O_h$ &  0.0 & 0.0 &   0.0 &  0.4 &  0.4 &  0.4  & 0.0 &  0.0  & 0.0&1.360\\
8 &$D_{6d}$&    0.0  &  0.1 & 0.0  &  3.8 &  9.6  &  9.5 &  0.1  & 0.1 &0.0&1.812\\
  &$D_{4h}$&    0.0 &  0.2 &  0.3  &  0.3 &  1.3  &  1.1  &  0.1  &  0.0 &  0.3&2.026\\
\end{tabularx}
\label{Tab_3}
\end{table}

\subsection{$\mu=1.5$}

\begin{table}[H]
\caption{Skyrmions of baryon number $B=1-4$ with pion mass value $\mu=1.5$. We list the energies $M_B$, the energy per baryon $M_B/B$, the diagonal elements of the inertia tensor $U_{ij},V_{ij},W_{ij}$ and the symmetries $G$ of the Skyrme solitons. Note that energies $M_B$ are given in units of $12\pi^2$.}
\begin{tabularx}{\textwidth}{cXXXXXXXXXXXX}
\hline\hline
 $B$& $G$  & $M_B$   &$M_B/B$&$U_{11}$ &$U_{22}$ &$U_{33}$ &$V_{11}$ &$V_{22}$ &$V_{33}$  & $W_{11}$ &$W_{22}$ &$W_{33}$  \\
\hline
1& $O(3)$ &  $1.530$ &  $1.530$ & 39.3 &  39.3 &  39.3&39.3 &39.3 &39.3 &39.3 &39.3&39.3\\
2& $D_{\infty h}$  &2.950  & 1.476 & 79.7 & 79.7 &57.1 &127.4 &127.4 &228.4 & 0.0 & 0.0 & 114.2\\
3& $T_d$   &   4.316     & 1.438 & 101.7&  101.7 & 101.7& 331.6 & 331.6 & 331.6 &70.3 &70.3 & 70.3\\
4& $O_h$  &  5.638& 1.410& 121.3 & 121.3 & 145.0 & 547.4 & 547.4 &547.4 & 0.0 & 0.0 &0.2\\
\end{tabularx}
\label{En_Sky_B1B4_m15}
\end{table}

\begin{table}[H]
\caption{We list the off-diagonal elements of the isospin $(U_{ij})$, spin $(V_{ij})$ and mixed $(W_{ij})$ inertia tensors,  the mean charge radii  $<r^2\!>^{1/2}$ and the symmetries $G$ of the Skyrme solitons. The mass parameter $\mu$ is chosen to be 1.5.}
\begin{tabularx}{\textwidth}{cXXXXXXXXXXc}
\hline\hline
 $B$& $G$  & $U_{12}$ &$U_{13}$ &$U_{23}$  & $V_{12}$ & $V_{13}$ & $V_{23}$&$W_{12}$ & $W_{13}$ & $W_{23}$ & $<r^2\!>^{1/2}$\\
\hline
1 & $O(3)$             &  0.0  &  0.0   &  0.0   & 0.0  & 0.0  & 0.0 &  0.0  &  0.0 &   0.0 &0.684\\ 
2 & $D_{\infty h}$  &  0.0  & 0.0    &  0.0  &  0.7 &  0.4 &   0.4 &   0.0  & 0.0 &  0.0& 0.926\\
3 & $T_d$              &   0.0 &  0.0   &  0.0 &   0.4 &  0.4  &  0.4 &  0.0  & 0.0 &  0.0 & 1.070\\
4 & $O_h$              &  0.0  & 0.0    & 0.0   &  0.8 &  0.8  &  0.8 &  0.0 &  0.0   & 0.0 & 1.185\\
\end{tabularx}
\label{Off_Sky_B1B4_m15}
\end{table}

\subsection{$\mu=2$}

\begin{table}[H]
\caption{Skyrmions of baryon number $B=1-4$ with pion mass value $\mu=2$. We list the energies $M_B$, the energy per baryon $M_B/B$, the diagonal elements of the inertia tensor $U_{ij},V_{ij},W_{ij}$ and the symmetries $G$ of the Skyrme solitons. Note that energies $M_B$ are given in units of $12\pi^2$.}
\begin{tabularx}{\textwidth}{cXXXXXXXXXXXX}
\hline\hline
 $B$& $G$  & $M_B$   &$M_B/B$&$U_{11}$ &$U_{22}$ &$U_{33}$ &$V_{11}$ &$V_{22}$ &$V_{33}$  & $W_{11}$ &$W_{22}$ &$W_{33}$  \\
\hline
1& $O(3)$ & 1.640 & 1.640 & 34.2 &  34.2  &  34.2 & 34.2 & 34.2 &  34.2 &34.2 & 34.2 &34.2\\
2& $D_{\infty h}$  &3.172 &1.587 &  69.2 & 69.2 & 49.6 &   110.9 &110.9  &  198.3 & 0.0 & 0.0 & 99.2\\
3& $T_d$ &  4.648 & 1.549 & 88.0 & 88.0 & 88.0 & 286.9 & 286.9 & 286.9 & 60.8 & 60.8 & 60.8\\
4& $O_h$  & 6.079 & 1.521 &  104.7 &  104.7 &   125.2  & 472.7  &  472.7  & 472.7 &0.0 & 0.0 & 0.3\\
\end{tabularx}
\end{table}

\begin{table}[H]
\caption{We list the off-diagonal elements of the isospin $(U_{ij})$, spin $(V_{ij})$ and mixed $(W_{ij})$ inertia tensors,  the mean charge radii  $<r^2\!>^{1/2}$ and the symmetries $G$ of the Skyrme solitons. The mass parameter $\mu$ is chosen to be 2.}
\begin{tabularx}{\textwidth}{cXXXXXXXXXXc}
\hline\hline
 $B$& $G$  & $U_{12}$ &$U_{13}$ &$U_{23}$  & $V_{12}$ & $V_{13}$ & $V_{23}$&$W_{12}$ & $W_{13}$ & $W_{23}$ & $<r^2\!>^{1/2}$\\
\hline
1& $O(3)$ &    0.0  &  0.0 &  0.0  & 0.0  & 0.0  &0.0  &  0.0 &  0.0 &  0.0 & 0.617\\
2& $D_{\infty h}$  &  0.0 &   0.0 &  0.0  &  1.2 &  0.6 &  0.6 & 0.0 &0.0 &  0.2 & 0.834\\
3& $T_d$ & 0.0  &  0.0 &  0.0  &  0.6 &   0.6 &   0.6  &  0.0 &  0.0 & 0.0 & 0.042\\
4& $O_h$ & 0.0&   0.0  &  0.0 &  1.2 &  1.2  &  1.2 &   0.0  & 0.0 & 0.0 &  1.062\\
\end{tabularx}
\label{Tab_7}
\end{table}


\end{document}